%% file: ms.tex
\documentclass[11pt,preprint]{aastex}

\usepackage{gensymb}

\input{mycommands}
\shorttitle{A Blind OH Survey with the GBT}
\shortauthors{Ronald J.\ Allen, David E. Hogg, \& Philip D. Engelke}

\begin{document}


\title{
THE STRUCTURE OF DARK MOLECULAR GAS IN THE GALAXY - I \\
A Pilot Survey for 18-cm OH Emission Towards $l \approx 105^{\circ}, b \approx +1^{\circ}$\\[0.2in]}

%  History of this document:
%  -------------------------
%    V1 early 2014 - RJA
%    V2 May 2014 - RJA
%    V3 Sept 2014 - RJA - added Philip's comments
%    thru V8 Nov 2014 - RJA - various changes and additions/subtractions
%    "Submit" Version - RJA - 14 Nov 2014 - figure file names changed for AJ
%    Rev1 in December 2014 - address referee's comments + Miller's comments
%    Rev2 in mid January 2015 - incorporate additional suggestions from Dave Hogg.
%    "SubmitRevision" - Rev2 re-submitted to AJ on Jan 16, 2015.

\author{Ronald J.\ Allen}
\affil{Space Telescope Science Institute, 3700 San Martin Drive, Baltimore, MD 21218, and \\
Center for Astrophysical Sciences, Department of Physics \& Astronomy, \\ The Johns Hopkins University}
\email{rjallen@stsci.edu}

%\and

\author{David E.\ Hogg}
\affil{National Radio Astronomy Observatory, 520 Edgemont Road, Charlottesville, VA 22903, USA}
\email{dhogg@nrao.edu}

%\and

\author{Philip D.\ Engelke}
\affil{Department of Physics \& Astronomy, The Johns Hopkins University}
\email{pengelk1@jhu.edu}

\begin{abstract}
We report the first results from a survey for 1665, 1667, and 1720 MHz OH emission over a small region of the Outer Galaxy centered at $l \approx 105.0^{\circ}, b \approx +1.0^{\circ}$. This sparse, high-sensitivity survey ($\Delta T_A \approx \Delta T_{mb} \approx 3.0-3.5$ mK rms in 0.55 \kmps\ channels), was carried out as a pilot project with the Green Bank Telescope (FWHM $\approx 7.6'$) on a $3 \times 9$ grid at $0.5^{\circ}$ spacing. The pointings were chosen to correspond with those of the existing \twCO(1-0) CfA survey of the Galaxy done at a similar resolution ($8.4'$). With 2-hr integrations, 1667 MHz OH emission was detected with the GBT at $\gtrsim 21$ of the 27 survey positions ($\geq 78\%$), confirming the ubiquity of molecular gas in the ISM as traced by this spectral line. With few exceptions, the main OH lines at 1665 and 1667 MHz appear in the ratio of 5:9 characteristic of LTE at our sensitivity levels.  No OH absorption features are recorded in the area of the present survey, in agreement with the low levels of continuum background emission in this direction. At each pointing the OH emission appears in several components extending over a wide range of radial velocity and coinciding with well-known features of Galactic structure such as the Local Arm and the Perseus Arm. In contrast, little CO emission is seen in the survey area; less than half of the $\gtrsim 50$ identified OH spectral features show detectable CO counterparts at the CfA sensitivity levels, and these are generally relatively faint. There are no CO features without corresponding OH emission in our survey. With very few exceptions, peaks in the OH profiles coincide with peaks in the GBT \HI\ spectra (obtained concurrently, FWHM $8.9'$), although the converse is not true. We conclude that main-line OH emission is a promising tracer for the ``dark molecular gas'' in the Galaxy discovered earlier in Far-IR and gamma-ray emission, although further work is needed to establish the quantitative details of the connection. Further aspects of the OH $\Leftrightarrow$ CO relation are revealed in a scatter plot of CO vs.\ OH line strengths. This plot suggests a rough proportionality between the bright envelope of the CO emission (when present) and level of the accompanying 1667 MHz OH emission. Finally, we note several cases of anomalous OH emission. One survey position shows several narrow OH spectral features which are not well correlated with the \HI\ spectrum; these features have been identified with a nearby known OH-IR star. Also, nine neighboring survey positions show enhanced emission at 1720 MHz, consistent with earlier observations and with models involving extended regions of elevated particle density.

\end{abstract}


\section{Introduction}
\label{sec:intro}

The 18-cm radio lines of the OH molecule have been widely observed in the ISM since the first detection more than 50 years ago in absorption against the bright continuum emission from Cas A \cite[][]{wbmh63}. The first major survey for OH in the Galaxy was also done in absorption \cite[][]{g68} against more than two dozen well-known thermal and nonthermal radio continuum sources in the northern sky. Within a year or two of the detection in absorption, OH emission from several discrete sources in the Galaxy was announced \cite[][]{wwdl65}, but it was the unexpected appearance of intense and narrow spectral lines of ``mysterium'' near the OH emission lines in several of these sources that captured the most attention at the time. We now recognize these features as arising from OH molecules in the ISM which are being excited to rotational levels above the ground state by localized non-LTE energetic processes, such as intense IR continuum from dust near hot young stars or high transient particle densities in supernova-driven shocks. Numerous studies of these bright ``OH maser'' sources in the Galaxy were undertaken in the years following the initial discoveries \cite[for a review, see e.g.][]{rm81}.

In contrast to studies of the relatively-bright discrete sources of OH maser emission, work on the extended OH emission in the ISM has been hampered by insufficient sensitivity. As summarized in \citet[][]{arb12}, early ``blind'' surveys for normal OH emission\footnote{In this context, ``normal'' (or ``thermal'') refers to emission with LTE line ratios of 5/9 in the main lines at 1665/1667 MHz. However, the excitation temperature of this emission is generally subthermal, i.e.\ below the kinetic temperature of the gas in which the OH molecules are embedded \cite[e.g.][]{g68}.} in the main lines of OH at 1665 and 1667 MHz were unsuccessful \cite[][]{pen64, kk72}, and by the early 1970's it appeared that, with the equipment available at the time, main-line OH emission from the extended ISM could be detected only in localized regions of high density in gas and dust \cite[][]{h68}. The first major extensive survey of the northern Galactic plane in all four OH lines was done by \citet[][]{t79} over more than a 6-year period with the 140-ft NRAO telescope beginning in 1970\footnote{That paper also includes in Table 1 a list of related OH surveys carried out in the period 1966-1976.}. Turner's survey was mainly directed at known sources from many different catalogs of stellar optical/IR objects or  emission/absorption nebulae; only a small fraction of the pointings were not directed towards specific targets. About half of the nearly 1800 pointings show detectable OH, often as a mix of emission and absorption. The large velocity extent and roughly-LTE intensity ratios of many main-line absorption features indicates that these arise in the general Galactic ISM subject to Galactic rotation and embedded in the extended Galactic nonthermal radio continuum emission. The emission features in Turner's data appear to be a more eclectic mix of narrower, brighter lines characteristic of various maser sources. There are perhaps only one or two dozen pointings among the 732 emission sources in Turner's survey for which the two main lines appear to be  visible and are plausibly in the expected LTE ratio of 1.8, although the profiles themselves are too faint to yield definitive results. In fact, we now know that the sensitivity of Turner's survey was too low to reveal main-line OH emission in LTE from the general ISM: Turner gives the $5\sigma$ sensitivity of his survey as 0.18 K in $\Delta T_A$, which becomes 240 mK in $\Delta T_{mb}$ units appropriate for extended sources observed with the 140-ft telescope. However, typical peak values for OH main-line profiles in the envelopes of molecular clouds are 50-100 mK \cite[][]{waf93} in the same units, and more recent observations of nearby high-latitude diffuse cirrus clouds show profiles with peak values of only 10-20 mK outside of the CO-bright regions \cite[][]{bjl10}.

Our understanding of the distribution of OH in the Galaxy has recently been greatly improved by the results of two recent surveys. First, the SPLASH survey \cite[][]{dwj14} mapped OH in an area of $\Delta l \times \Delta b = 10^{\circ} \times 4^{\circ}$ along the plane of the Inner Galaxy in the southern sky with the Parkes Telescope, and found that OH was widely detected in all four of the transitions which were observed. Because the brightness of the radio continuum emission concentrated in the Galactic plane is often close to the OH excitation temperature, the profiles show a mix of emission and absorption. Features of Galactic structure were observed, but it was not possible to detect diffuse OH beyond the central CO-bright regions of molecular complexes to a typical sensitivity limit of $\Delta T_{mb} \approx 30$ mK ($3\sigma$). The detailed observation in the four transitions allowed the authors to detect and identify many OH masers, and to evaluate the excitation conditions in the diffuse OH gas.

The second survey \cite[][]{arb12, arb13} examined the diffuse 1667 MHz OH emission in a small $\Delta l \times \Delta b = 1.0^{\circ} \times 4^{\circ}$ region of the Outer Galaxy with a greater sensitivity of $\Delta T_{mb} \approx 15$ mK ($3\sigma$). The radio continuum background from the Galaxy is significantly fainter here. The correspondence between features in the diffuse OH emission and the neutral hydrogen at the same position was good, but fewer than 10\% of the OH emission features were found in the CfA Survey of the \twCO(1-0) emission in the Galaxy \cite[][]{dht01}. This may be significant because of reports over that past decade of evidence for ``dark gas'' in the Galaxy first reported by \citet[][]{gct05}. In its apparent association with \HI, the overall morphology of the OH emission resembled that of the dark gas \cite[for a recent summary, see][]{lvp14}, a potentially important connection since the original tracers of that dark gas (Far-IR and gamma-ray emission) did not have discrete spectral features and hence could not provide distance information based on Doppler velocities and Galactic rotation.

The present survey has been undertaken as a pilot project with the NRAO Green Bank Telescope (GBT) in part to further explore the apparent connection of the extended OH emission to the dark gas in the Galaxy, and in part to begin a study of the 3D spatial structure of this component of the molecular ISM. Owing to spectrometer limitations, and to the presence of low-level interference in the spectra, the previous blind survey carried out at Onsala by \citet[][]{arb12, arb13} was limited to the 1667 MHz OH line and to distances within $\approx 2 $ kpc of the Sun. The present GBT survey, centered on a slightly different direction but still towards the Outer Galaxy, records both main lines and takes advantage of the superior sensitivity and spectral coverage of the GBT receivers as well as the low interference levels of the National Radio Quiet Zone where the GBT is located. Finally, the exact correspondence of our OH survey pointing positions to those of the CfA CO survey, plus the nearly-identical angular resolutions of the two data sets, allows a straightforward observational comparison of these two tracers of molecular gas in the Galactic ISM.

\section{Observing Program}
\label{sec:observations}

\subsection{Survey construction}

Our pilot survey is located in an area close to the Galactic plane in the second quadrant. It is a \textit{blind} survey, not directed at any specific previously-identified astronomical objects such as \HII\ regions, supernova remnants, etc. However, the location has been chosen for several other reasons. First, the line of sight traverses several distinct features of Galactic spiral structure including the Perseus and Outer arms. Second, the relation between radial velocity and distance from the Sun is very nearly linear, allowing unambiguous determinations of distances and sizes from the mean velocities and velocity widths of spectral features. Third, the ambient Galactic radio continuum emission near 18 cm wavelength is very faint in this direction ($< 1$K) and generally below the range of excitation temperatures observed for OH in emission ($\approx 4-10$K; \citet[][]{h69,rwg76,c79}), thereby optimizing the detectability of faint OH\footnote{More recent work \cite[][]{ll96} has shown that the OH detected primarily in absorption profiles has significantly lower excitation temperatures, close to the 2.8 K cosmic background. However, for better or worse, this component contributes very little to the OH emission. The fraction of OH in this state is presently unknown.}. Finally, there are areas of the sky here where the \twCO(1-0) emission is very faint or even absent from the CfA survey, enhancing the opportunity to learn more about the ``CO-dark'' component of molecular gas in the Galaxy.

As described in \S \ref{sec:intro}, any program aimed at a quantitative measurement of OH spectral features will require long integration times in order to achieve sufficient sensitivity for parameter estimation on profiles with very modest S/N. As an example, for their study of OH main-line emission in the translucent high-latitude cloud MBM40 with the GBT, \citet[][]{cmw12} spent about 1/2 hour of integration at each pointing, but even then they did not achieve enough S/N to use the 1665 MHz profiles. An improvement in S/N of at least a factor of 2 is required to render the 1665 data useful, providing additional independent estimates of profile parameters and helping to identify any profiles exhibiting anomalous excitation. We conclude that the integration time requirement is at least $\approx \! 2$ hours per pointing for the main-line OH spectra with the GBT. Without further improvements in receiver sensitivity, even longer integration times would be required to measure LTE-level emission from the 1612 and 1720 MHz satellite lines, and we have not considered that possibility here. However, the GBT receivers are indeed capable of simultaneous recording of one or even both of these satellite lines. We have therefore obtained spectra of the 1720 MHz line as well, since enhanced emission from this line has been observed widely in the Galactic ISM \cite[][]{t82}. The results of the Onsala survey also showed that the OH emission has an extent similar to the \HI\ both in space and in velocity; hence \HI\ profiles at the same pointing positions are also of interest. Fortunately, the GBT receivers provide that capability, although the 21-cm \HI\ line is so bright that integration times of only a few minutes per pointing were sufficient and were therefore carried out separately.

In the absence of prior information on the typical linear scale sizes of OH features, the grid spacing of $0.5^{\circ}$ is a compromise among competing factors including sky coverage, integration time, and the limited clock time available on the telescope for such programs. An important result of the present survey will be information on the spatial structure of OH emission which we detect and a comparison of that with the SPLASH survey \cite[][]{dwj14}, which is providing structural information on the combination of OH emission and absorption at smaller angular scales ($\approx \! 15'$) with somewhat lower sensitivity in the Southern Hemisphere (Inner Galaxy). The number of pointings is similarly time-limited, but here there are some guidelines available from the Onsala results. A reasonable goal is to obtain $\approx \! 100$ OH spectral components for a statistically-meaningful comparison with the corresponding \twCO\ profiles. The mini-survey by \citet[][]{arb12} consisted of 81 profiles, each with several sub-components. However, the separation of profile sub-components at each pointing was severely compromised by low S/N; the few high S/N pointings in that mini-survey show as many as 3 OH components. With sufficient S/N at each pointing, $\approx \! 100$ OH components could perhaps be obtained from $\approx \! 30$ pointings.
%
%----------------------------------
\begin{figure*}[ht!] % order of placement preference: here, top
%\epsscale{0.9}
\vspace{0.1in}
\begin{center}
\includegraphics[width=0.4\textwidth, angle=0]{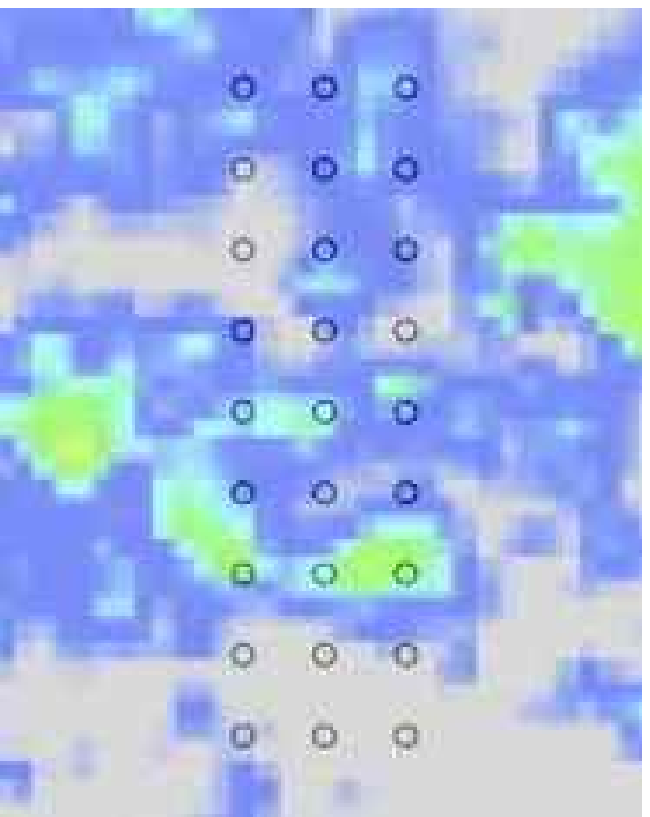} \hspace{0.2in} \includegraphics[width=0.410\textwidth, angle=0]{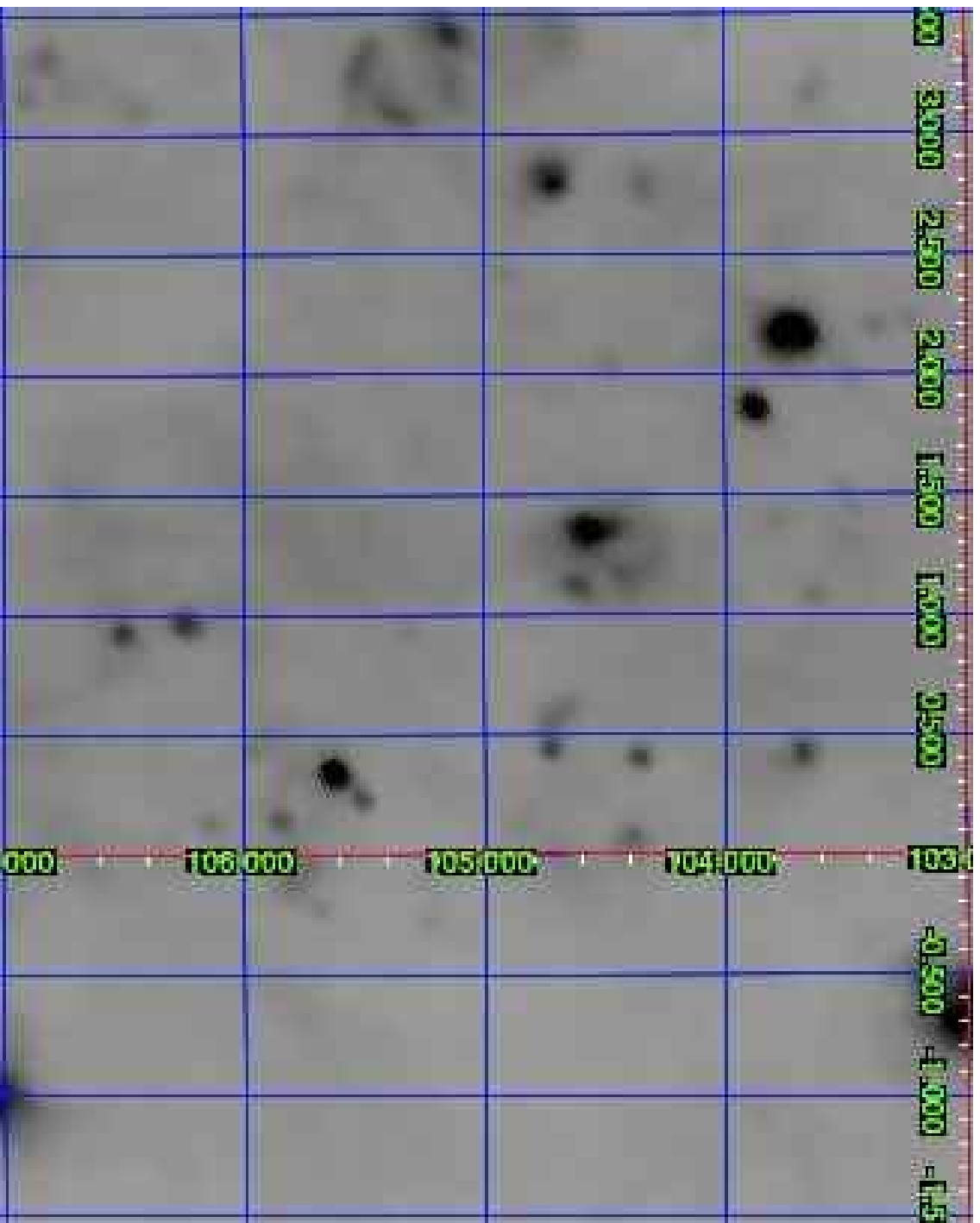} \vspace{-0.1in} 
\caption{
The survey area in two tracers of Galactic emission. Left: The observing position grid is shown superposed on the relevant parts of the CfA \twCO(1-0) emission survey of \citet[][see their Figure 2]{dht01}. The $3 \times 9$ grid is centered at $l = 105.0^{\circ}, b = +1.0^{\circ}$ with spacing $0.5^{\circ}$. The diameters of the circles approximately indicate the telescope beam FWHM. Right: The survey area in 100 $\mu$ IRAS emission at the same scale, from the IRIS re-processing data base at IRSA \cite[][http://irsa.ipac.caltech.edu]{ml05}. The brightest source in this field is $\approx 640$ MJy/sr at $l = 103.74^{\circ}, b = +2.17^{\circ}$. The source nearest to a survey position is $\approx 410$ MJy/sr at  $l = 104.58^{\circ}, b = +1.35^{\circ}$. Grid lines are at intervals of $\Delta l = 1.00^{\circ}, \Delta b = 0.50^{\circ}$.
\label{fig:survey}
}
\end{center}
\end{figure*}
%-------------------------------------------

\vspace{-0.2in}

With the previous considerations in mind, the parameters of our pilot survey are as follows: the survey covers a $3 \times 9 = 27$-point grid centered at $l = 105.0^{\circ},\ b = +1.0^{\circ}$ with a grid spacing of $0.5^{\circ}$. The total integration time is 2 hours per pointing for the OH lines and 5 minutes for \HI. Figure \ref{fig:survey}a shows the survey grid positions superposed on a small portion of the CfA \twCO(1-0) survey. The small circles at each grid point show the approximate FWHM angular resolution of all 3 data sets discussed here, OH ($7.6'$), CO ($8.4'$), and \HI\ ($8.9'$). Figure \ref{fig:survey}b shows the $100\mu$ IRAS emission in the area.

\subsection{Observations}

The observations were made with the Robert C. Byrd Green Bank Telescope (GBT), using the Gregorian receiver system operating in the the frequency band 1.15 - 1.73 GHz (``L-band''). The main beam efficiency of this instrument is 0.95 in this frequency band, as determined by NRAO staff \cite[][Figure 1]{m09}, and hence the antenna and main-beam brightness temperatures $T_{mb} = T_A/0.95$ are virtually identical. Results given in this paper use both scales, but the choice will be clarified in the text, caption, or axis legend. The receiver is a single-beam, dual-polarization instrument, having an effective system temperature in the range 16 - 20 K, depending upon the weather and the level of the Galactic background continuum emission. The signal is fed to the control room by means of an IF system for which we chose a bandwidth of 12.5 MHz in order to minimize the possibility of harmful radio interference. The signal from the IF system was copied and directed to two sections of the GBT Spectrometer, one centered at 1666.4 MHz, the other at 1720.5 MHz. Because of the anticipated weakness of the OH line at 1612 MHz, and because of the greater likelihood of interference at that frequency, we did not observe that transition. Each 12.5 MHz spectral window supported both (linear) polarizations, and the sampling of the autocorrelation data at 9 levels and subsequent Fourier transformation provided 8192 channels of spectral data sampled at a frequency interval of 1.526 kHz with a frequency resolution of 1.846 kHz. The IF band centered at 1666.4 MHz is sufficiently wide to record both of the ``main'' lines of OH at 1665.4018 and 1667.3590 MHz; these were separated later during the data reduction phase.

Because of the expectation that there was likely to be widespread OH emission at a low level everywhere in the survey area \cite[][]{arb12}, we chose to use frequency-switching as the basis for removing the instrumental signature from the data rather than position-switching. The switching was made by moving the center frequency of the 12.5 MHz band by $\pm 2$ MHz from the nominal center frequency and subtracting the shifted spectrum. The wide IF bandwidth in use meant that with this scheme the signal spectrum appears in both switching cycles, but in different channel numbers. The spectra from the two cycles were therefore shifted and inverted (``folded'') before subtraction so as to average the signal data from both parts of the switching cycle. While optimizing the S/N and avoiding the possible subtraction of signal still present in a reference field, this technique samples a different part of the IF bandpass during the two parts of the switching cycle and hence can leave residual baseline structure in the output spectrum. We explored the use of an additional reference position located at a high Galactic latitude, but this required significant time and movement of the telescope and little improvement to the baselines was obtained from it. In addition, this runs the risk of inadvertently subtracting signals from very local gas. We did balance the receivers before each 10 minute scan, and this appeared to offer some improvement in spectral baseline stability. Scans were accumulated such that the total integration time for each survey position in the OH lines was $\approx 2$ hours. A separate observation of \HI\ at each pointing position was made; the instrumental configuration was the same as for the OH observations, except that the center frequency was set to be 1420.4 MHz, and only one scan of 5 minutes duration was required because of the great strength of the \HI\ signal. The main observing program was completed in 14 sessions of $\approx \! 5$ hours each between September 2013 and January 2014. The pointing of the telescope was checked at the start of each session using standard procedures for the GBT.

\subsection{Data reduction}
\label{sec:datareduction}

The \HI\ observations required no correction for baselines at the levels of interest here. Also in general only one of the two polarizations was reduced since the \HI\ line is known to be unpolarized. The OH data required more careful treatment, and each polarization in each 10-minute OH scan was reviewed for the presence of interference or other instrumental problems. In general, the quality of the data was very high, and little editing was required. The data for each polarization was separately averaged and archived. The data for the two polarizations were then averaged, a baseline was subtracted, the resulting profile was smoothed by two channels with a Gaussian convolution (without decimation), and the final result was archived for analysis. The baseline was determined with a polynomial fit over the velocity range of $ - 320$ to $+ 200$ \kmps\ (approximately 1900 channels) in segments chosen to avoid signal peaks. The order of the polynomial was progressively increased until the improvement in the fit was negligible; a value of 5 was typical for the 1667 spectra. The resultant rms noise level in the OH spectra is typically 3.0 - 3.5 mK in $T_{mb}$ units. The final spectral resolution of 3.05176 kHz corresponds to 0.64, 0.55, and 0.53 \kmps\ at the \HI, 1665/1667 MHz OH, and  the 1720 MHz OH lines, respectively. Note that the channel spacing in the final smoothed but undecimated spectra is half the quoted spectral resolution.

\section{Survey Results}
\label{sec:results}

\subsection{General features of the \HI, OH, and \twCO(1-0) line profiles}

Figure \ref{fig:stackedplots1} shows a mosaic of results from this pilot survey along with the corresponding \twCO\ spectra from the CfA survey \cite[][]{dht01}. For each position in the $3 \times 9 = 27$-point grid we present 3 spectra aligned in radial velocity above each other in a single ``stacked-plot'' panel, with \HI\ at the top, 1667 MHz OH in the middle, and \twCO(1-0) at the bottom. As indicated on the plots, the y-axis amplitude scales are in units of $T_{A} = 0.95 \times T_{mb}$ for the GBT \HI\ and OH spectra, and $T_{mb}$ for the CfA CO spectra. This figure shows several major features of large-scale Galactic structure at this $l, b$ location including the Local Arm, the Perseus Arm, and (only in \HI) the Outer Arm. As a specific example, consider the three spectra for $l = 104.5^{\circ},\ b = +2.0^{\circ}$ on the right side of Fig.\ \ref{fig:stackedplots1}c. The \HI\ profile (top panel) extends over almost the entire velocity range observed, from $-130 \lesssim V_{LSR} \lesssim +15$ \kmps, with five distinct peaks near $V_{LSR} \approx -2$ \kmps\  (the immediate solar vicinity), $V_{LSR} \approx -20$ \kmps\ (the Local Arm), $V_{LSR} \approx -50$ \kmps\ and $V_{LSR} \approx -65$ \kmps\ (the Perseus Arm), and $V_{LSR} \approx -100$ \kmps\ (the Outer Arm). The 1667 MHz OH spectrum shows features one can identify with the solar vicinity, the Local Arm, and the Perseus Arm. Noteworthy is the fact that in general every OH peak has an associated \HI\ peak at essentially the same radial velocity, although the converse is not true at present levels of OH sensitivity. Cases where OH peaks appear without a corresponding \HI\ peak are rare and can often be identified with OH maser sources, as we shall describe in \S \ref{sec:OHIRstar}. Note also that, at this particular pointing, faint CO emission features can be identified near 0, -20, and -50 \kmps. We will discuss the quantitative comparison of OH and CO in \S \ref{sec:OHandCOcomparison}.

\vspace{-0.2in}

\subsection{Variations in profile structure on scales of $0.5^{\circ}$}

The profiles displayed in Figure \ref{fig:stackedplots1} show significant changes over the angular scale of $0.5^{\circ}$ used in the survey sampling. The changes are less pronounced for the features at $V_{LSR} \approx 0$ \kmps\ because even the cores of molecular clouds would be expected to subtend angles of $0.5^{\circ}$ if they are at distances of 200 pc. The features which arise from molecular clouds in the Perseus Arm show much greater variation. As an example, the strong feature seen at approximately -50 km/s at position $l = 105.0^{\circ},\ b = +1.0^{\circ}$ is weaker by a factor of four in three of the four adjoining pixels. It continues to be visible at $l = 105.0^{\circ},\ b = +0.5^{\circ}$, but is actually stronger at $l = 104.5^{\circ},\ b = +0.5^{\circ}$, and a second feature has appeared at $-65$ \kmps\ in the latter spectrum. Recent determinations \cite[][]{chr14} measure the distance to the inner part of the Perseus Arm in this direction to be 3.2 kpc. The grid spacing in the present OH data therefore corresponds to a linear scale of $\approx 28$ pc, sufficient to resolve broad structure in a giant molecular cloud, but too coarse to resolve core structures in such clouds. In order to compare the OH molecular clouds with the more familiar CO molecular clouds, it will be necessary to map selected OH features on a finer grid, perhaps even down to the Nyquist interval of the GBT beam.

\subsection{Gaussian fits to discrete features in the 1667 MHz OH spectra}

Gaussian fits were found for all distinct OH features, although such fits were often a poor representation since many of the profile features are too complex to be effectively fitted by even multiple Gaussians. However, the Gaussian fits do provide rough values for profile parameters that are useful for discussion and for a general comparison with other ISM components. The 1667 MHz Gaussian fits have amplitudes ranging from 7 mK to 60 mK, with the brightest recorded profile feature peaking at 100 mK. The average amplitude is near 25 mK with a standard deviation of 15 mK. Profile widths show a wide range from  $1 - 10$ \kmps, with an average of $\approx 3.9$ \kmps\ and a standard deviation of 2.5 \kmps. We will discuss the line strengths based on direct integration of the profiles in \S \ref{profileanalysis}.

\subsection{1665/1667 OH emission line strength ratios}
\label{sec:lineratios}

The ratio of the main OH lines at 1665 and 1667 MHz is an important indicator for the excitation mechanism of OH molecules in the ISM, and may also be used as a measure of optical depth. The satellite lines at 1612 and 1720 MHz are quite sensitive to anomalous excitation by Far-IR photons and shocks, but the main lines at 1665 and 1667 MHz usually appear in the LTE ratio of 5:9 characteristic of optically-thin collisional excitation. Larger main-line ratios (up to 1:1) have been  interpreted as indicating higher optical depth \cite[][]{h69}, but cases of non-LTE behavior of the main lines have also been reported \cite[][]{c79}. However, accurately determining line ratios by comparing the velocity-integrated line strengths can be difficult with weak emission lines, as the result is sensitive not only to receiver noise but also to uncertainties in the baseline subtraction process. We will discuss this problem in more detail in \S \ref{baselineuncertainties}. Figure \ref{fig:stackedplotdiffs} illustrates a qualititative but in some ways more robust method for making this comparison. In this Figure we present as an example the OH spectra recorded at 3 adjacent pointings in the grid located at $l = 105.50^{\circ}$ (left column), $l = 105.00^{\circ}$ (middle column), and $l = 104.50^{\circ}$ (right column), but all at the same latitude of $b = +0.50^{\circ}$. This presentation is a variation on the ``stacked-plot'' presentations of Figure \ref{fig:stackedplots1}, only here each column shows the 1665 spectrum along the upper row, the 1667 MHz spectrum along the middle row, and the scaled difference spectrum assuming low-opacity emission lines in LTE (symbolically, $1665 - 1667/1.8$) along the bottom row. Note that the scaled difference spectra do not show features that echo the main line profiles (either $+$ or $-$), thereby strongly suggesting that the main lines are indeed generally in LTE even though the baseline levels may not be rigorously zero over the relevant velocity ranges. This is true even for profiles which show significant substructure on small scales in radial velocity, such as the example shown in the right column of Figure \ref{fig:stackedplotdiffs}. Those few pointings where the difference spectra clearly show relatively narrow features above the noise will be discussed in sections \ref{sec:OHIRstar} and \ref{sec:brightseventeentwenty}. Only one feature of the more than 50 in our survey shows evidence for high optical depth, and a second may be enhanced by a nearby source of Far-IR emission; these cases will be discussed in \S \ref{sec:non-lte-main-lines}.

%---------------------------
% Note that the mosaic of survey results in Figure 2 = fig:stackedplots1 appears as a float at the end of this manuscript. The figure numbering that follows here must therefore continue with figure 3. This is set by a "setcounter" statement.

\setcounter{figure}{2}  % this will get incremented to 3 at the next statement
%---------------------------

%---------------------------
\begin{figure*}[h!]
%\epsscale{1.0}
%\vspace{-0.1in}
%\begin{center}
\includegraphics[width=2.0in, angle=+0]{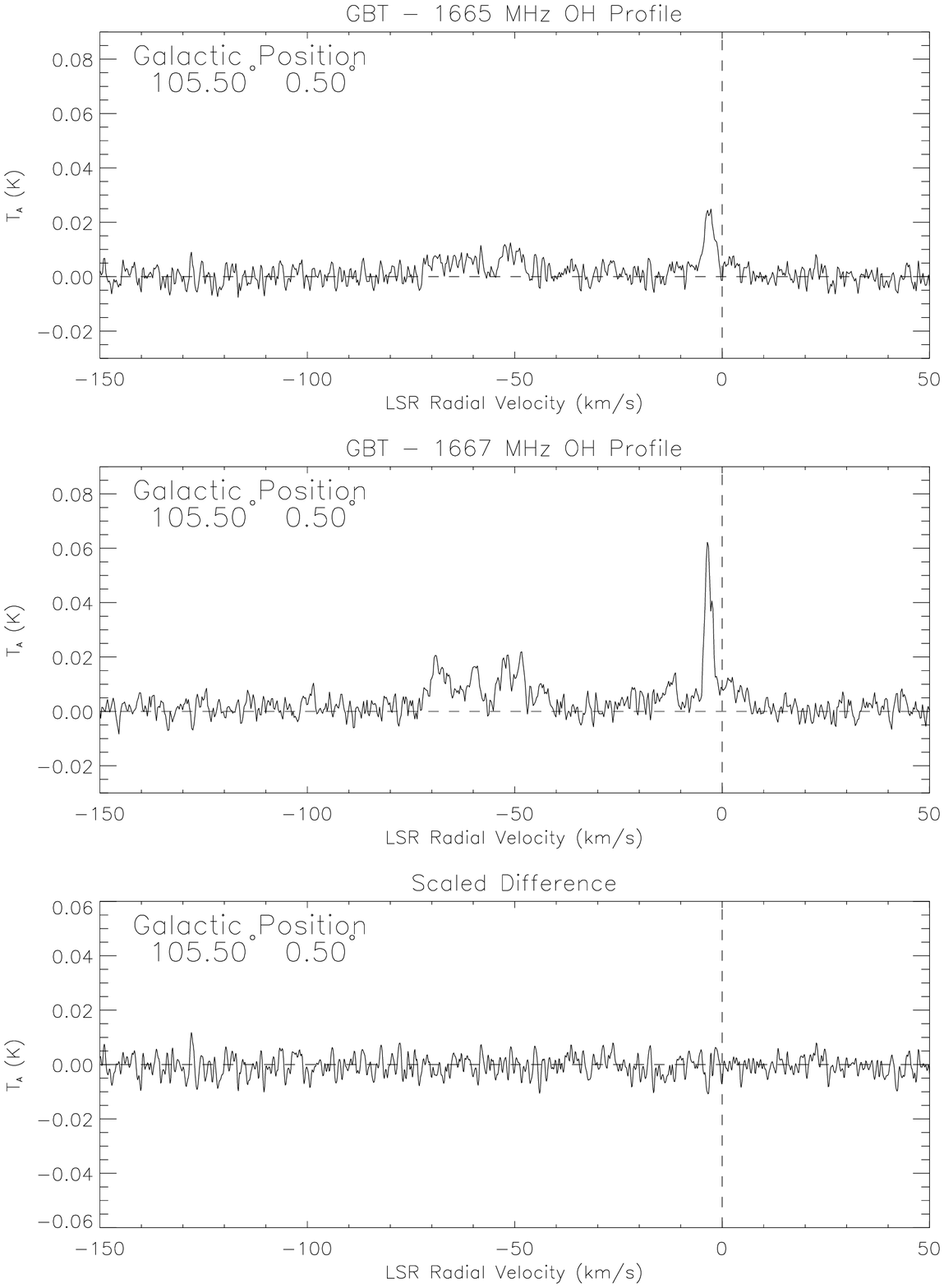}
\includegraphics[width=2.0in, angle=+0]{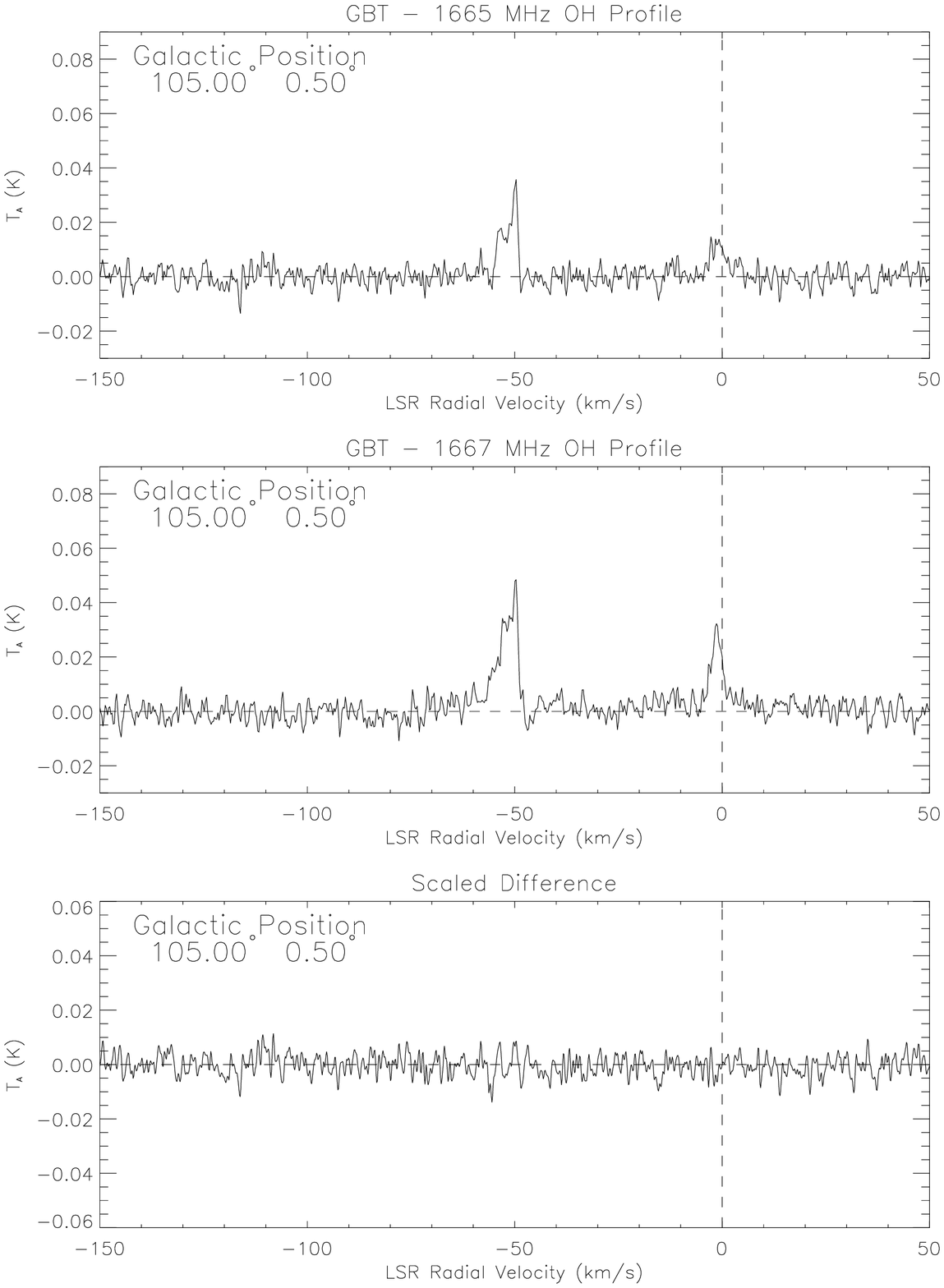}
\includegraphics[width=2.0in, angle=+0]{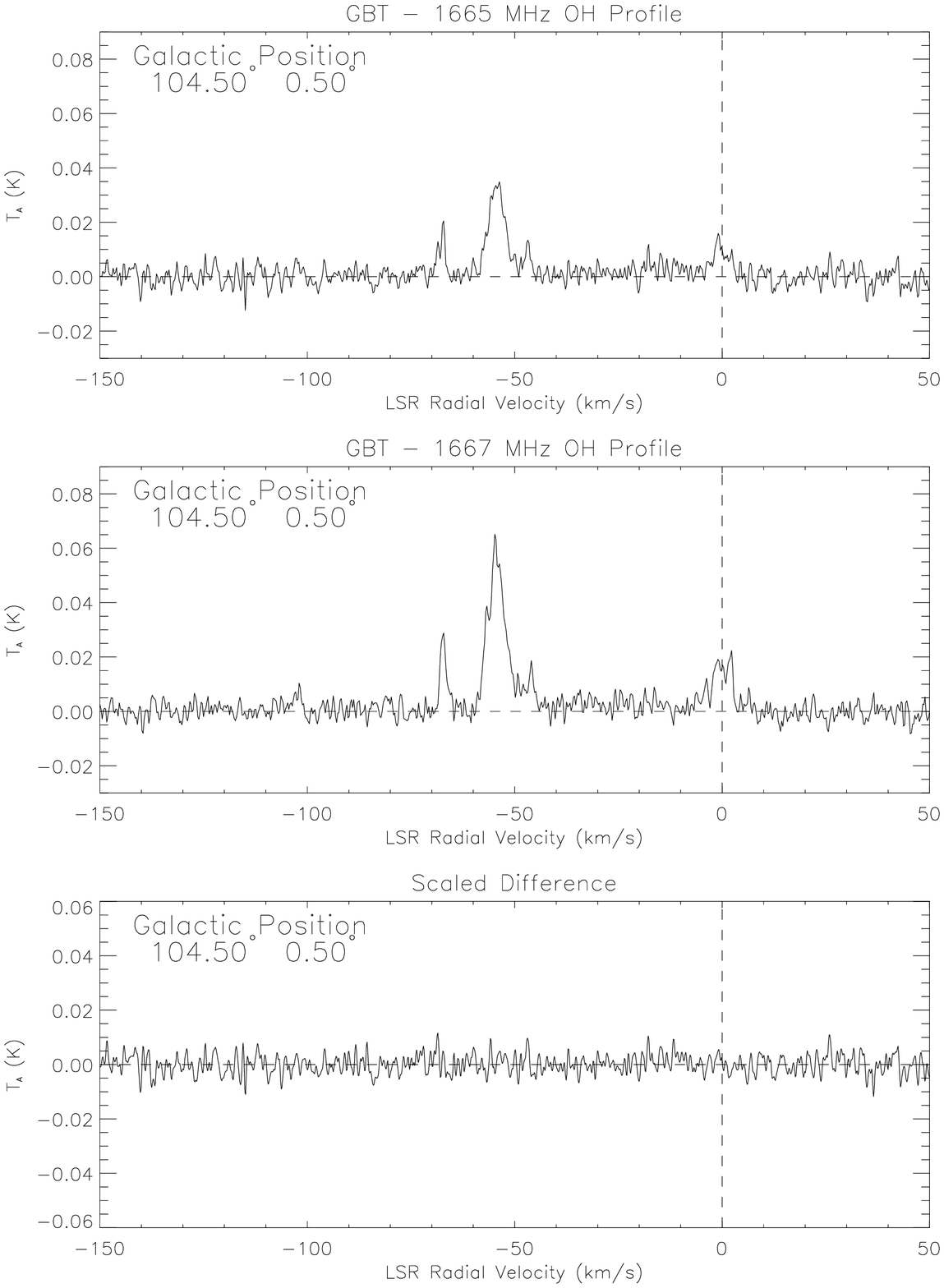}
\vspace{-0.1in}
\caption{\small Main-line OH intensity ratios for three adjacent pointings at $b = +0.50^{\circ}$ in the observed grid: $l = 105.50^{\circ}$, (left column), $l = 105.00^{\circ}$ (middle column); and $104.50^{\circ}$ (right column). The spectra shown are: 1665 MHz (top row), 1667 MHz (middle row), and the scaled profile differences $1665 - 1667/1.8$ (bottom row). Note that these scaled differences show no discrete features which mimic the original line profiles. \normalsize}
\label{fig:stackedplotdiffs}
%\end{center}
%\vspace{-0.1in}
\end{figure*}
%-----------------------------------------------------

\subsection{Miscellaneous noteworthy spectra}

\subsubsection{A deep exposure at $l = 105.5^{\circ},\ b = +3.0^{\circ}$}
\label{sec:deepexposure}

\begin{figure*}[t!]
%\epsscale{1.0}
%\vspace{-0.1in}
\begin{center}
\includegraphics[width=2.0in, angle=+0]{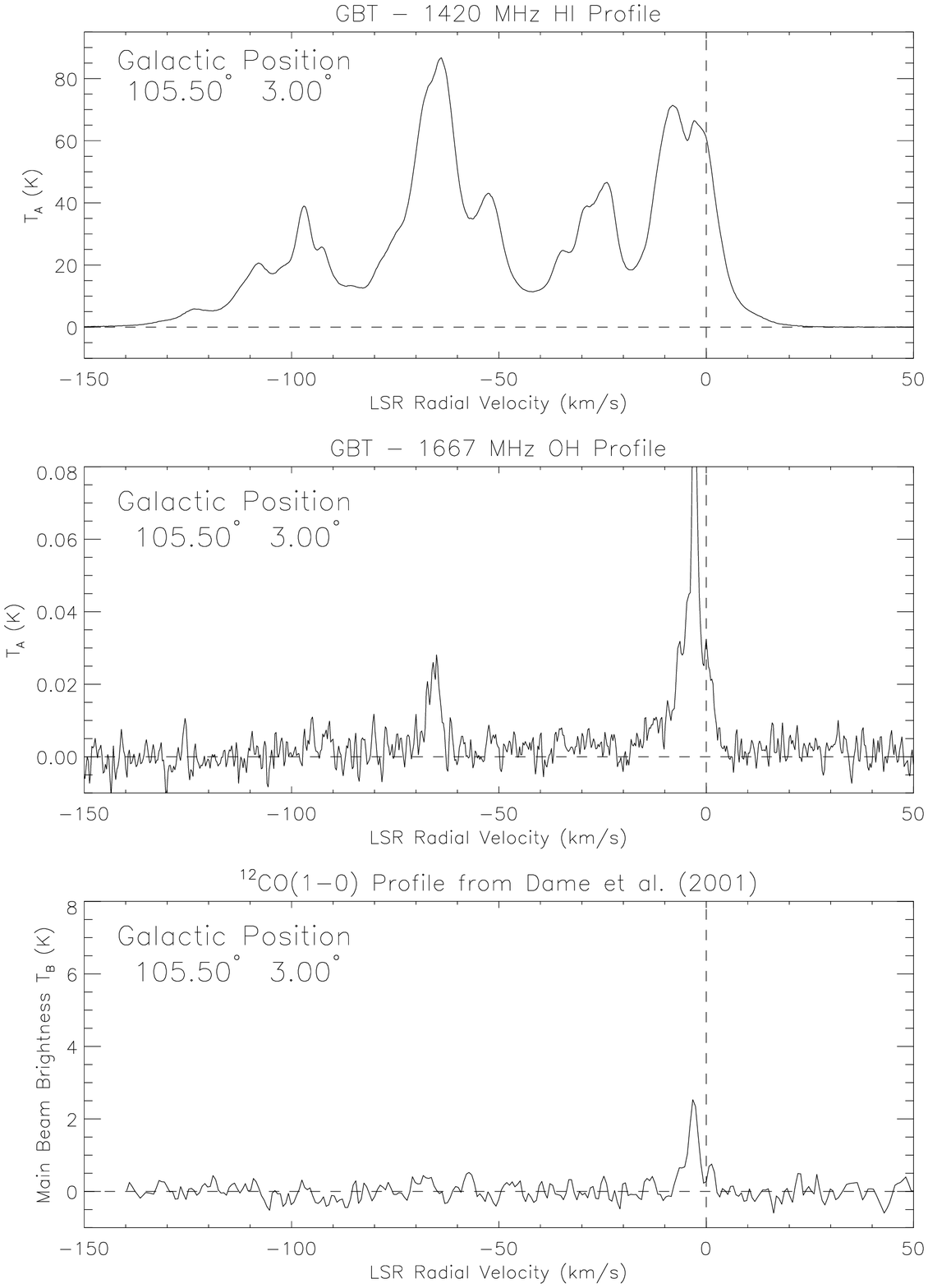}
\includegraphics[width=2.0in, angle=+0]{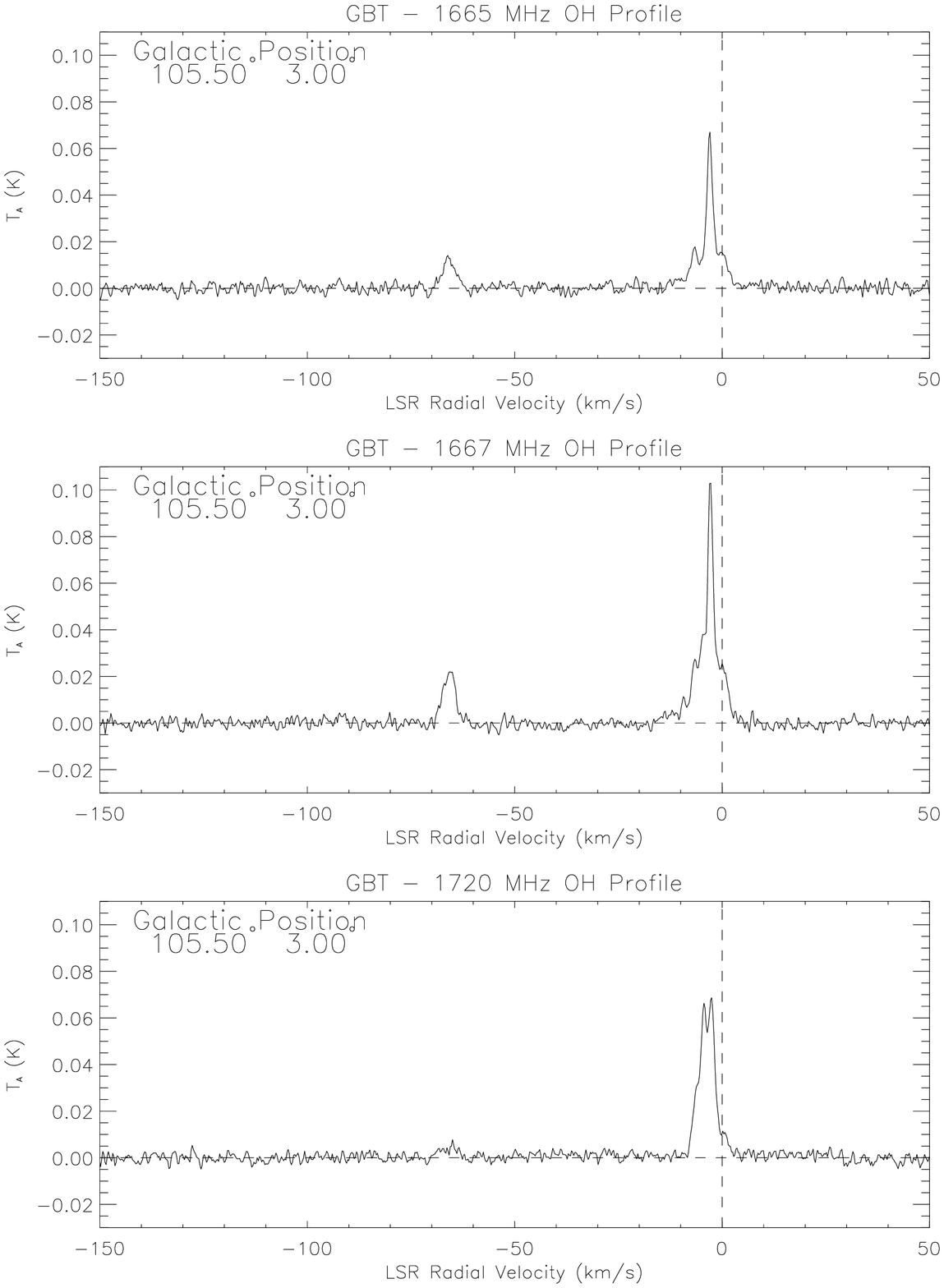}
\vspace{-0.1in}
\caption{\small A long 403-min exposure spectrum (right column of 3 panels) is shown next to the standard 1-hour \HI-OH-CO stacked-plot survey result at $l = 105.50^{\circ},\ b = +3.00^{\circ}$ (left column). The long exposure data in the right column is shown in a stacked plot of 1665 OH (upper panel), 1667 OH (middle panel; compare to the standard 1-hr survey result directly to the left), and 1720 OH (lower panel). Note that this latter spectrum clearly shows a feature at 1720 MHz in the Perseus Arm near -65 \kmps.  \normalsize}
\label{fig:noteworthy1}
\end{center}
\vspace{-0.1in}
\end{figure*}

In an attempt to extend the OH detections to distances as far as possible from the Sun, we have made a longer exposure at the pointing $l = 105.50^{\circ},\ b = +3.00^{\circ}$ (located in the upper left corner of the observing grid in Figure \ref{fig:survey}) with the aim of recording OH emission in the Outer Arm of the Galaxy near $- 100$ \kmps. The \HI-OH-CO stacked plot for this pointing (left column in Figure \ref{fig:stackedplots1}a) is reproduced here in the left column of Figure \ref{fig:noteworthy1} for comparison. The right column of Figure \ref{fig:noteworthy1} shows all three long-exposure spectra in a 1665-1667-1720 MHz stacked-plot display.
%
% ---------------------------------------------------
%-----------------------------------------------------
%
The total integration time for the spectra in the right column is 403 min, and the final rms noise after baseline subtraction is $\approx 1.8$ mK. The improvement over our usual 2-hr integrations (left column, middle panel) is very nearly a factor 2, as expected, indicating that receiver instability is in general not the limiting factor in our observations. Unfortunately we have not succeeded in detecting the Outer Arm of the Galaxy in OH emission; whether we have just been unlucky with this particular pointing or whether the Outer Arm is less rich in OH emission remains an open question. However, this long exposure reveals interesting features at 1720 MHz (right column, lower panel) near zero velocity (local) and near -65 \kmps\ (Perseus Arm) which will be discussed in \S \ref{sec:brightseventeentwenty}.

\subsubsection{An OH-IR star near $l = 105.0^{\circ},\ b = +2.5^{\circ}$}
\label{sec:OHIRstar}

\begin{figure*}[t!]
%\epsscale{1.0}
%\vspace{-0.1in}
\begin{center}
\includegraphics[width=2.0in, angle=+0]{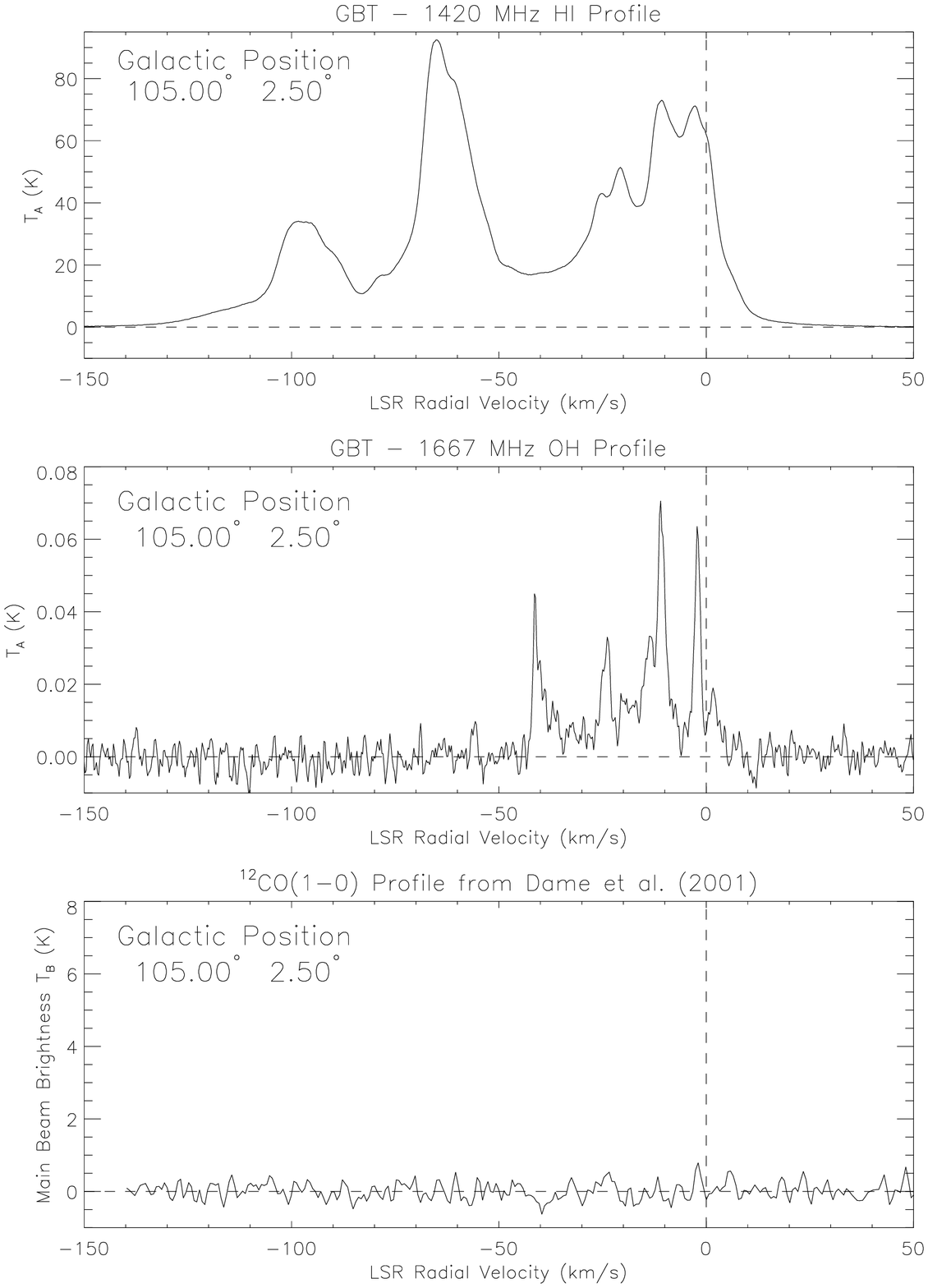}
\includegraphics[width=2.0in, angle=+0]{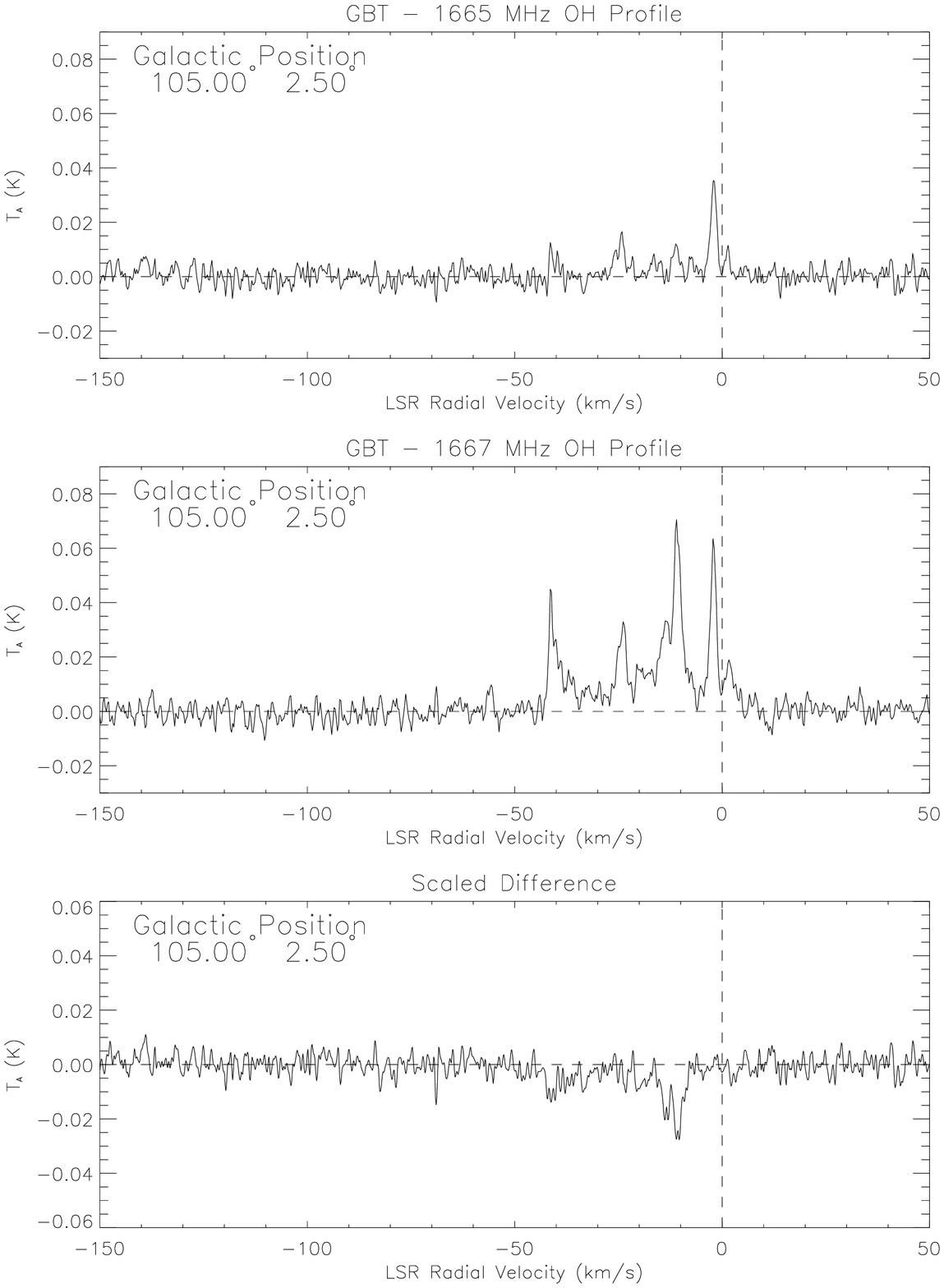}
\includegraphics[width=2.0in, angle=+0]{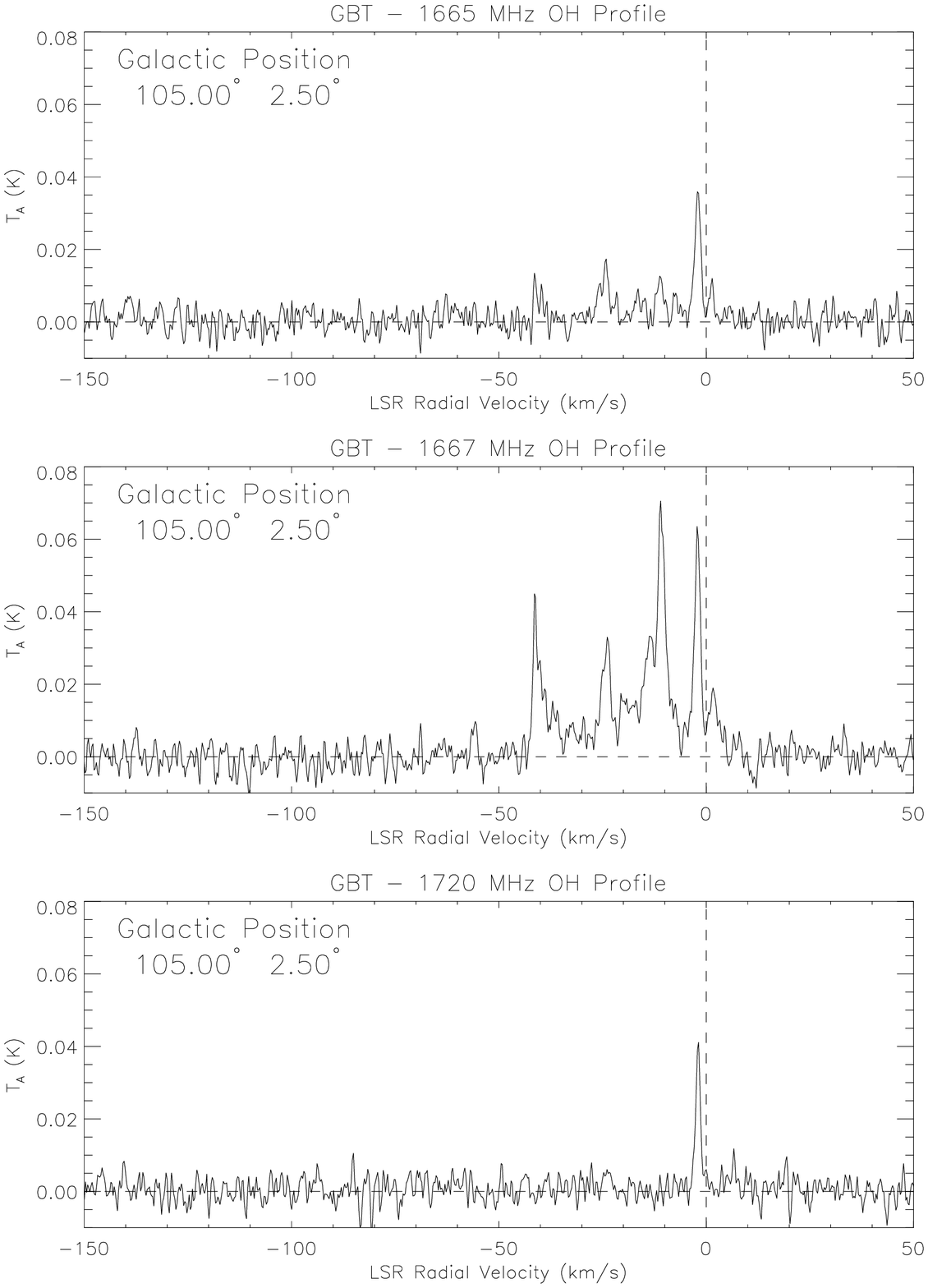}
\vspace{-0.1in}
\caption{\small Survey spectra at $l = 105.0^{\circ},\ b = +2.5^{\circ}$. In the left column note that narrow OH features are observed without corresponding \HI\ features, something that is not seen at any other survey position. In the lower plot of the middle column, the scaled difference of 1665 and 1667 MHz spectra show two residual features, suggesting that these features are not in LTE. These anomalous spectra can be explained by the presence of the OH-IR star 104.9+2.4 within the telescope field of view.  See text \S \ref{sec:OHIRstar} for a discussion. \normalsize}
\label{fig:noteworthy2}
\end{center}
\vspace{-0.1in}
\end{figure*}

\begin{figure*}[t!]
%\epsscale{1.0}
\vspace{-0.1in}
\begin{center}
\includegraphics[width=3.0in, angle=+0]{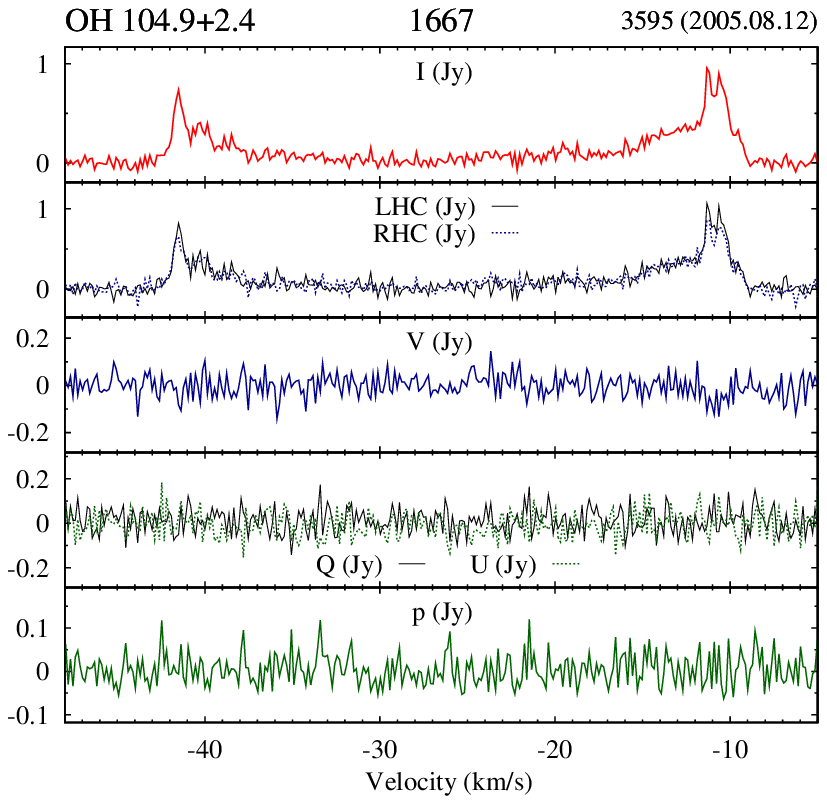}
\vspace{-0.1in}
\caption{\small Spectrum of OH 104.9+2.4 at 1667 MHz with the Nan\c{c}ay radio telescope, from \citet[][]{wsg12}.  \normalsize}
\label{fig:wsg12fig213}
\end{center}
\vspace{-0.1in}
\end{figure*}

Figure \ref{fig:noteworthy2} reproduces in the left column the stacked \HI-OH-CO plot of Figure \ref{fig:stackedplots1}c from our survey pointing at $l = 105.0^{\circ},\ b = +2.5^{\circ}$. The OH spectrum in the center panel of this column is of a nature not seen at any other survey pointing, showing bright ``spikey'' features that often do not correspond with \HI\ peaks. Referring to the 1665-1667-1720 MHz stacked plot in the right column of Figure \ref{fig:noteworthy2}, these features are especially bright at the following approximate velocities: -2, -33 and -40 \kmps\ at 1665 MHz, -2, -11, -33, and -41 \kmps\ at 1667 MHz, and -2 \kmps\ at 1720 MHz. In contrast to the usual situation, the \HI\ profile at this position (Figure \ref{fig:noteworthy2}, left column, top panel) does not show a peak corresponding with the -41 \kmps\ OH feature, suggesting an association at least of this OH feature with a compact ``non-ISM'' source. Further confirmation of anomalous features in the OH results at this pointing are shown in the three panels in the center column of Figure \ref{fig:noteworthy2}, where the scaled difference spectrum for the main OH lines is shown as the lower panel. Here it is now clear that not only does the -41 \kmps\ feature show an anomalous OH main-line ratio,  the -11 \kmps\ feature is also anomalous. It is natural to think that these two spikes might be associated, and indeed they are. 
%
% ---------------------------------------------------
%-----------------------------------------------------
%
An examination of archival material reveals the presence of a known OH-IR star at $l = 104.9^{\circ},\ b = +2.4^{\circ}$; this source is listed in the ``Hamburg Database of Circumstellar OH Masers'' \cite[][www.hs.uni-hamburg.de/maserdb]{eh11} and is close enough to be detected on the flank of the GBT point spread function ($0.13^{\circ}$ FWHM). The spectrum recorded on this object with the Nan\c{c}ay Radio Telescope at 1667 MHz by \citet[][]{wsg12} is reproduced in Figure \ref{fig:wsg12fig213}. It shows the characteristic ``double-horned'' profile of an OH-IR star, with narrow features at both -11 and -41 \kmps, and accounts for half of the bright narrow spikes seen in at 1667 MHz in Figure \ref{fig:noteworthy2}. OH-IR stars are usually stronger at 1612 MHz, but unfortunately we did not obtain a spectrum at that frequency during these observations.
%
%-----------------------------------------------------
%-----------------------------------------------------

Curiously, the narrow line (FWHM $\approx 1.5$ \kmps) at -2 \kmps\ in the 1667 MHz spectrum of Figure \ref{fig:noteworthy2} (center column) appears to be in LTE as far as the main lines are concerned, although the bottom panel in the right column of Figure \ref{fig:noteworthy2} shows a bright 1720 MHz feature at this velocity. This feature does not seem to be associated with the OH-IR star, but rather is an example of several survey pointings with enhanced emission at 1720 MHz which will be presented and discussed further in \S \ref{sec:brightseventeentwenty}.
%-----------------------------------------------------

\subsubsection{Main OH line Non-LTE features}
\label{sec:non-lte-main-lines}

\begin{figure*}[t!]
%\epsscale{1.0}
%\vspace{-0.1in}
\begin{center}
\includegraphics[width=2.0in, angle=+0]{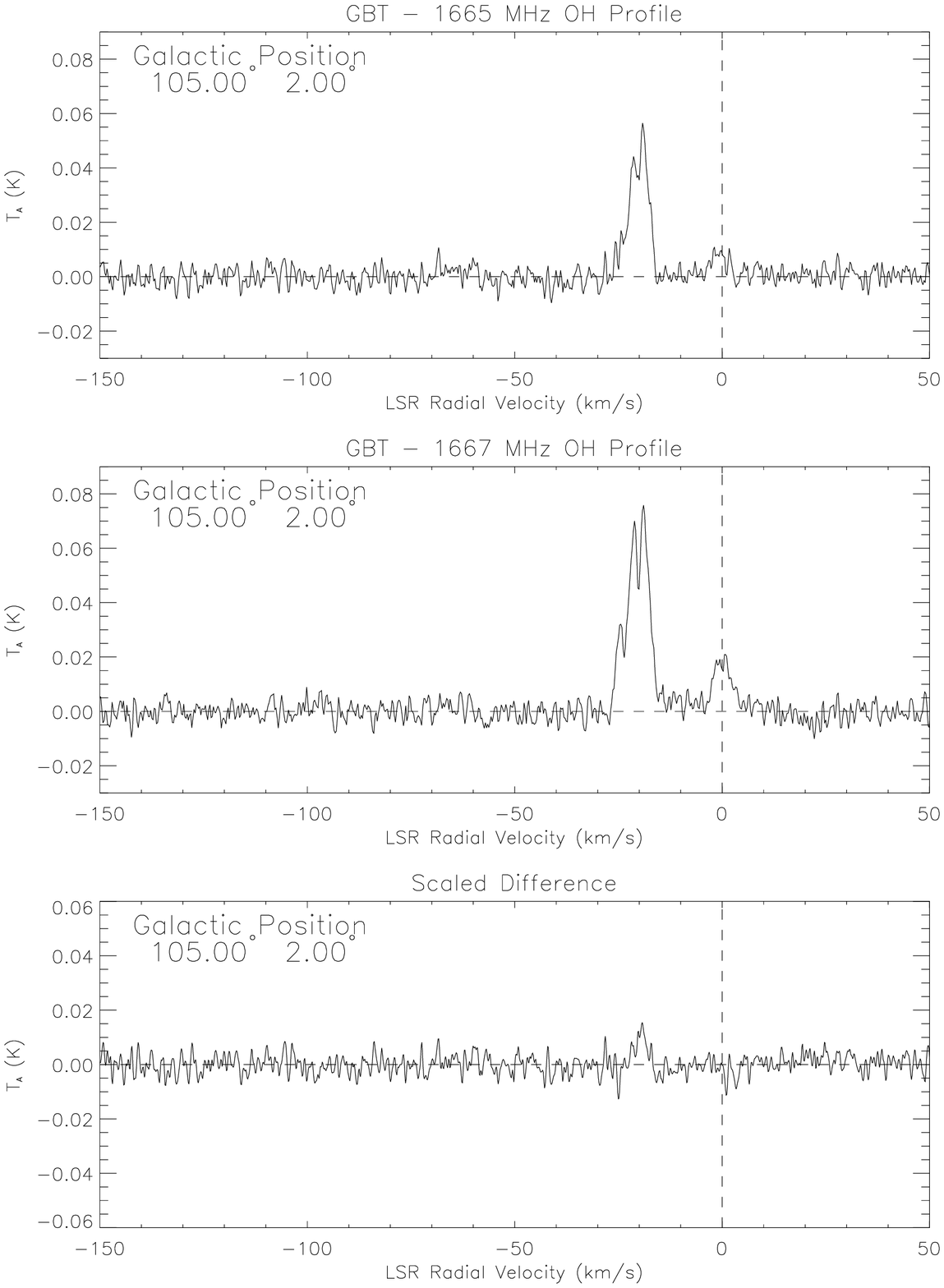}
\includegraphics[width=2.0in, angle=+0]{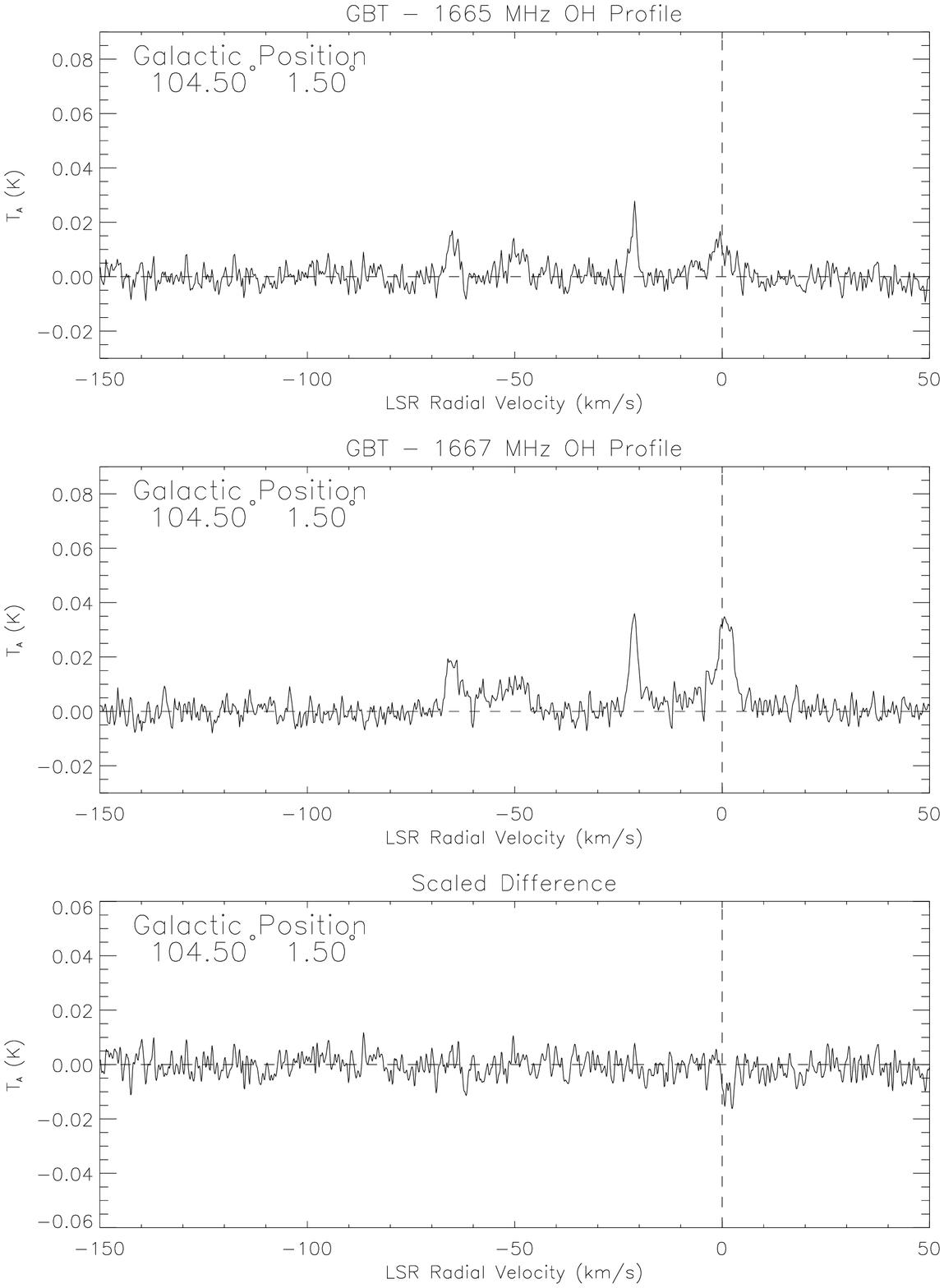}
\vspace{-0.1in}
\caption{\small Scaled difference spectra ($1665 - 1667/1.8$) for two positions that show departures in the main 1665/1667 OH line ratios from the LTE value of 5:9. Left panel: The 1665 MHz feature near -20 \kmps\ appears to be too bright, and leaves a positive residual in the scaled different spectrum shown in the lower row. Right panel: The 1667 feature near 1.5 \kmps\ appears to be too bright, leaving a negative residual. See text \S \ref{sec:non-lte-main-lines} for further discussion.  \normalsize}
\label{fig:opticallythick}
\end{center}
\vspace{-0.1in}
\end{figure*}

Figure \ref{fig:opticallythick} shows scaled difference stacked plots of 1665 and 1667 main-line spectra at $l = 105.00^{\circ}$, $b = +2.00^{\circ}$ (left column) and $l = 104.50^{\circ}$, $b = +1.50^{\circ}$ (right column). Each of these two spectra shows one feature that appears to deviate slightly from the expected LTE ratio. The first feature is near -20 \kmps\ in the left column, where the 1665 line is apparently too bright and the scaled difference spectrum shown in the lower panel has a positive residual. This could occur if the 1667 MHz feature is somewhat optically thick; in that case the usual optically-thin 1665/1667 MHz LTE line ratio of 5/9 trends towards 1/1. Such a phenomenon was first noted in the early OH observations of dense dust clouds in the Galaxy by \citet[][]{h69}, who used the measured deviation of the line ratio from 5:9 to estimate the opacity. However, this procedure assumes that the excitation temperature of the two transitions is the same. In a careful study of absorption and emission features in and around a number of continuum sources, \citet[][]{rwg76} found that the excitation temperatures measured for the 1665 lines appeared to be a degree or two higher than those for the 1667 lines, although the authors did not consider such differences to be significant at the time. \citet[][]{c79} specifically addressed this point with new, more sensitive observations of several features in two continuum sources and concluded that such differences were indeed real, and furthermore could also occur in the opposite sense, with the 1667 excitation temperature exceeding that of the 1665 line. The profiles shown in the right panel of Figure \ref{fig:opticallythick} may be an example of such an anomaly; the spectra at this pointing show a feature near 1.5 \kmps\ where the 1667 MHz emission is slightly stronger than expected for LTE, leaving a negative residual.

The origin of these small main-line anomalies is unclear. The models computed by \citet[][]{ger78} generally show that the excitation temperatures of the main lines are very nearly equal; however, there is one model involving warm dust (see their Figure 9) when the 1667 excitation temperature can exceed that of the 1665 line. It is possible that both high optical depth (favoring the 1665 line) and warm dust (favoring the 1667 line) play roles in accounting for the main-line anomalies. A tantalizing fact in the present case is that the IRSA data \cite[][]{ml05} reveal a $100 \mu$ source in close proximity to the line of sight for the pointing illustrated in the right panel of Figure \ref{fig:opticallythick}. This pointing  appears to have an overly-bright 1667 line; however, we presently have no evidence that the IR source is in fact physically close to the OH-emitting gas we have observed.

% ---------------------------------------------------
%-----------------------------------------------------

\subsubsection{Pointings with enhanced 1720 MHz emission}
\label{sec:brightseventeentwenty}

\begin{figure*}[h!]
%\epsscale{1.0}
%\vspace{-0.1in}
\begin{center}
\includegraphics[width=2.0in, angle=+0]{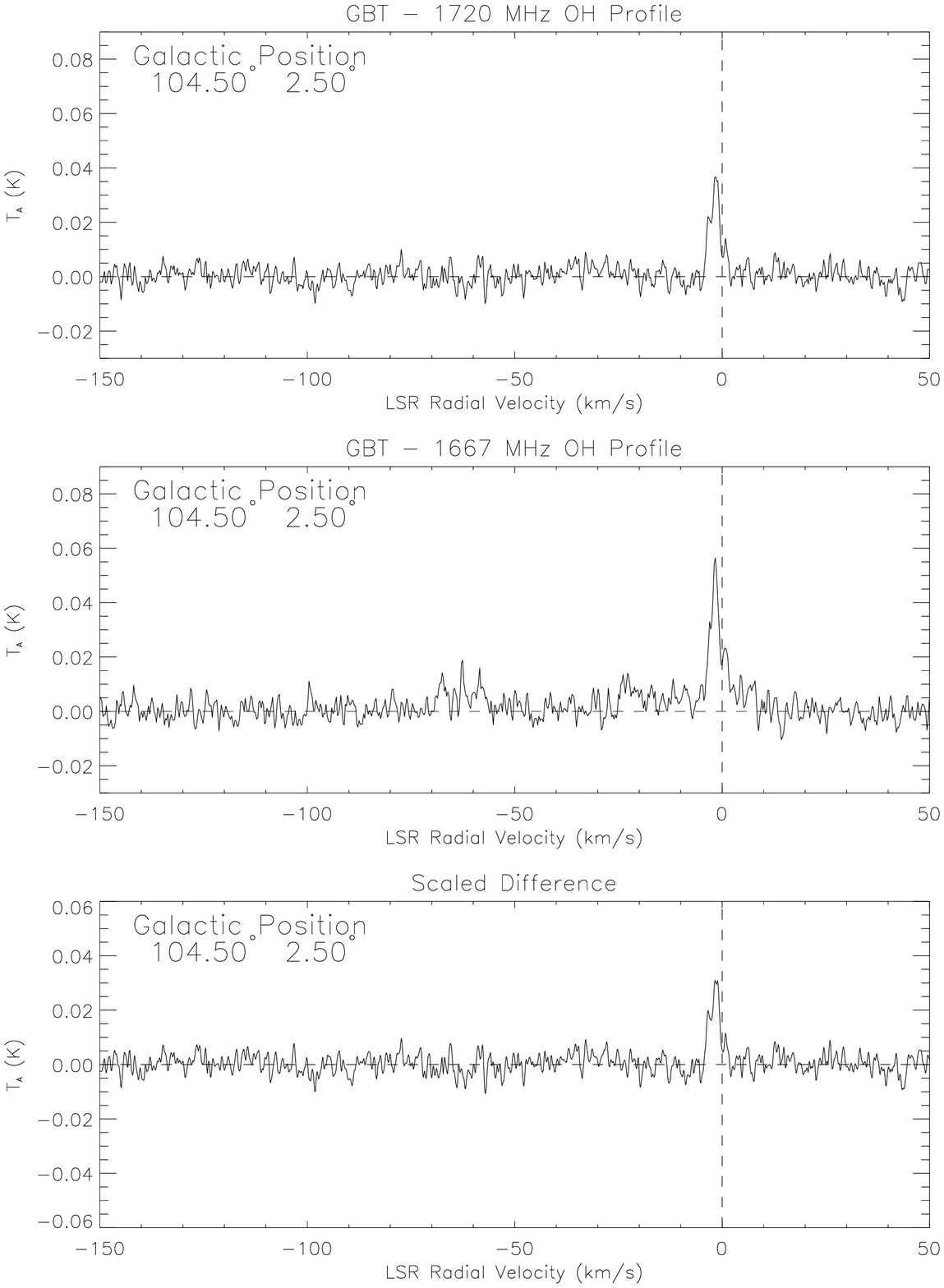}
\includegraphics[width=2.0in, angle=+0]{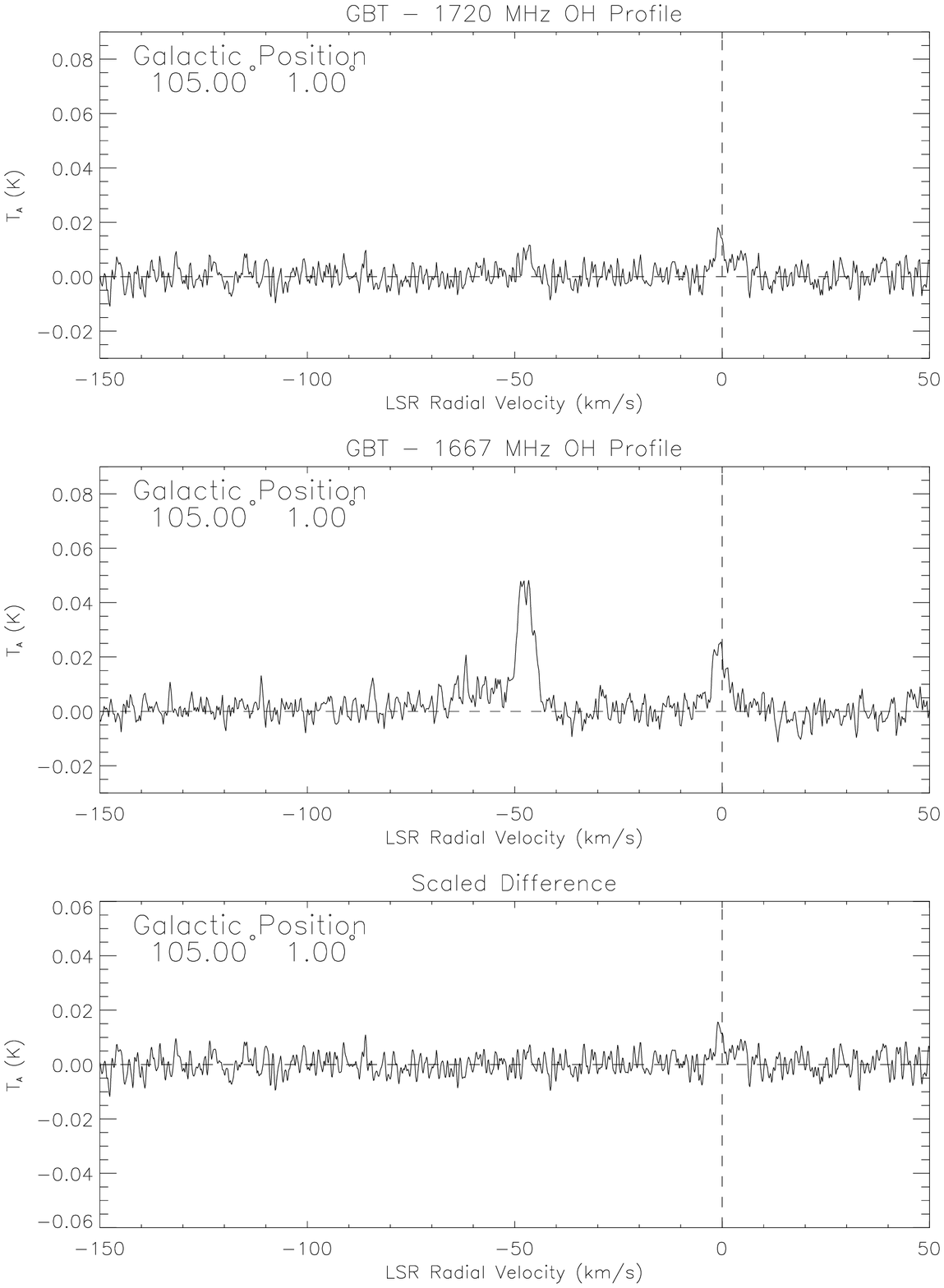}
\includegraphics[width=2.0in, angle=+0]{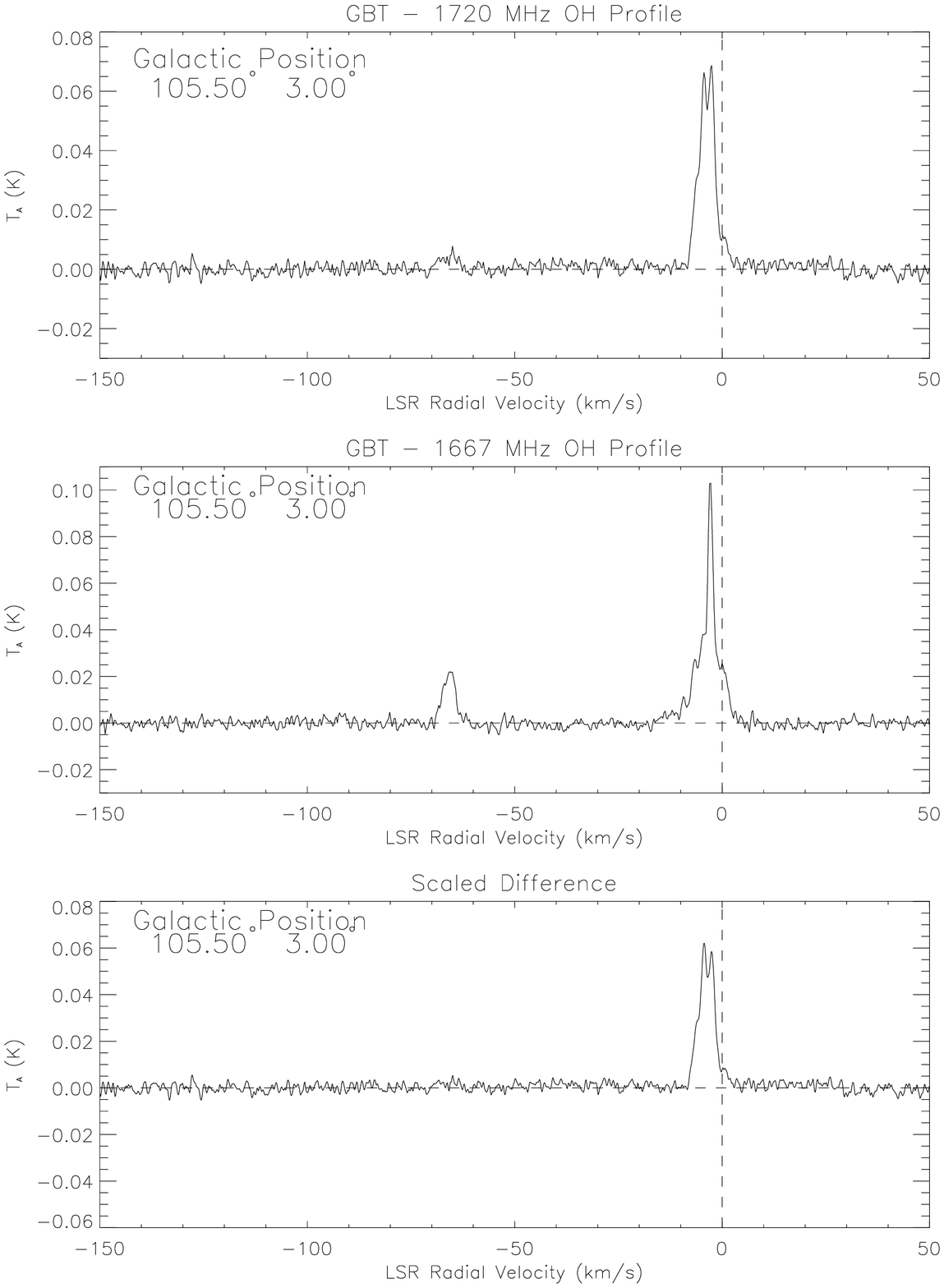}
\vspace{-0.1in}
\caption{\small
Three examples of unusually strong 1720 MHz OH emission in our blind survey. The 1720 MHz spectrum is shown in the top row, 1667 MHz in the middle row, and the scaled difference (1720) - (1667)/9 along the bottom. Left column: a residual feature near $V \approx 0$ \kmps; Middle column: a feature near $V \approx 0$ \kmps, but the Perseus Arm feature near -50 \kmps appears to be in LTE at this sensitivity; Right column: our most sensitive 1720 MHz scaled difference spectrum with $2 \times$ better S/N showing an enhanced 1720 MHz feature near $V \approx -5$ \kmps. See text \S \ref{sec:brightseventeentwenty} for further discussion.
\normalsize}
\label{fig:maserstackedplotdiffs}
\end{center}
%\vspace{-0.1in}
\end{figure*}

\begin{figure*}[!ht] % order of placement preference: here, top
%\epsscale{0.9}
%\vspace{-0.1in}
\begin{center}
\includegraphics[width=0.4\textwidth, angle=0]{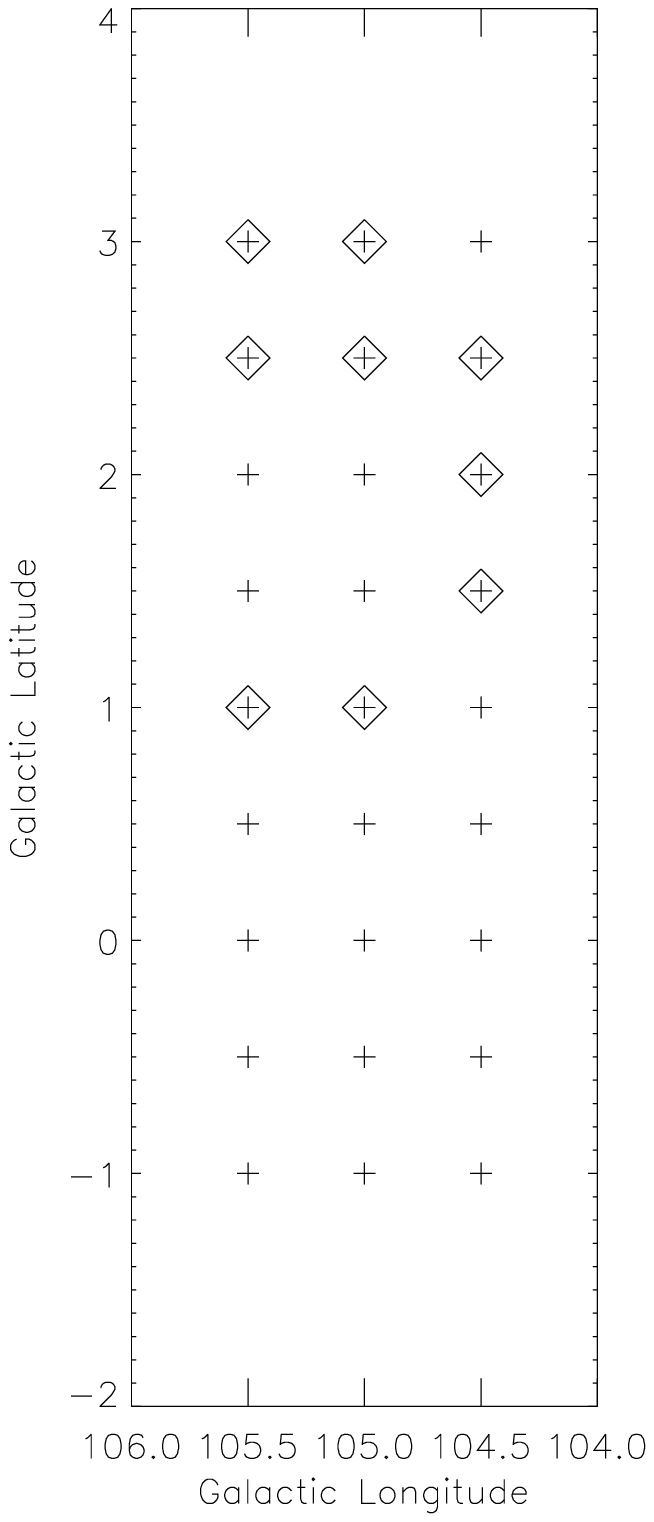}
\caption{\small
Sketch of the survey area with the locations of enhanced 1720 MHz emission near $V_{LSR} \approx 0$ shown with diamonds. See text section \ref{sec:brightseventeentwenty} for additional discussion.
\normalsize}
\label{fig:brightseventeentwenty}
\end{center}
%\vspace{-0.1in}
\end{figure*}

Although a large majority of the 1665 and 1667 spectral features at the positions we observed were found to be in the LTE ratio or close to it, there are at least nine pointings with 1720 MHz features that appear to be significantly stronger than the LTE ratio of 1/9 expected when compared to the corresponding 1667 MHz emission at those positions. Some examples are shown in Figure \ref{fig:maserstackedplotdiffs}. These positions show emission features up to 7.7 times stronger than expected for LTE conditions, with a mean of $4.4 \pm 1.1$. It is noteworthy that these unusually bright 1720 MHz emission features in our survey are all found in the velocity range $-10 \gtrsim V_{LSR} \gtrsim 0$ \kmps, suggesting that these features are very local to the Sun. Furthermore, the features appear to be clustered into a sub-region of our survey area; Figure \ref{fig:brightseventeentwenty} shows the pointings in our survey grid where these relatively strong 1720 MHz features are found.
%
%---------------------------
%-----------------------------------------------------
%
%----------------------------------
%
%-------------------------------------------
%
%
Abnormally bright OH 1720 MHz emission has been known since the early days of 18-cm OH observations of the ISM; see \citet[][]{gje73} for a brief history and the results of an important early study. In a seminal paper, \citet[][]{rwg76} showed that the excitation temperature of local gas along the line of sight to the radio galaxy 3C123 is negative and $\leq -6$K, indicating a weak maser. These authors also provided an early pumping model to explain the emission involving far-infrared radiation and collisions with neutral and charged particles. The models were developed further by \citet[][]{e76} who emphasized the important role of particle collisions (and hence elevated density) in pumping the 1720 line, and \citet[][]{t82} subsequently suggested that these features might be good tracers of spiral arms. More recently, \citet[][]{lge99} have shown that this line can be strongly excited by SN-generated shocks propagating into a molecular cloud, and can provide tight constraints on physical conditions in the shocked ISM. This model has been successfully applied by several authors to explain enhanced 1720 MHz OH emission in several known SNR \cite[][]{hyw06, hyw08, pfs08}. These studies emphasize the utility of this emission line as an ISM diagnostic, and we will focus more attention on it in our future work.

%\clearpage

\section{Line Strength Determinations}
\label{sec:OHandCOcomparison}

\subsection{Profile analysis}
\label{profileanalysis}

The line strengths (profile integrals) of all discrete features which could be visually identified on the 27 1667 MHz OH spectra in the survey were calculated, along with the corresponding line strengths on the \twCO(1-0) spectra obtained over the same velocity ranges. The numerical expression used is $ S =  \Sigma T_{mb} \times \Delta V$ in units of  $^{\circ}$K $\ast$ \kmps, where $T_{mb}$ is the main-beam brightness temperature, $\Delta V$ is the channel spacing of the data, and the sums are done numerically over the channels containing measurable signal for each feature in a spectrum. This latter determination is made by visually displaying the 1667 MHz OH spectra and interactively choosing the velocity ranges over which the spectrum is to be integrated.  Thereafter the \twCO(1-0) spectrum is extracted from the CfA data base and integrated over the same velocity ranges. The procedure was repeated until all visually-identifiable discrete 1667 MHz OH features at each pointing were processed.

\subsection{Error analysis}
\label{baselineuncertainties}

The $1 \sigma$ dispersions in the line strength values were estimated as a combination of receiver noise and baseline uncertainty.  The cumulative error owing to random receiver noise was computed in the usual fashion by determining the value of the rms noise in $^{\circ}$K per independent spectral channel, multiplying by the square root of the number of such channels in each profile integral, and finally multiplying by the channel spacing in \kmps. We have done a 2-channel gaussian smoothing on all spectra before archiving, including the spectra on which the line integrals are based. This has reduced the rms noise in the OH spectra down to the level of 3.0 - 3.5 mK which we quoted in \S \ref{sec:datareduction} of this paper. However, we did not decimate the final OH spectra; hence neighboring channels in the archived spectra at intervals of 0.275 \kmps\ are not statistically independent; only every second channel at intervals of 0.55 \kmps\ qualifies for that. The formula for calculating the receiver noise contribution  $E_{noise}$ to the error in the line strength $S$ reckoned over the velocity interval $V_{lo} - V_{hi}$ is therefore $E_{noise}^{OH} = 0.0035 \times \sqrt{(V_{lo} - V_{hi})/0.55} \times 0.275$ in units of $^{\circ}$K $\ast$ \kmps. By contrast, the CfA \twCO\ spectra have not been decimated, hence the appropriate equation is: $E_{noise}^{CO} = 0.32 \times \sqrt{(V_{lo} - V_{hi})/0.65} \times 0.65$ where 0.65 is the channel interval in \kmps\ of the CfA CO spectral data.

At the other extreme, the baseline fitting procedure used on every spectral feature  ensures that the contribution of the baseline uncertainties to the errors in the line strength is essentially completely correlated over the entire velocity interval covered by that feature. In order to determine this error, the following observational approach was adopted. Sections of empty baseline were chosen with a number of spectral channels comparable to that covering a typical emission feature. For all OH and CO spectra the sections were integrated and divided by the exact number of channels used in order to obtain a per-channel value, and the standard deviations of these values were calculated for a set of typical baseline determinations. These standard deviation values are 0.0005 $^{\circ}$K per 0.275 \kmps\ channel for all of the OH spectra, and 0.05 $^{\circ}$K per 0.65 \kmps\ channel for the CfA \twCO\ spectra.  The total baseline error is obtained by multiplying this per-channel value by the number of channels in that particular profile, and then by the channel spacing in \kmps.  The expressions used are: $E_{base}^{OH} = 0.0005 \times \left[ (V_{lo} - V_{hi})/0.275 \right] \times 0.275$ for the OH spectra, and $E_{base}^{CO} = 0.05 \times \left[ (V_{lo} - V_{hi})/0.65 \right] \times 0.65$ for the CO spectra. The final rms errors $\Delta S_{rms}$ on the line strengths $S$ are calculated as the quadrature sum of the two error contributions, $\Delta S_{rms} = \sqrt{E_{noise}^2 + E_{base}^2}$ for each of the two tracers OH and CO.

\subsection{A comparison of 1667 MHz OH and \twCO(1-0) line strengths}
\label{sec:linestrengthresults}

\begin{figure*}[!ht] % order of placement preference: here, top
%\epsscale{0.9}
\begin{center}
\includegraphics[width=0.7\textwidth, angle=0]{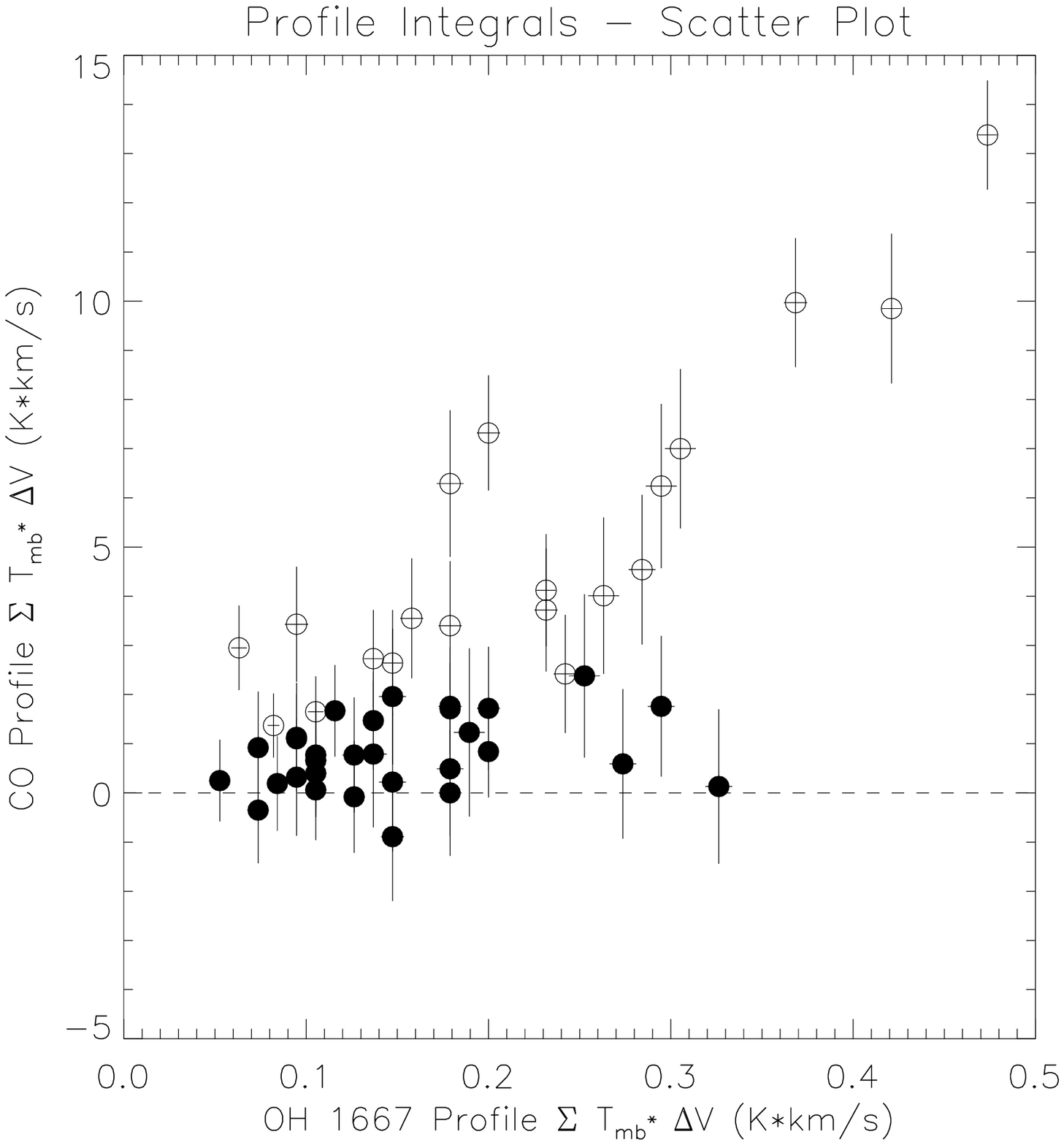}
\caption{\small
Scatter plot of OH 1667 MHz emission line strength (X-axis) and the \twCO(1-0) line strength (Y-axis) integrated over the same velocity ranges. The open and filled circles together show all OH features with line strengths in excess of $8 \sigma$. Points denoting \twCO(1-0) data for which the CfA CO line strength is within $2\sigma$ of the CO zero line are shown as filled circles.}
\normalsize
\label{fig:scatterplot}
\end{center}
\end{figure*}

A total of 53 features were visually identified as showing significant 1667 MHz OH emission over the 27 pointings of the blind survey. The two features associated with the OH-IR star and a third feature possibly associated with a Far-IR source were removed from the sample. The scatter plot of Figure \ref{fig:scatterplot} shows the line strengths of the 50 remaining OH features plotted along the X-axis with their $\pm 1 \sigma$ (horizontal) error bars, many of which are too small to be discerned. The corresponding \twCO(1-0) line strengths with their associated (vertical) error bars are plotted along the Y-axis. A horizontal dashed line shows the zero point for the CO data.

The first point to note about the scatter plot of Figure \ref{fig:scatterplot} is that every OH point plotted has a S/N in excess of 8; the OH data in this plot is essentially error-free. The second point is that the scatter plot appears to show two classes of CO emission: the first is roughly correlated with OH, and the second clusters around zero. To enhance this separation we have plotted the 30 low S/N CO features (those less than $2 \sigma$ above the zero line) as filled circles. The third point to note about Figure \ref{fig:scatterplot} is that an improvement in the \twCO(1-0) sensitivity (such as would be available if we had used the FCRAO survey of \citet[][]{hbs98} instead of the CfA survey of \citet[][]{dht01} in the Outer Galaxy) would undoubtedly reduce the size of the error bars on the CO data, but the mean values would in general be at or below the upper limits shown with the error bars in Figure \ref{fig:scatterplot}. Improved CO data is therefore unlikely to move this second class of CO data points with line strengths near zero (filled circles) upwards into the first class of points that are more proportional to the OH emission (open circles). We conclude that the \twCO(1-0) emission accompanying the 1667 MHz OH features we have measured falls into two roughly-equal categories: CO features that show an approximate correlation with OH emission, and CO features that are faint or absent even when appreciable S/N OH emission is present. This quantifies the impression obtained from the appearance of the stacked-plot spectra in Figure \ref{fig:stackedplots1}.

It is apparent that the 1667 MHz OH data is revealing a substantially larger fraction of molecular gas in the ISM than is indicated by the \twCO(1-0) line, and this gas is in different places. Possible reasons for this behavior include a selective CO photodissociation mechanism which leaves OH molecules intact in these regions of the ISM (possibly requiring special conditions that may be unrealistic), or insufficient collisional excitation of the CO(1-0) rotational line as could occur in regions of low volume density. A more quantitative discussion including estimates of the implied \Htwo\ column densities is in preparation for a subsequent publication.  

%
%----------------------------------
%
%-------------------------------------------

%\clearpage

\section{Conclusions}
\label{sec:conclusions}

Using the Green Bank Telescope, we have carried out a blind, high-sensitivity survey for OH emission over a sparsely-sampled grid of 27 pointings centered at $l = 105.0^{\circ}, b = +1.0^{\circ}$. OH emission at 1667 MHz has been detected in 53 separate features distributed over a total of $\gtrsim 21$ of the 27 pointings. The results confirm and extend the conclusions first reported by \citet[][]{arb12, arb13}, namely, that main-line OH emission is ubiquitous in the Outer Galaxy, and that this emission is often not accompanied by \twCO(1-0) emission. Additional results from the present observing program include:
\begin{itemize}
\item No spectral features are found in absorption, in agreement with the low levels of 18-cm continuum background emission in this direction in the Galaxy.
\item Observations at both 1665 and 1667 MHz show that the main OH lines are generally found with the optically-thin, LTE ratio of 5:9 on a detailed, channel-by-channel basis. This would be expected if these lines are excited by collisions in a low-density medium.
\item The spectra show several components which are well separated in velocity. These components appear to be associated with well-known large-scale features of Galactic Structure including the Local Arm and the Perseus Arm.
\item Substantial changes in the structure of the OH profiles can be seen on the $0.5^{\circ}$ angular scale of this survey. These changes are especially large for features located in the Perseus Arm, where the survey grid spacing corresponds to 28 pc.
\item Gaussian fits to OH emission-line features exhibit amplitudes of $\langle T_{mb} \rangle \approx 25 \pm 15$ mK, and line widths of $\langle \Delta V\rangle  \approx 3.9 \pm 2.5$ \kmps, where the $\pm$ values given are indicative of the $1\sigma$ dispersions in the quoted means. However, many observed features cannot be fitted in a satisfactory manner with single gaussians.
\item Consistent with previous work, extended emission is seen from the 1720 MHz satellite line of OH. This enhanced emission extends over $\approx 1/3$ of our survey pointings and appears to form a large structure (or structures) on the sky near $V_{LSR} \approx 0$ \kmps.
\item Narrow-band anomalous emission in both main lines of OH is found at a single pointing out of the total of 27, and has been identified \textit{post facto} with a known OH-IR star at $l = 104.9^{\circ},\ b = +2.4^{\circ}$.
\item Evidence for other non-LTE main-line ratios has been found at only two positions of the 27 in the survey. The reasons for these enhancements are not clear; in the first case the 1667 line may be enhanced by a nearby Far-IR source, and in the second case the 1667 line may be diminished if the optical depth is sufficiently large. 
\item Emission line strengths were determined for 50 distinct 1667 MHz OH spectral features in excess of $8 \sigma$, and compared with the corresponding \twCO(1-0) line strengths calculated from the CfA survey data. 20 of these OH features showed detectable CO emission with line strengths $\gtrsim 2 \sigma$ in the CfA survey. The line strengths of these relatively-bright CO components appear to be roughly correlated with the corresponding OH features.
\item The remaining $\approx \! 30$ bright OH features identified in the present blind survey show no detectable \twCO(1-0) emission at the $2 \sigma$ level of the CfA survey. While higher-sensitivity CO data would undoubtedly turn many of the CO non-detections into measurements, such data is not likely to recover the ``missing'' CO emission. In particular the filled circles in Figure \ref{fig:scatterplot}, which now cluster near the dashed zero line of CO line strength, are not likely to move upwards into the region presently occupied by the open circles, at least not for OH line strengths in excess of $\approx 0.2$ \degrees K $\ast$ \kmps.
\end{itemize}

The results described here indicate that high-sensitivity observations of the 18-cm emission lines from the OH molecule are a promising new tracer for studying large-scale features in the Galactic molecular ISM and for revealing the presence of \Htwo\ in amounts and in locations not well traced by the 3-mm line of \twCO(1-0). More sensitive and more extensive blind surveys of the Galactic sky in OH emission are likely to expand and consolidate this result. The spatial structure of the OH emission can be measured especially well in the Perseus Arm, and surveys carried out with the GBT in the Outer Galaxy at pointing intervals smaller than that used here would provide new insight into the scale sizes and physical conditions of this component of the molecular ISM. Differences in the excitation mechanisms and optical depths of OH and CO can provide new diagnostic tools for studying structural features of molecular clouds. Bl ind surveys which include the satellite lines of OH could identify hitherto unknown OH-IR stars and help us to understand the origin of the line enhancements. The appearance of OH emission in locations where \HI\ is abundant but the \twCO(1-0) line is weak or absent confirms an association of OH with ``dark molecular gas''. It also suggests a degree of mixing of the neutral atomic and molecular phases in the low-density ISM which deserves further scrutiny. Translating the results of observations such as those reported here into quantitative measures of \Htwo\ column density requires careful consideration of a number of details including OH line excitation temperatures and relative abundances. These topics will be addressed in future publications.

\section*{Acknowledgments}

We are grateful to the staff at the Green Bank Observatory for their advice and assistance with the operation of the GBT, in particular Karen O'Neil, Jay Lockman, Toney Minter, Ron Maddalena, and Amanda Kepley, and for the development and support of the GBTIDL data analysis software, especially Jim Braatz and Bob Garwood. Marcus Hughes assisted with the data reduction during a summer internship at STScI. PDE is partially supported by a Student Observing Support Award from the NRAO and partially by the Director's Discretionary Research Fund at STScI. We thank Miller Goss and Carl Heiles for sharing their considerable knowledge about OH obtained from many years of observing this molecular tracer in the Galactic ISM. Miller also provided important clarifications on the early history of OH work as well as comments on an early version of this paper. We are especially grateful to the referee Jo Dawson; her generous comments have helped us to correct a number of misconceptions and led to significant improvements in this paper. This research has benefited from the IRAS/IRIS archives at IPAC/IRSA, and from the \twCO(1-0) all-sky survey archives and the SAOImage software provided by the CfA.

%\vspace{-0.2in}

%--------------------------------

%\end{document}

\clearpage

%=================================
%
% The following 9 figure environments present the stacked plots for HI, OH 1667, and CO in 3 groups of 3 plots per page printed over 3 pages. 
%
%-----------------------------------
% The figure number needs to be artificially set here so it comes out to be "2". This is done with a "setcounter" command.
%-----------------------------------
%
\setcounter{figure}{1} % this will get incremented at the next statement
\begin{figure*}[h!]
%\epsscale{1.0}
\vspace{-0.1in}
%\begin{center}
\includegraphics[width=2.0in, angle=+0]{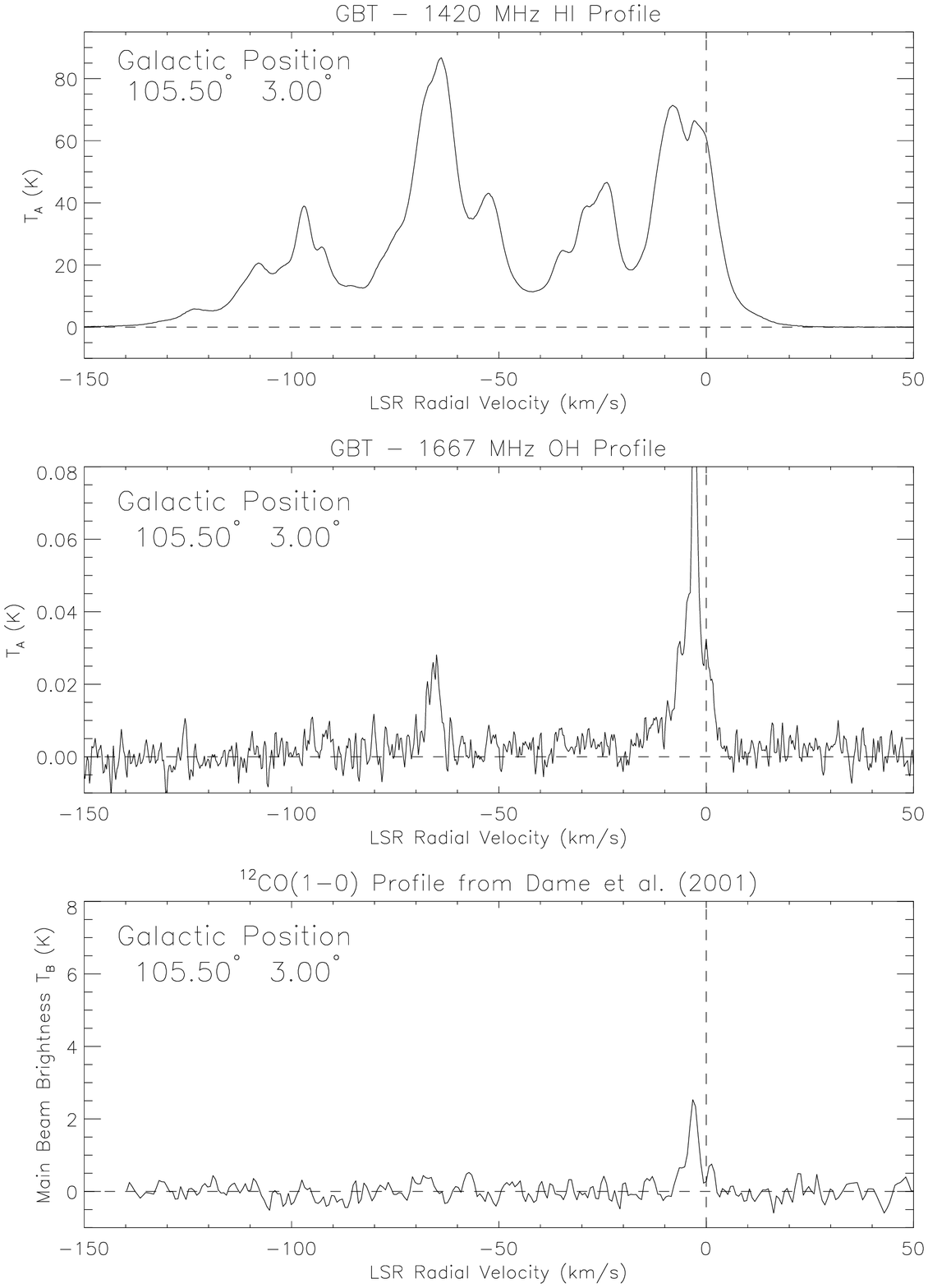}
\includegraphics[width=2.0in, angle=+0]{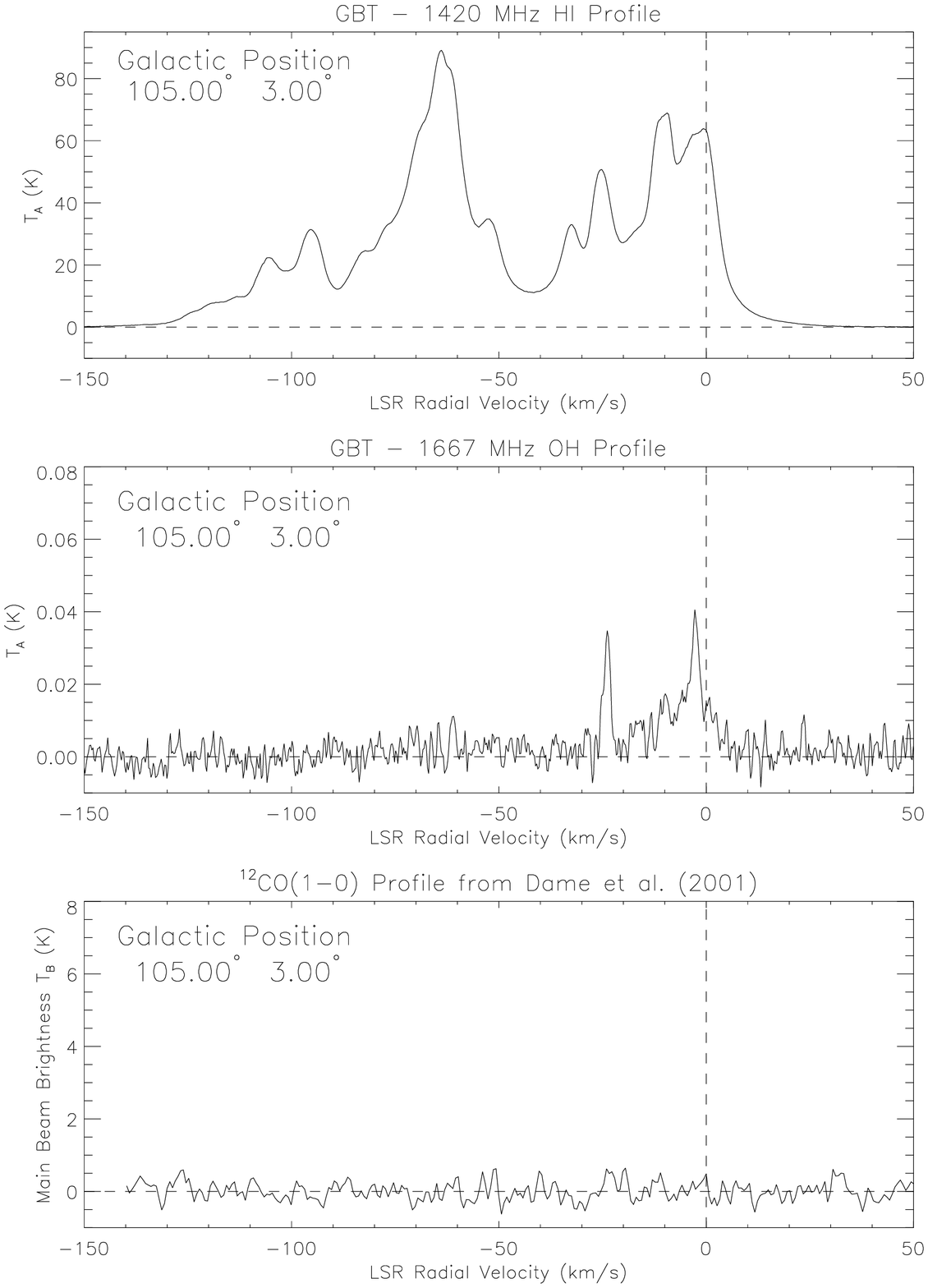}
\includegraphics[width=2.0in, angle=+0]{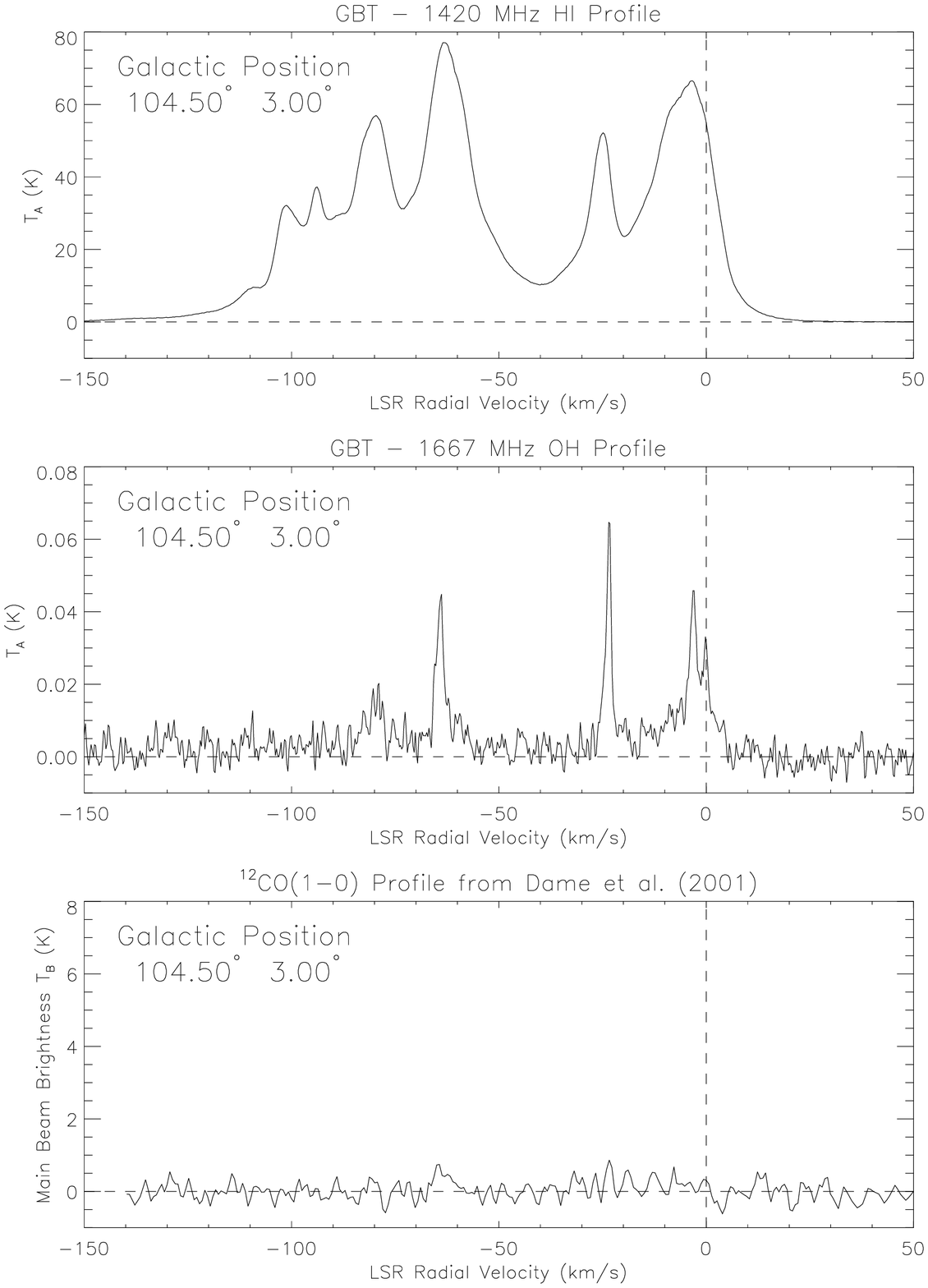}
\vspace{-0.1in}
\caption{\small a). Profiles at $b = +3.00^{\circ}$ for $l = 105.50^{\circ}$ (left), $105.0^{\circ}$ (middle), and $104.50^{\circ}$ (right).
% Top Row: \HI; middle row: 1667 MHz OH; bottom row, \twCO(1-0).
\normalsize}
\label{fig:stackedplots1}
%\end{center}
\vspace{-0.1in}
\end{figure*}
%--------------------------------
\setcounter{figure}{1}  % this will get incremented at the next statement
\begin{figure*}[h!]
%\epsscale{1.0}
%\vspace{-0.1in}
%\begin{center}
\includegraphics[width=2.0in, angle=+0]{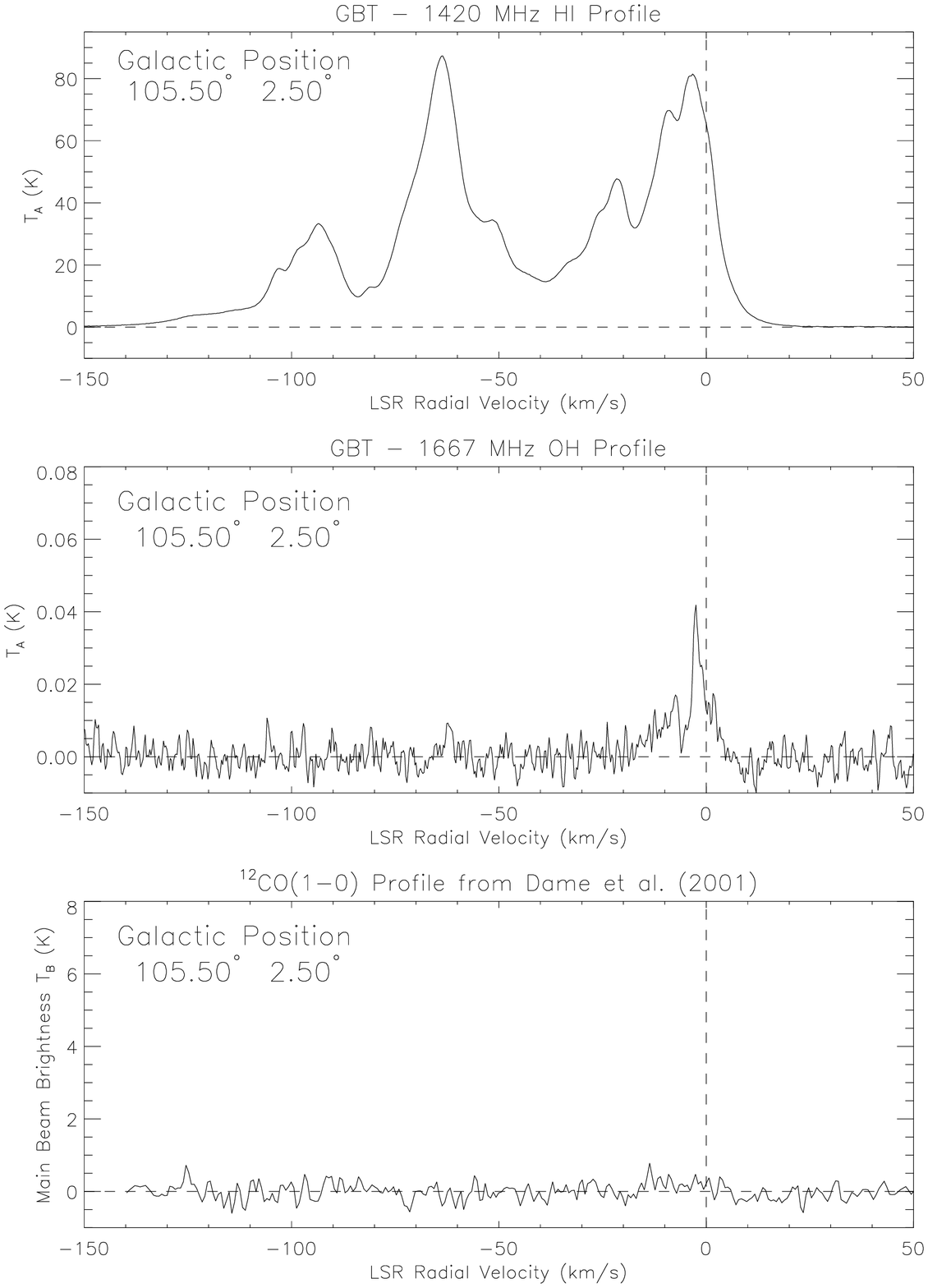}
\includegraphics[width=2.0in, angle=+0]{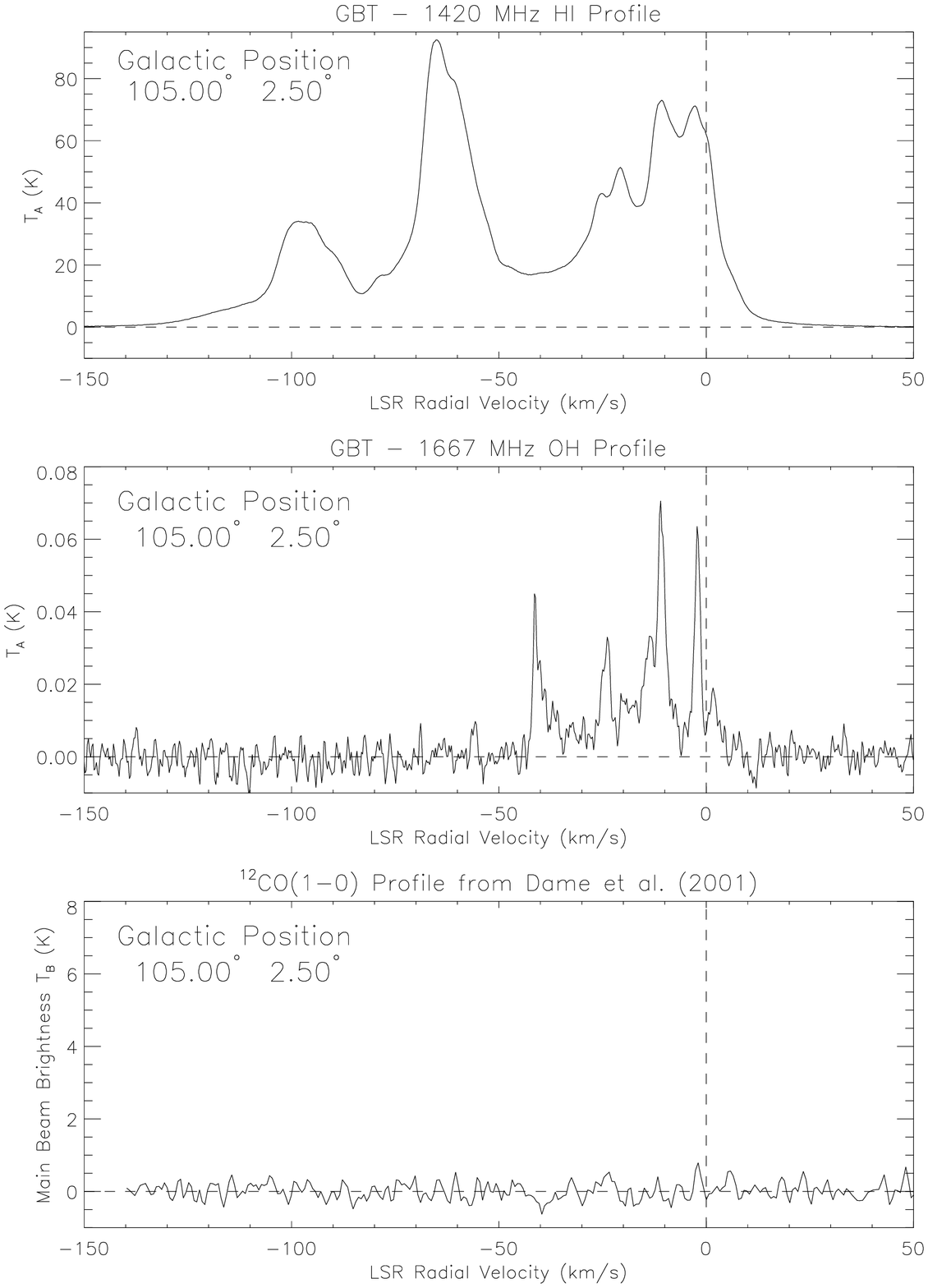}
\includegraphics[width=2.0in, angle=+0]{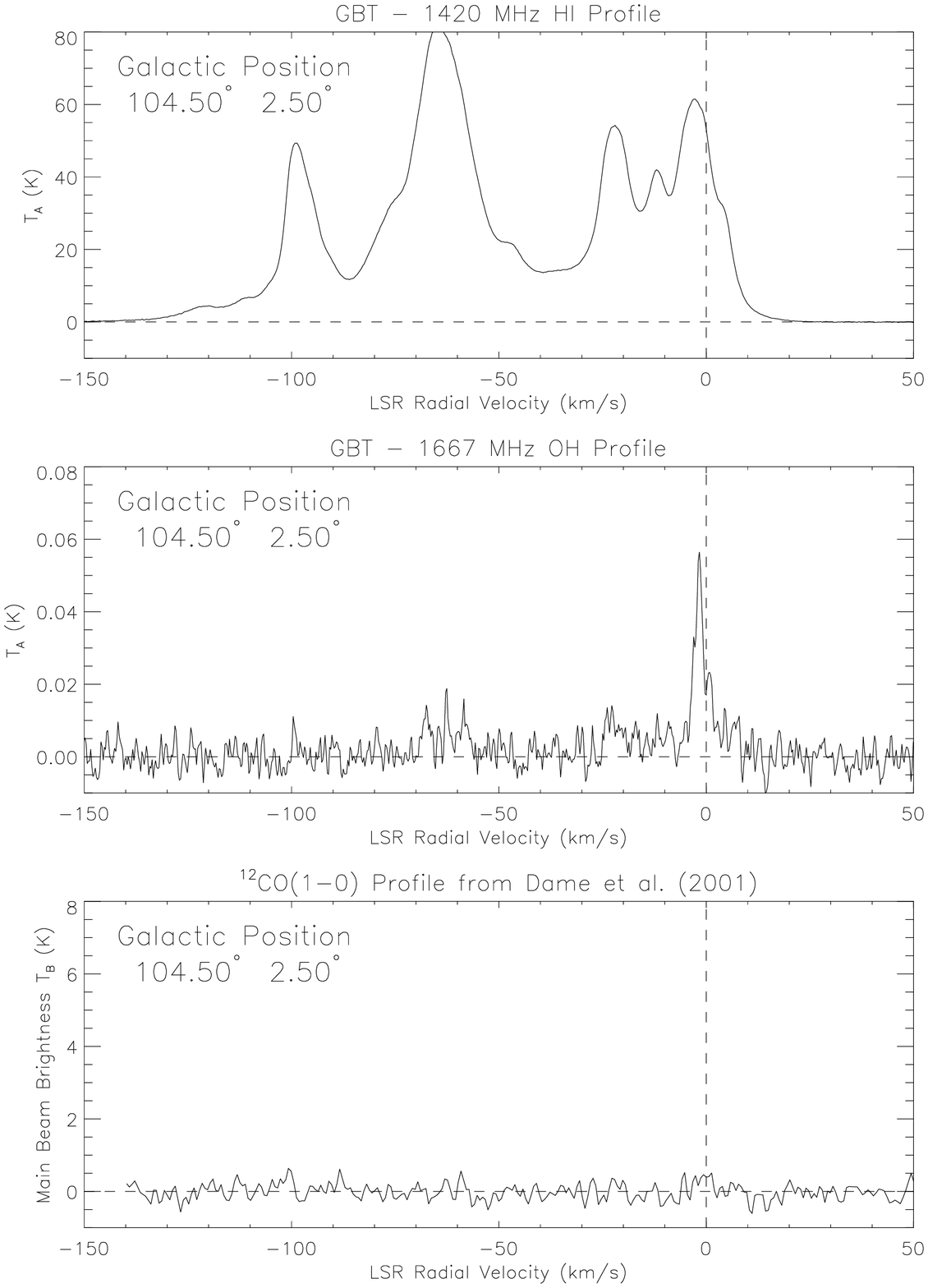}
\vspace{-0.1in}
\caption{\small b). As in Fig \ref{fig:stackedplots1}a, for $b = +2.50^{\circ}$. \normalsize}
%\end{center}
\vspace{-0.1in}
\end{figure*}
%--------------------------------
\setcounter{figure}{1}  % this will get incremented at the next statement
\begin{figure*}[h!]
%\epsscale{1.0}
%\vspace{-0.1in}
%\begin{center}
\includegraphics[width=2.0in, angle=+0]{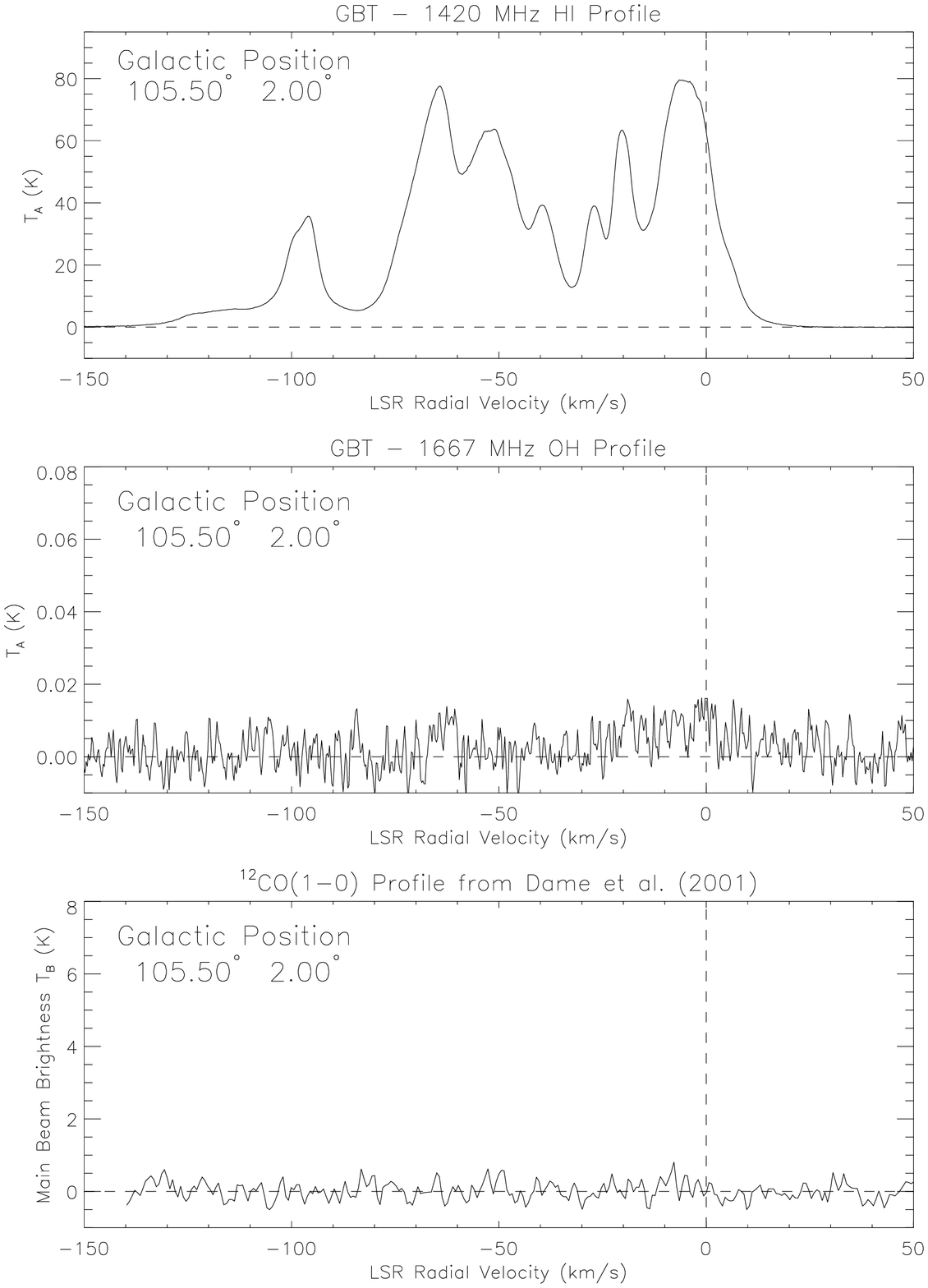}
\includegraphics[width=2.0in, angle=+0]{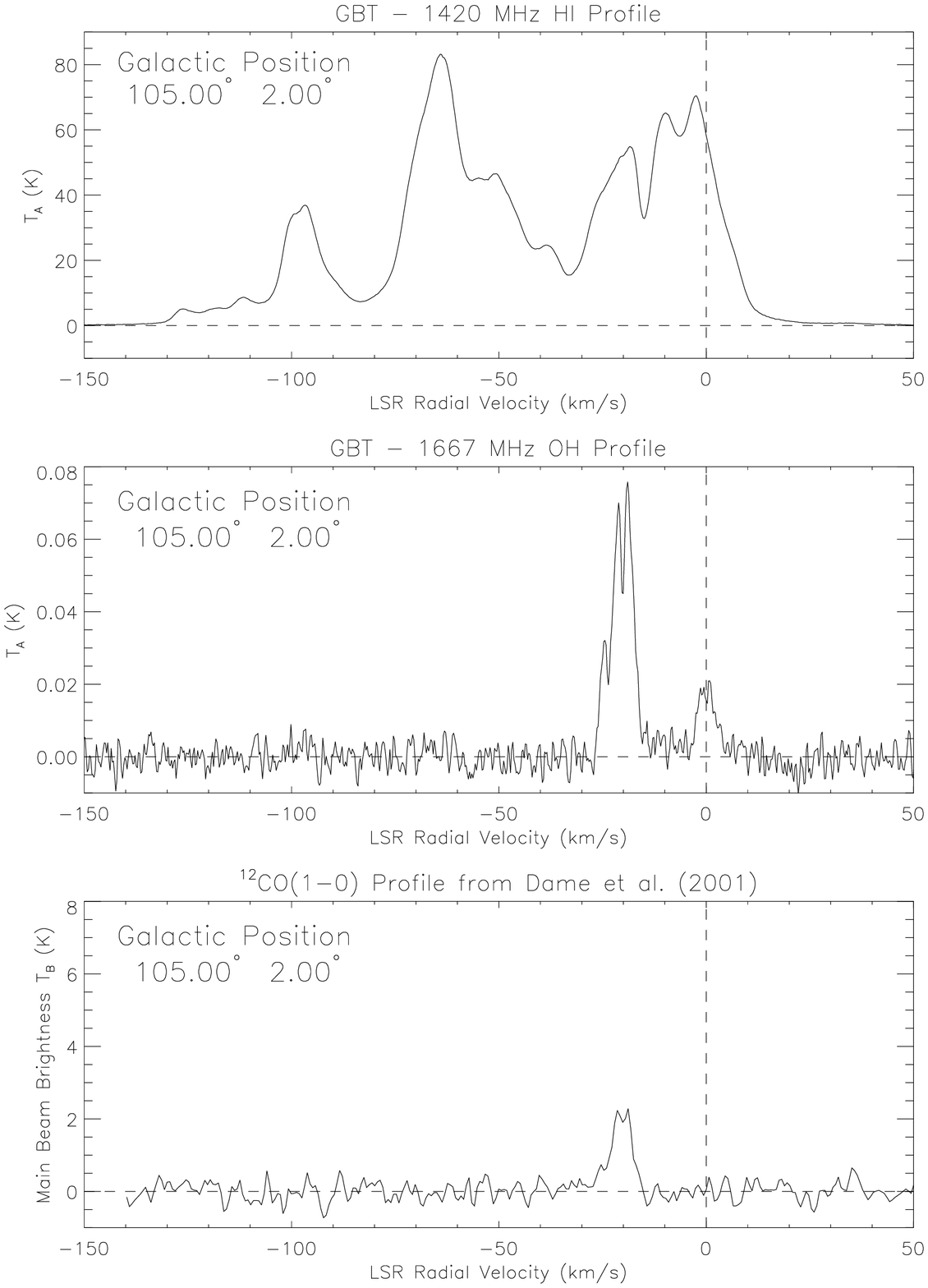}
\includegraphics[width=2.0in, angle=+0]{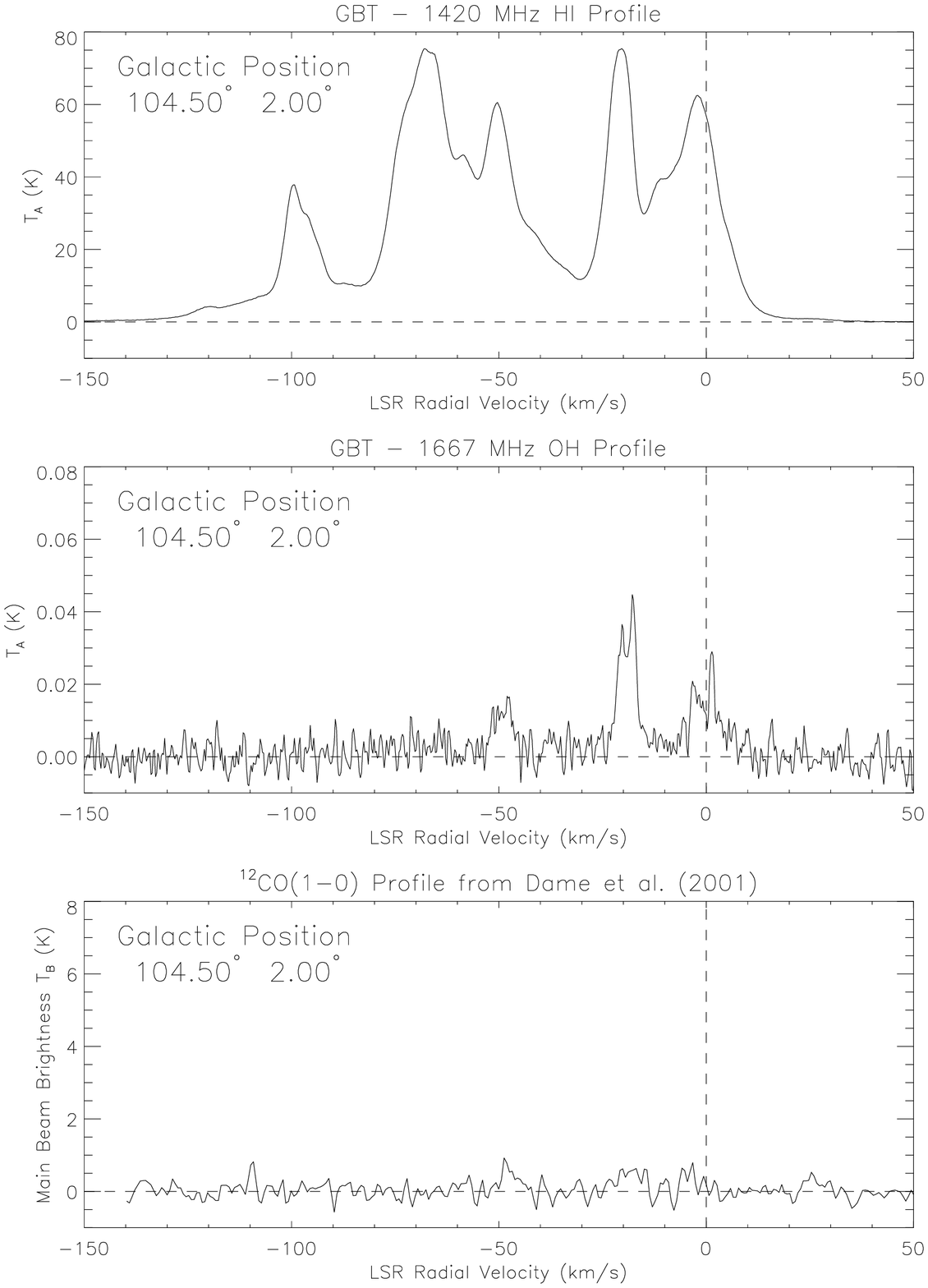}
\vspace{-0.1in}
\caption{\small c). As in Fig \ref{fig:stackedplots1}a, for $b = +2.00^{\circ}$. \normalsize}
%\end{center}
\vspace{-0.1in}
\end{figure*}
%--------------------------------
%\newpage
% ------------------------------
% Figure 2, PAGE 2
%--------------------------------
\setcounter{figure}{1}  % this will get incremented at the next statement
\begin{figure*}[h!]
%\epsscale{1.0}
\vspace{-0.1in}
%\begin{center}
\includegraphics[width=2.0in, angle=+0]{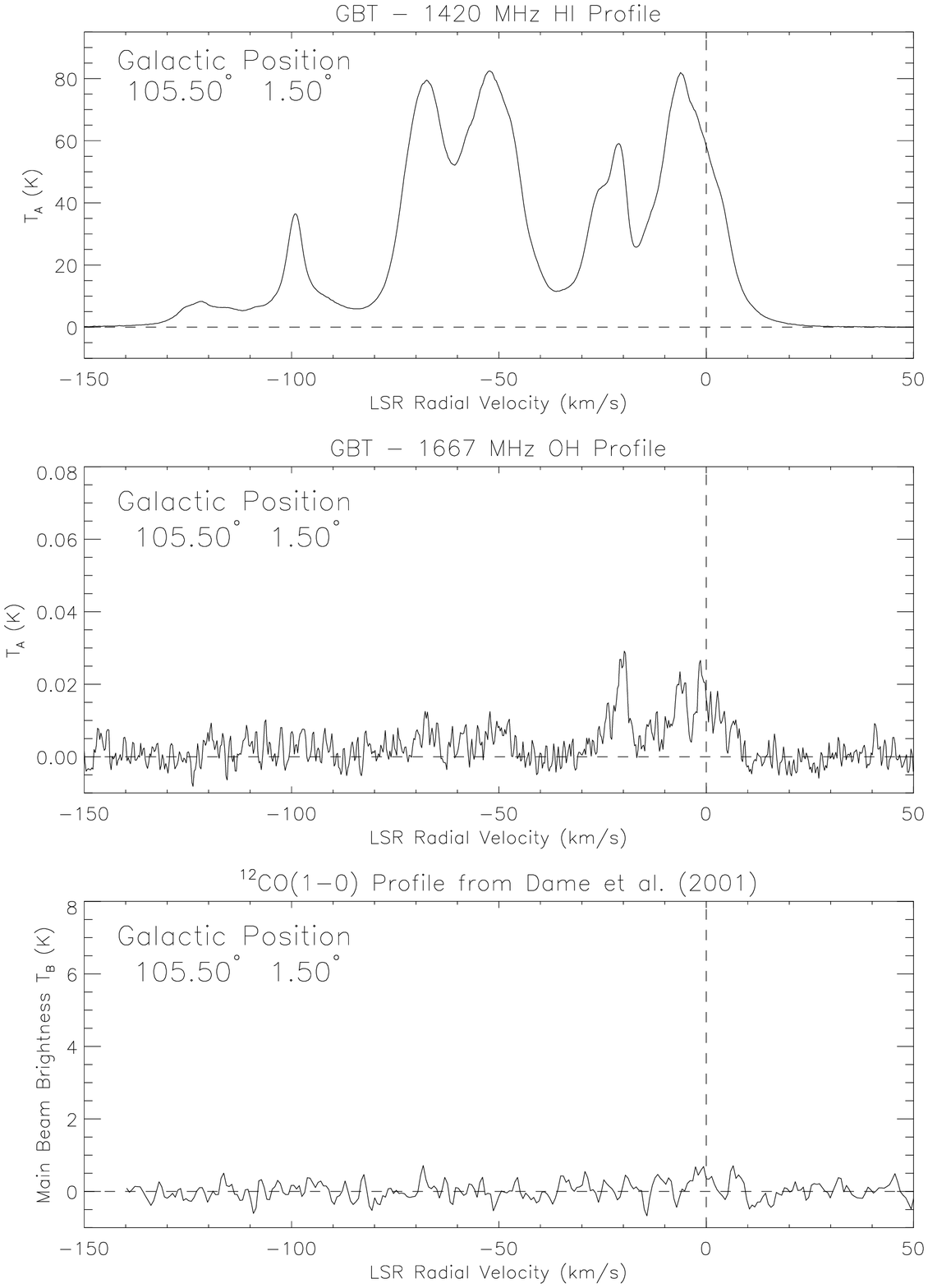}
\includegraphics[width=2.0in, angle=+0]{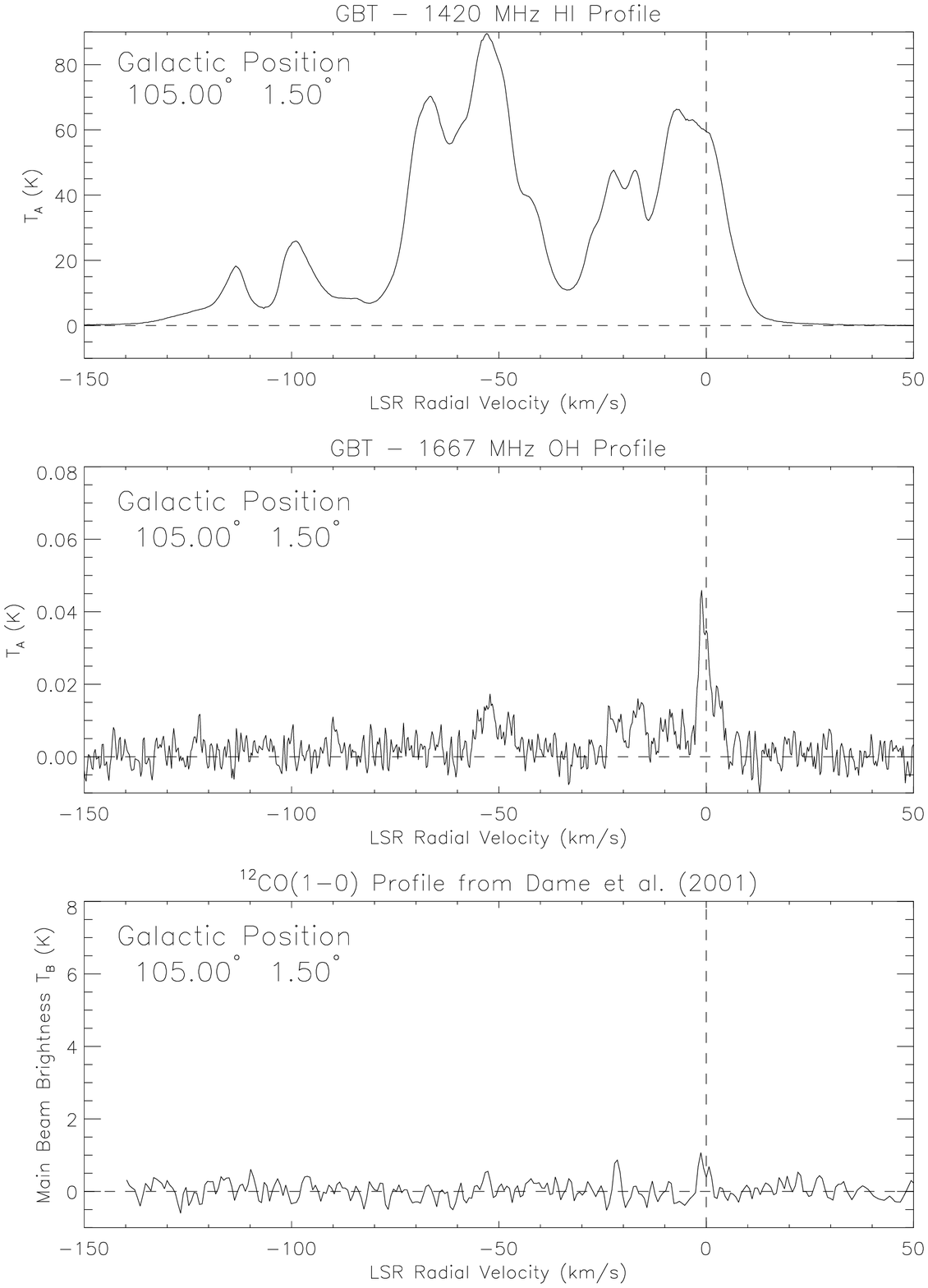}
\includegraphics[width=2.0in, angle=+0]{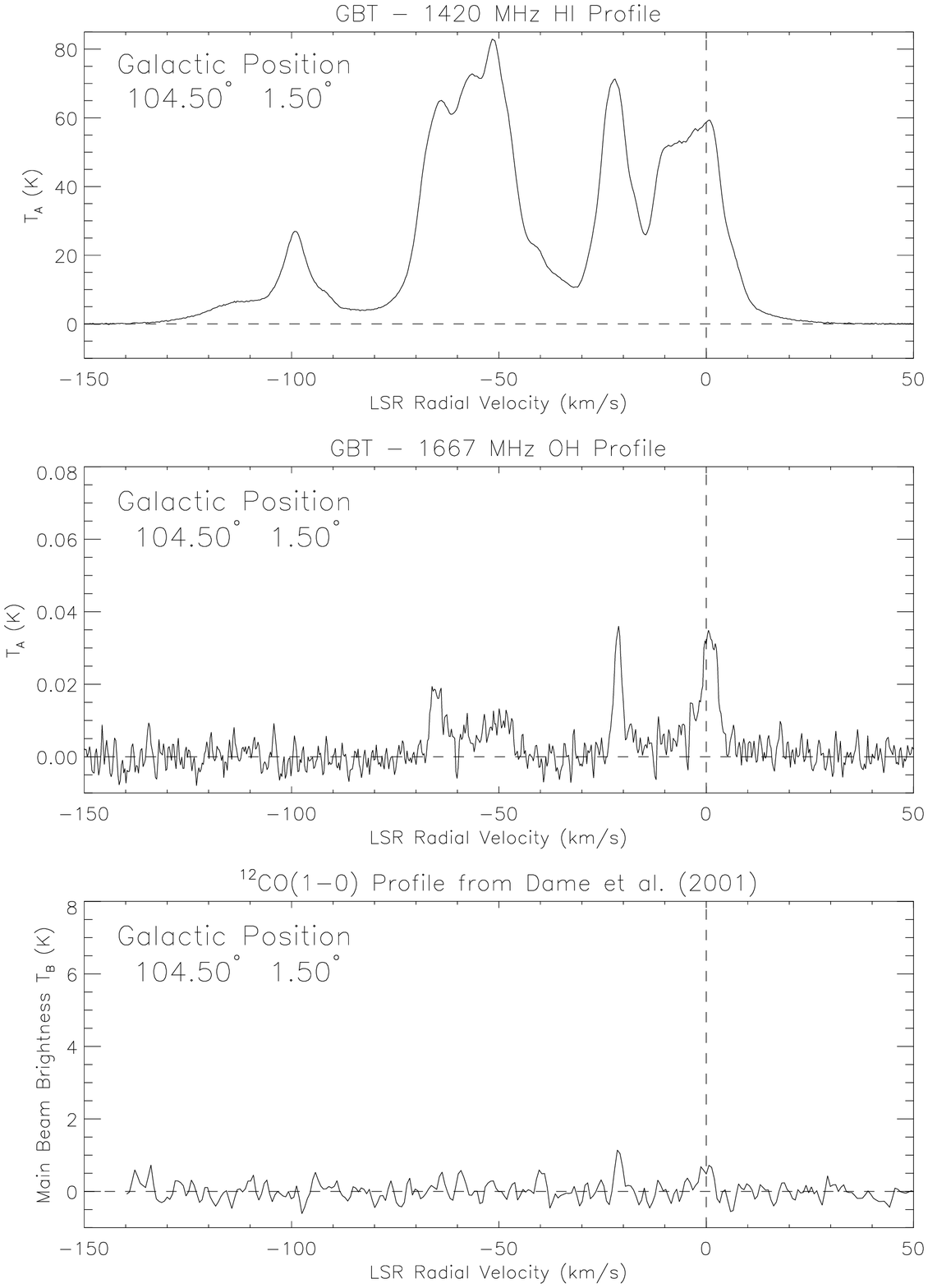}
\vspace{-0.1in}
\caption{\small d). As in Fig \ref{fig:stackedplots1}a, for $b = +1.5^{\circ}$. 
\normalsize}
%\end{center}
\vspace{-0.1in}
\end{figure*}
%-------------------------------- 
\setcounter{figure}{1}  % this will get incremented at the next statement
\begin{figure*}[h!]
%\epsscale{1.0}
%\vspace{-0.1in}
%\begin{center}
\includegraphics[width=2.0in, angle=+0]{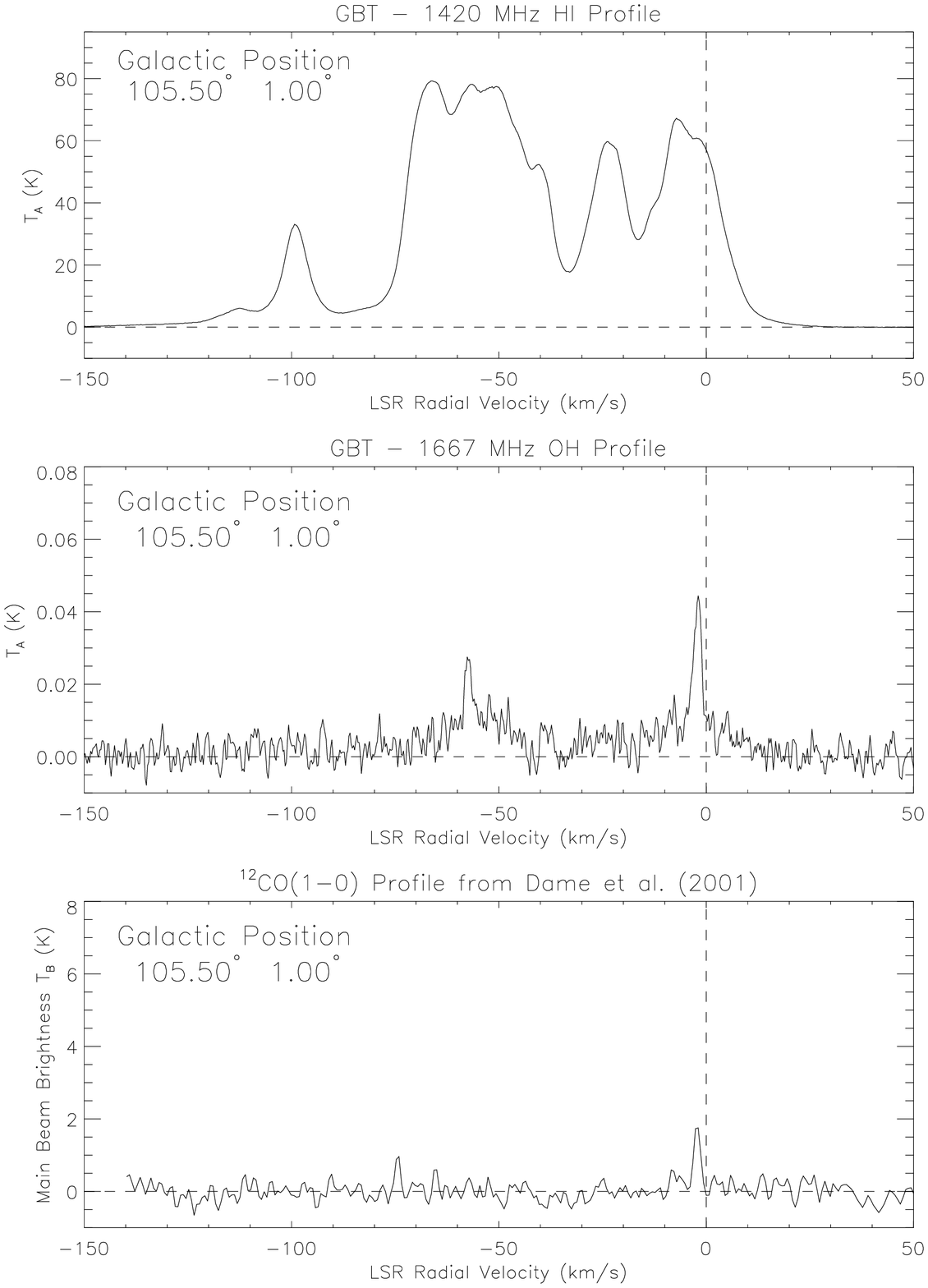}
\includegraphics[width=2.0in, angle=+0]{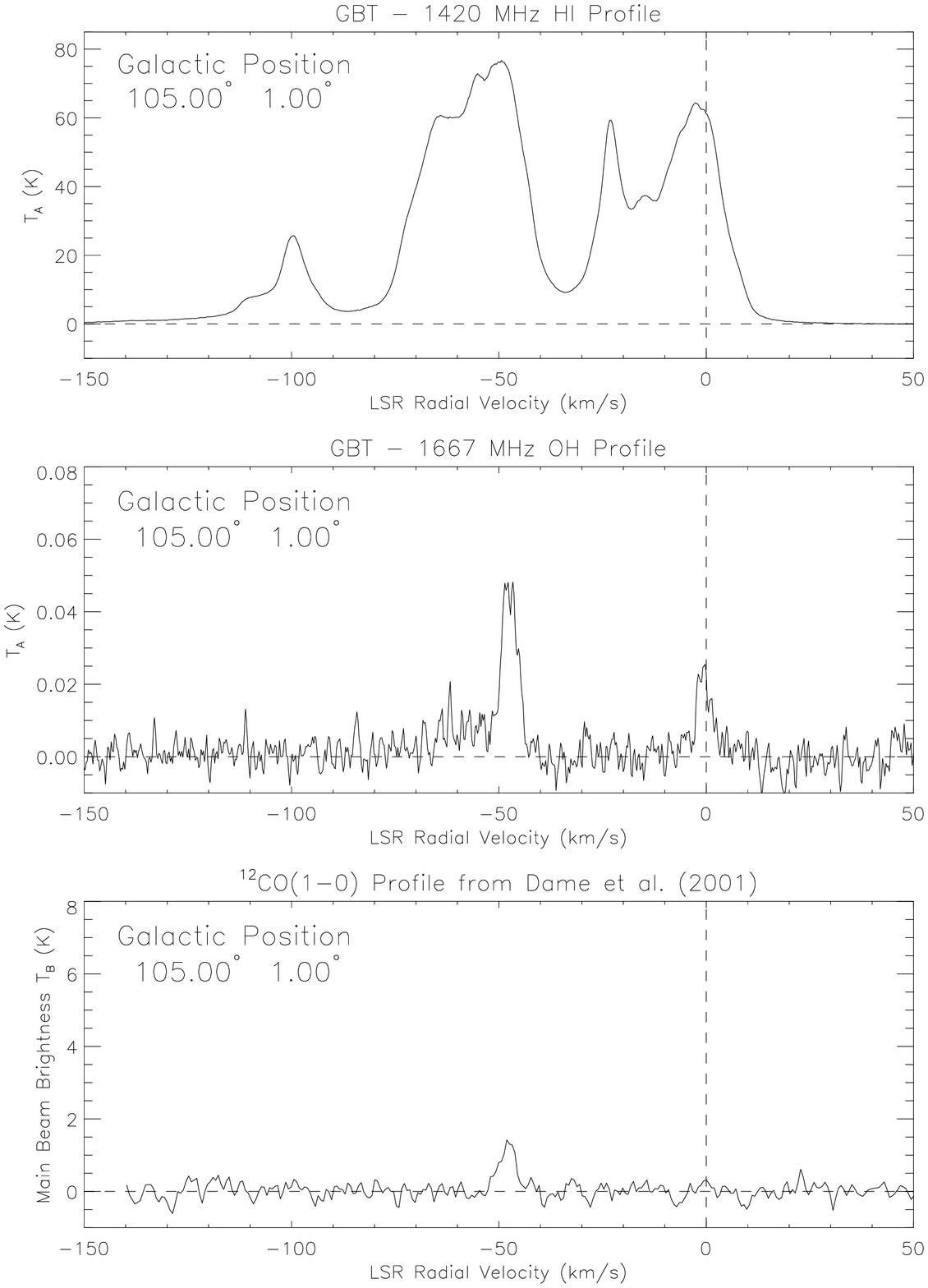}
\includegraphics[width=2.0in, angle=+0]{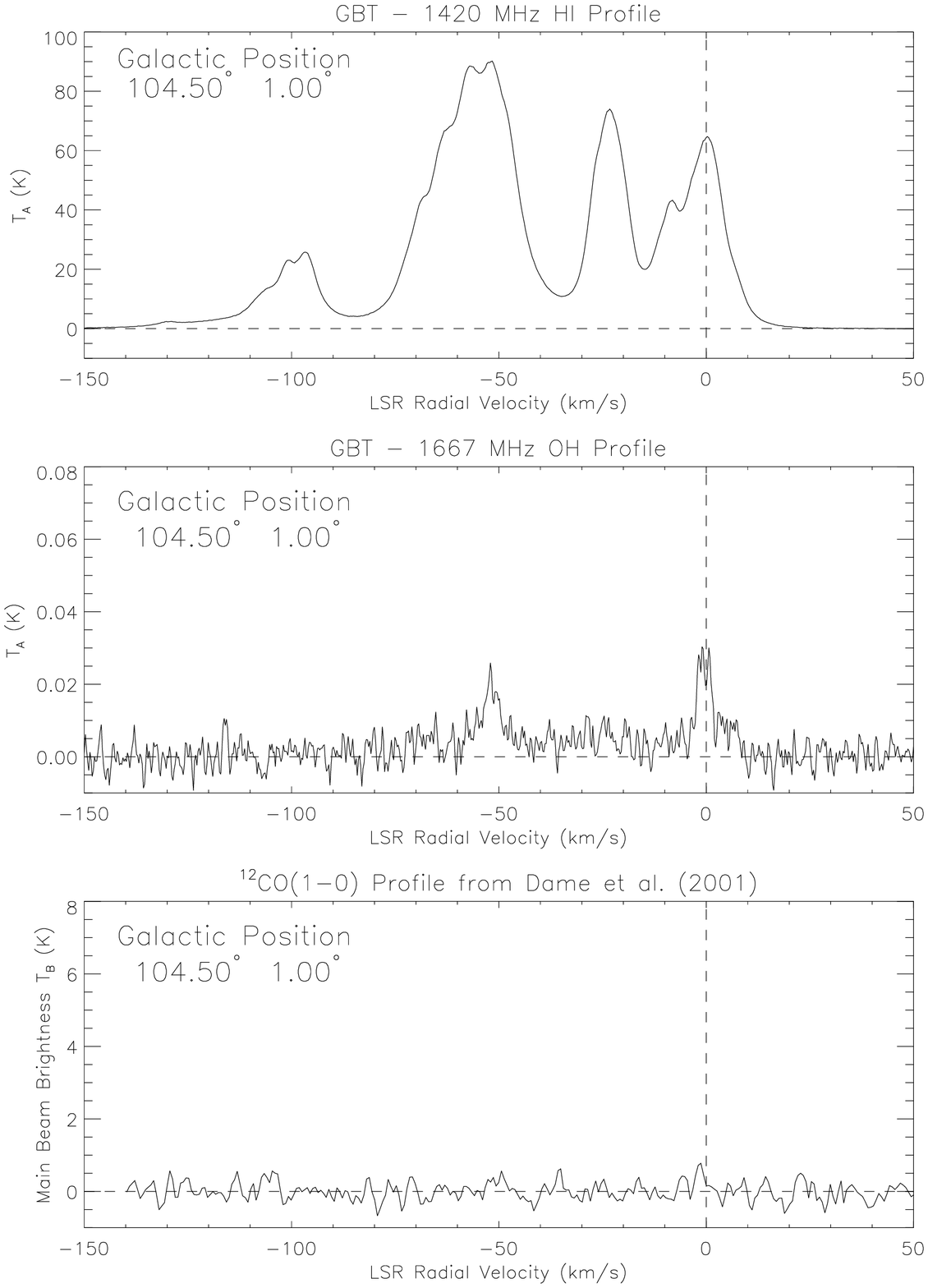}
\vspace{-0.1in}
\caption{\small e). As in Fig \ref{fig:stackedplots1}a, for $b = +1.00^{\circ}$. \normalsize}
%\label{fig:stackedplotsB}
%\end{center}
\vspace{-0.1in}
\end{figure*}
%------------------------------
\setcounter{figure}{1}  % this will get incremented at the next statement
\begin{figure*}[h!]
%\epsscale{1.0}
%\vspace{-0.1in}
%\begin{center}
\includegraphics[width=2.0in, angle=+0]{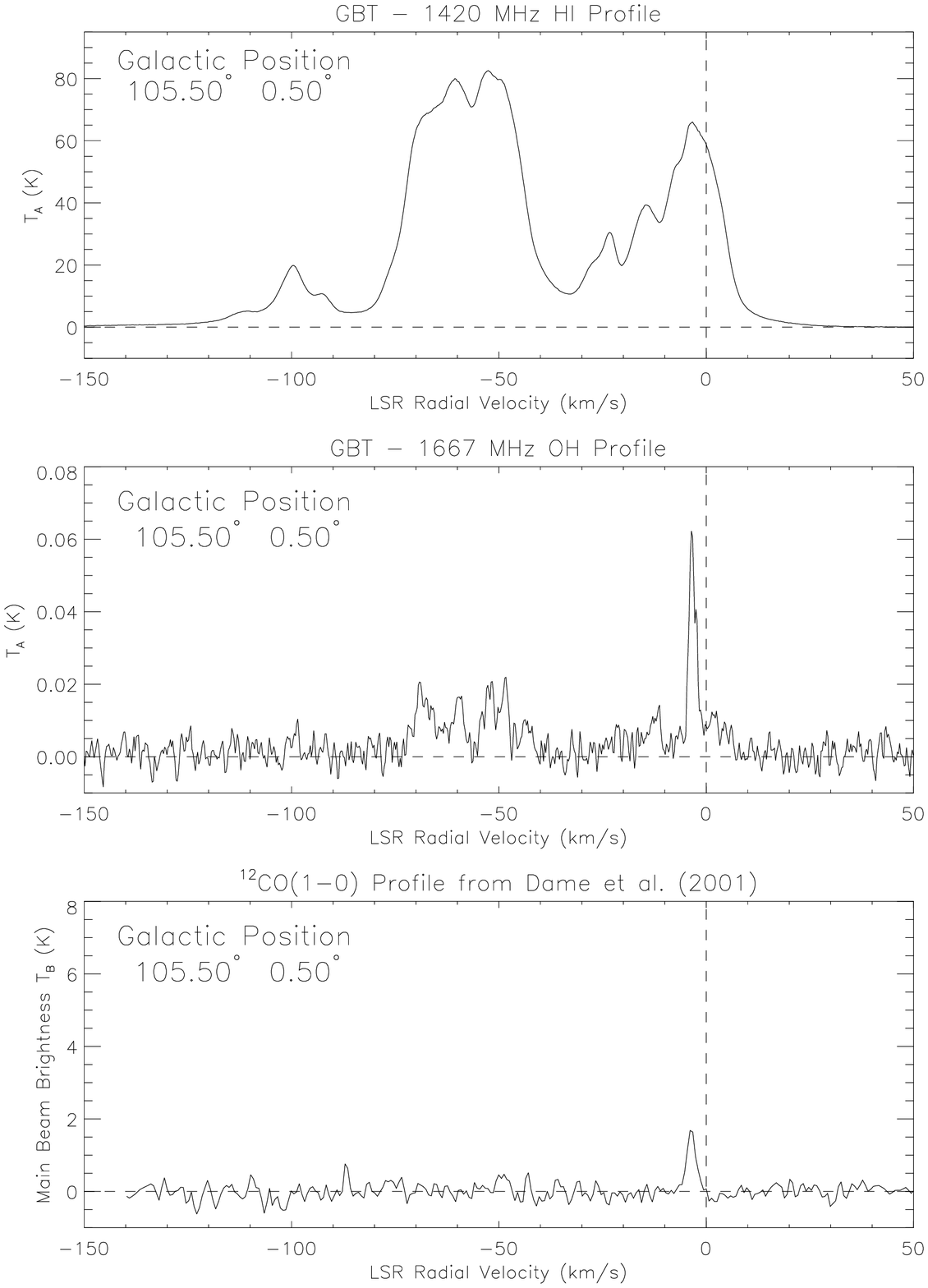}
\includegraphics[width=2.0in, angle=+0]{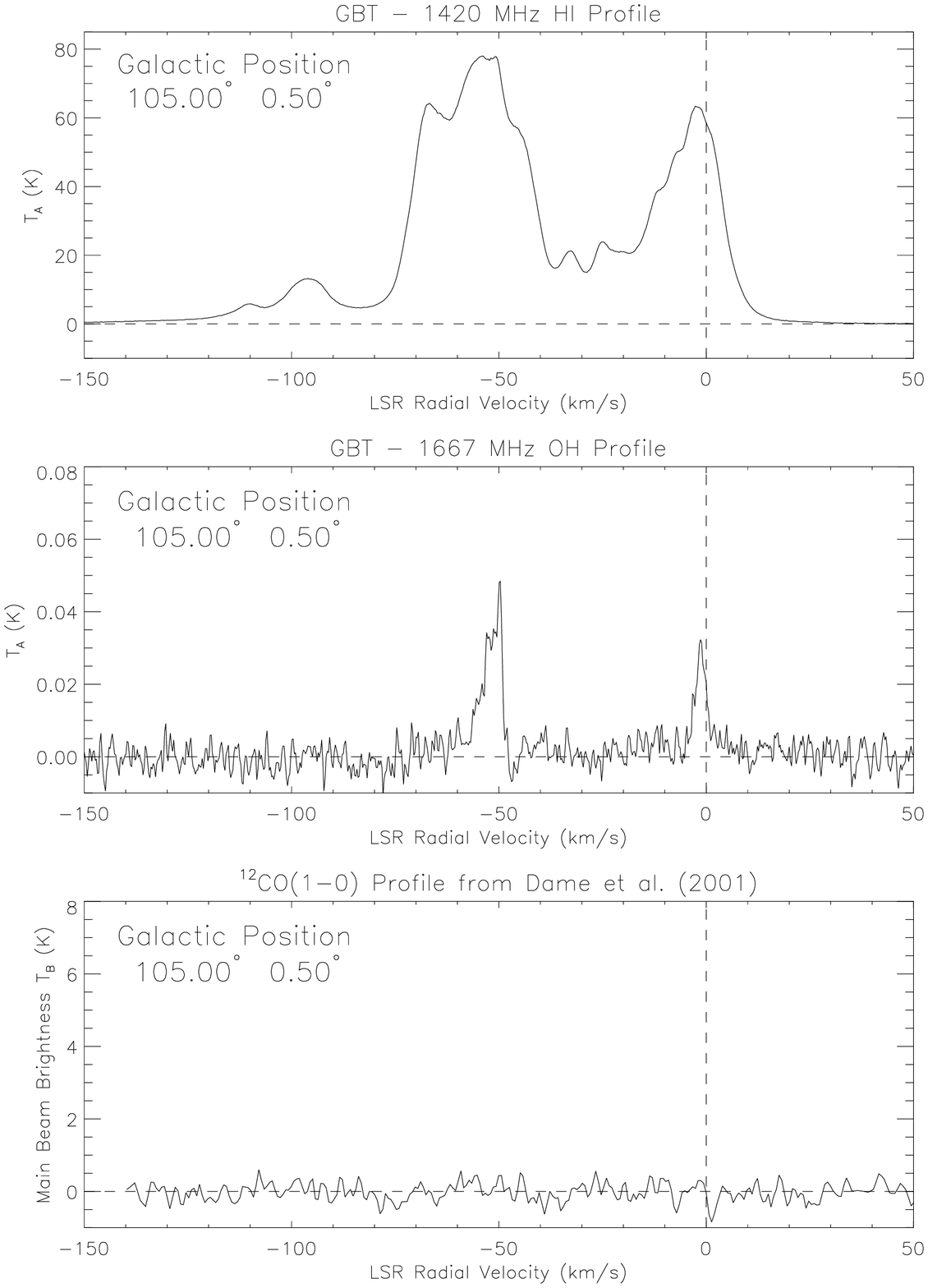}
\includegraphics[width=2.0in, angle=+0]{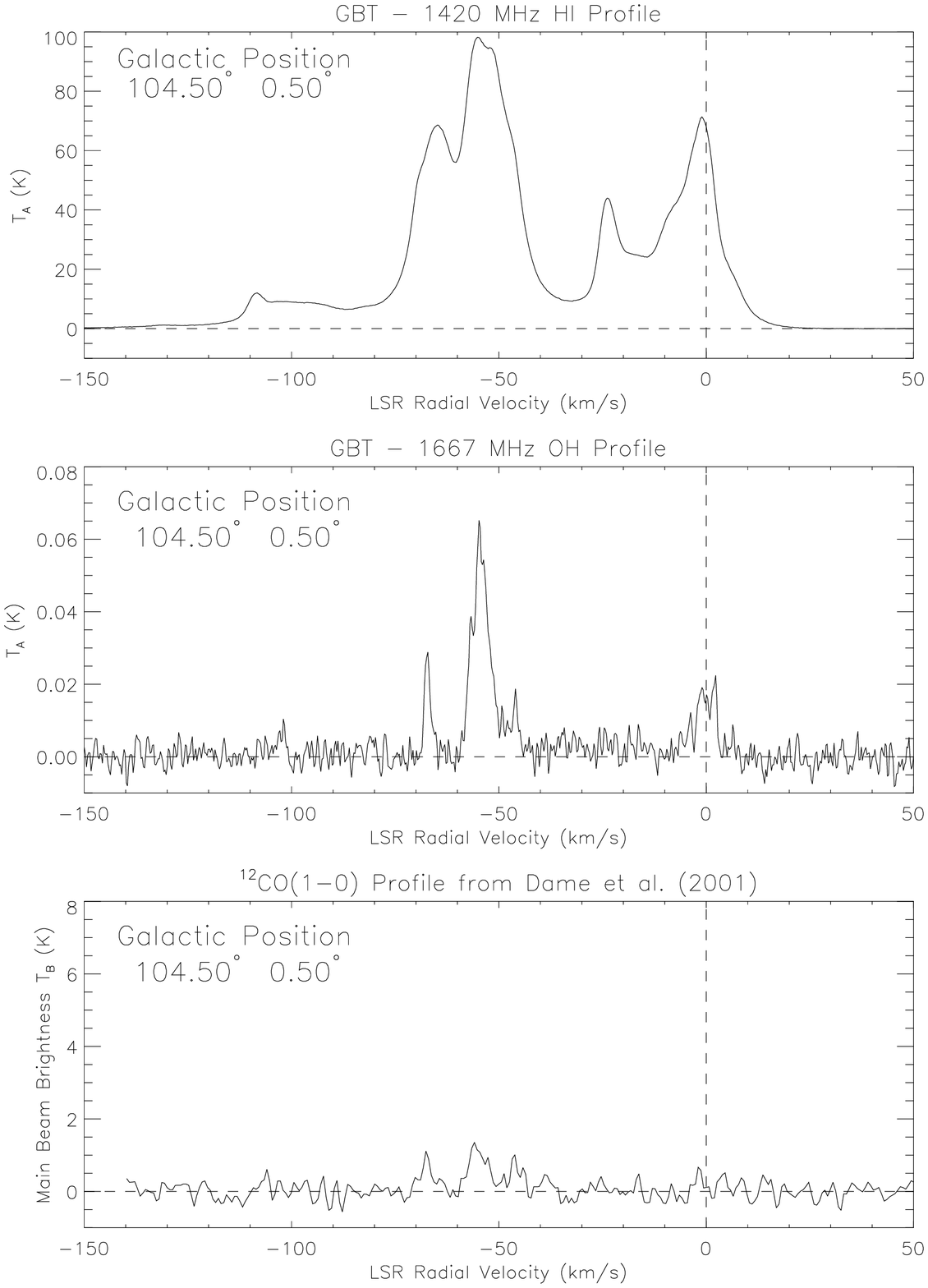}
\vspace{-0.1in}
\caption{\small f). As in Fig \ref{fig:stackedplots1}a, for $b = +0.50^{\circ}$. \normalsize}
%\label{fig:stackedplotsC}
%\end{center}
\vspace{-0.1in}
\end{figure*}
%--------------------------------
%
%--------------------------------
% Figure 2, PAGE 3
%-----------------------------------------------------
\setcounter{figure}{1}  % this will get incremented at the next statement
\begin{figure*}[h!]
%\epsscale{1.0}
\vspace{-0.1in}
%\begin{center}
\includegraphics[width=2.0in, angle=+0]{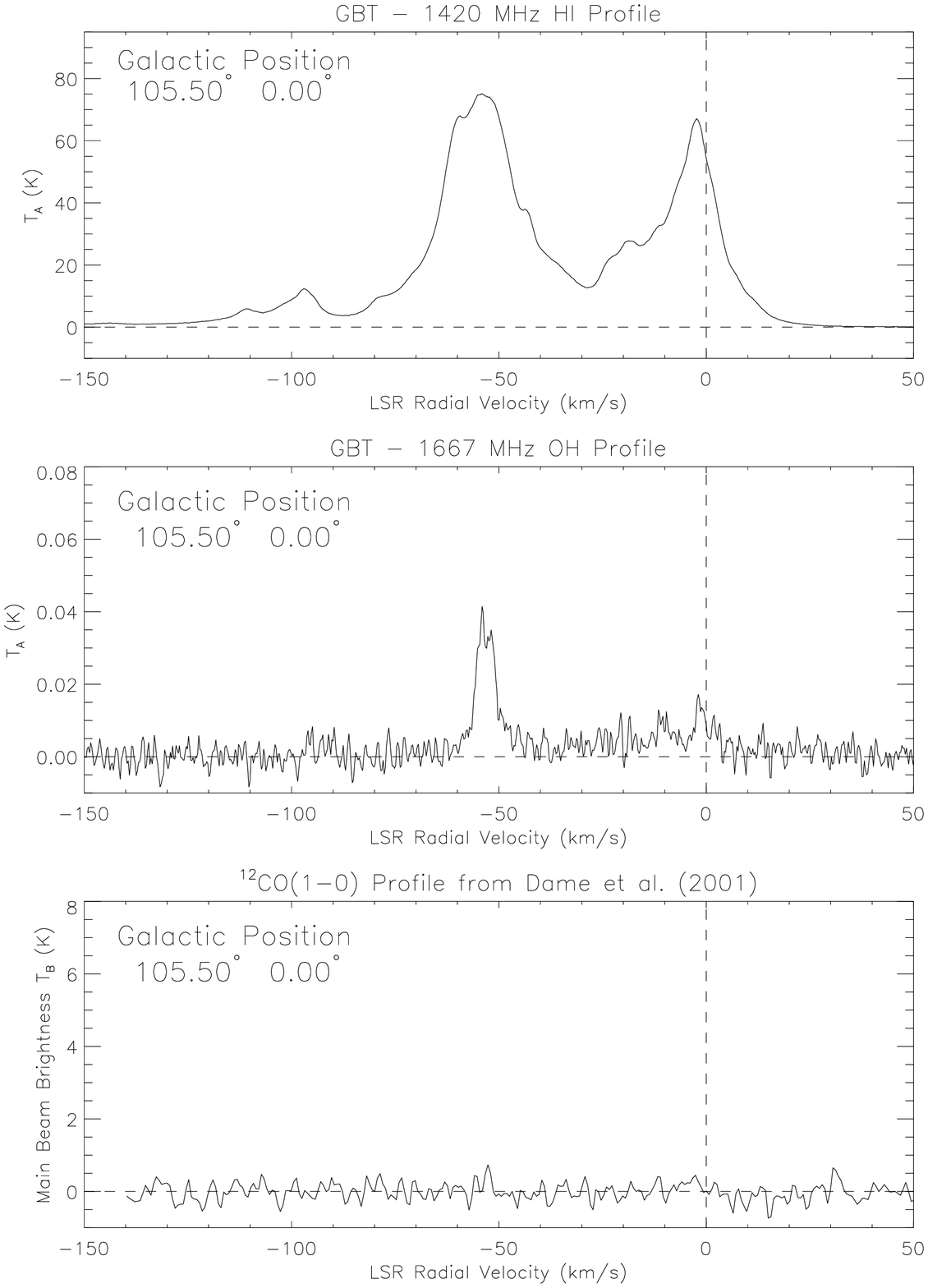}
\includegraphics[width=2.0in, angle=+0]{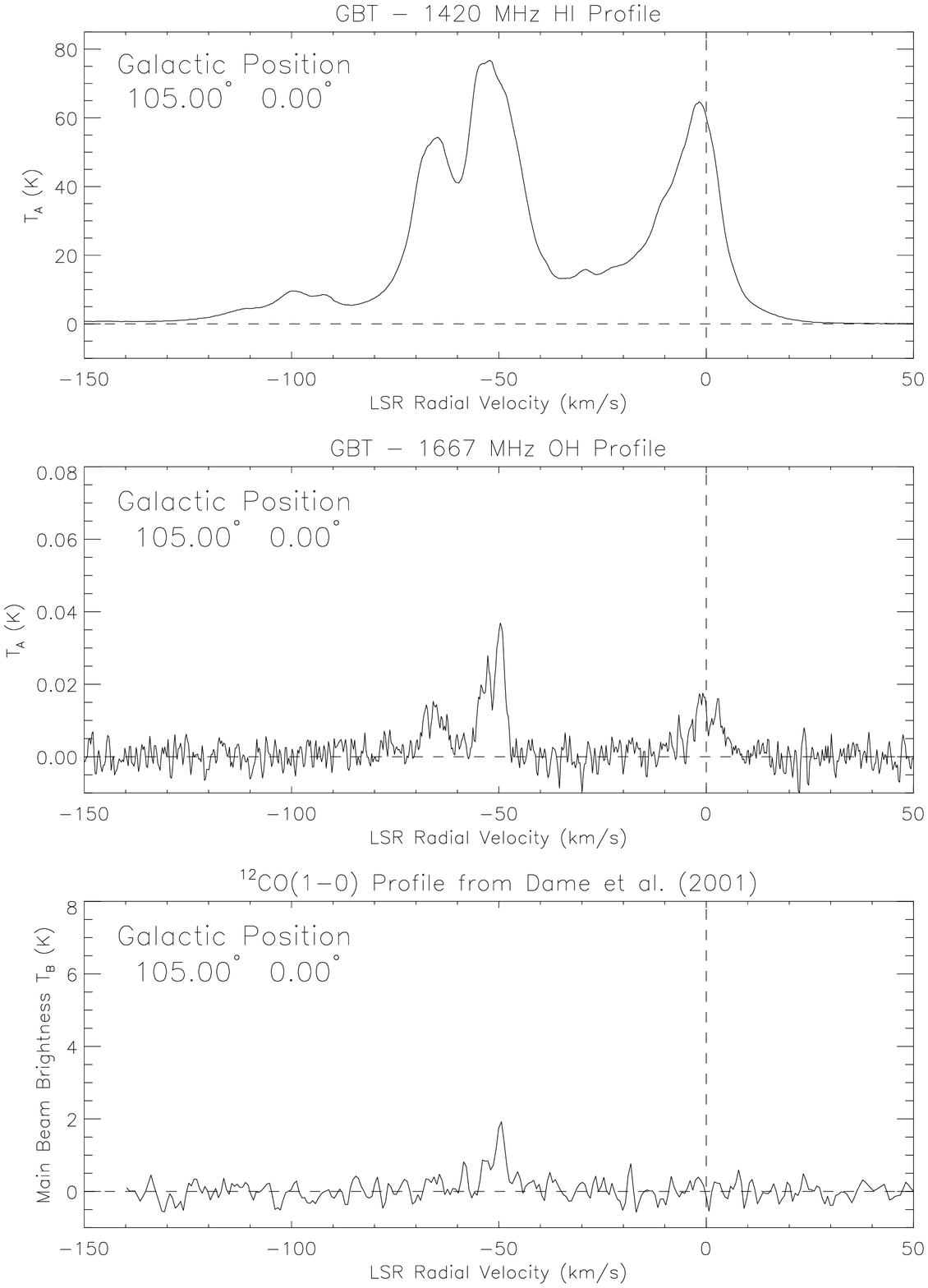}
\includegraphics[width=2.0in, angle=+0]{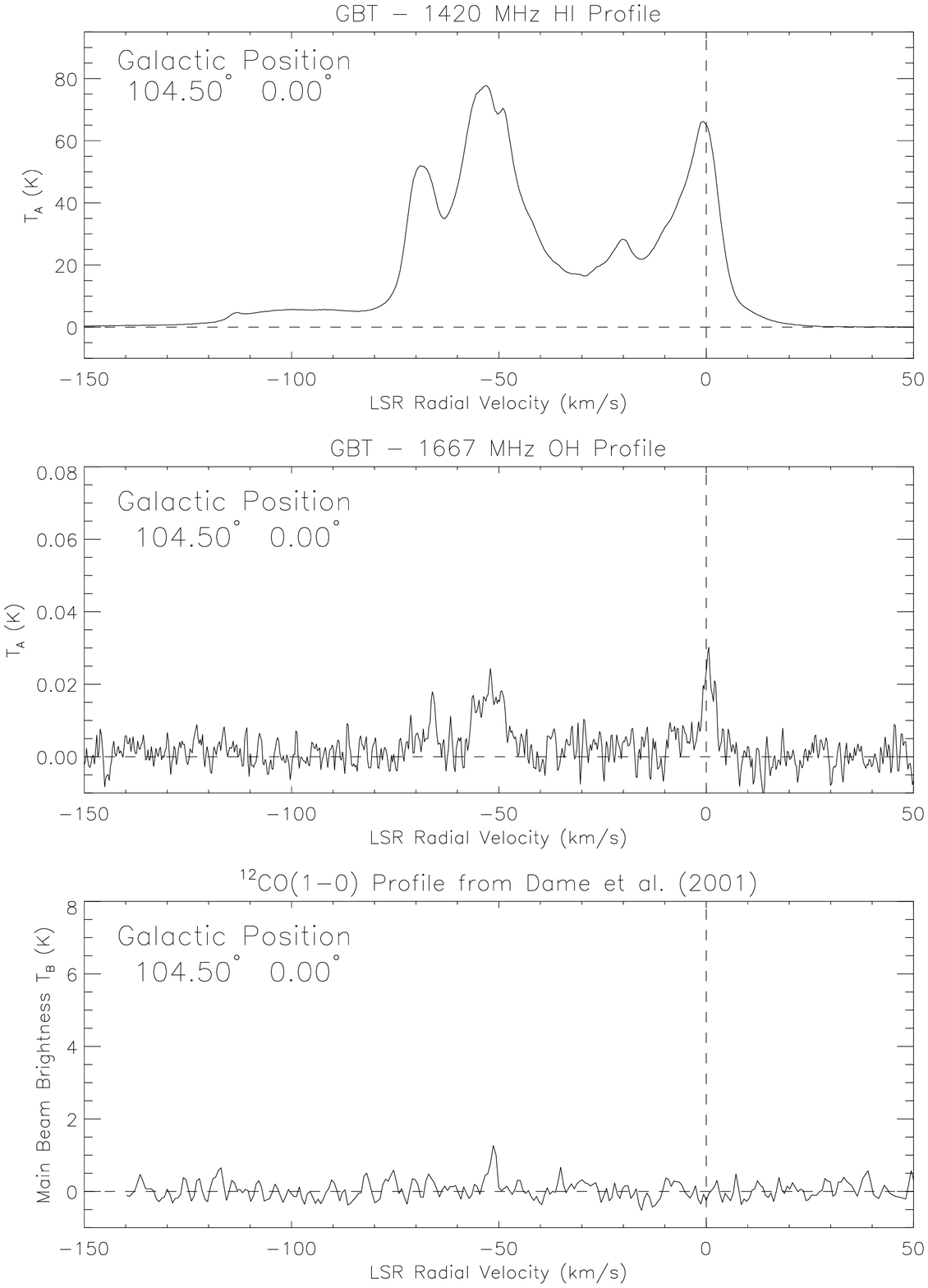}
\vspace{-0.1in}
\caption{\small g). As in Fig \ref{fig:stackedplots1}a, for $b = +0.00^{\circ}$. \normalsize}
%\label{fig:stackedplotmosaics}
%\end{center}
\vspace{-0.1in}
\end{figure*}
%-----------------------------------------------------
\setcounter{figure}{1}  % this will get incremented at the next statement
\begin{figure*}[h!]
%\epsscale{1.0}
%\vspace{-0.1in}
%\begin{center}
\includegraphics[width=2.0in, angle=+0]{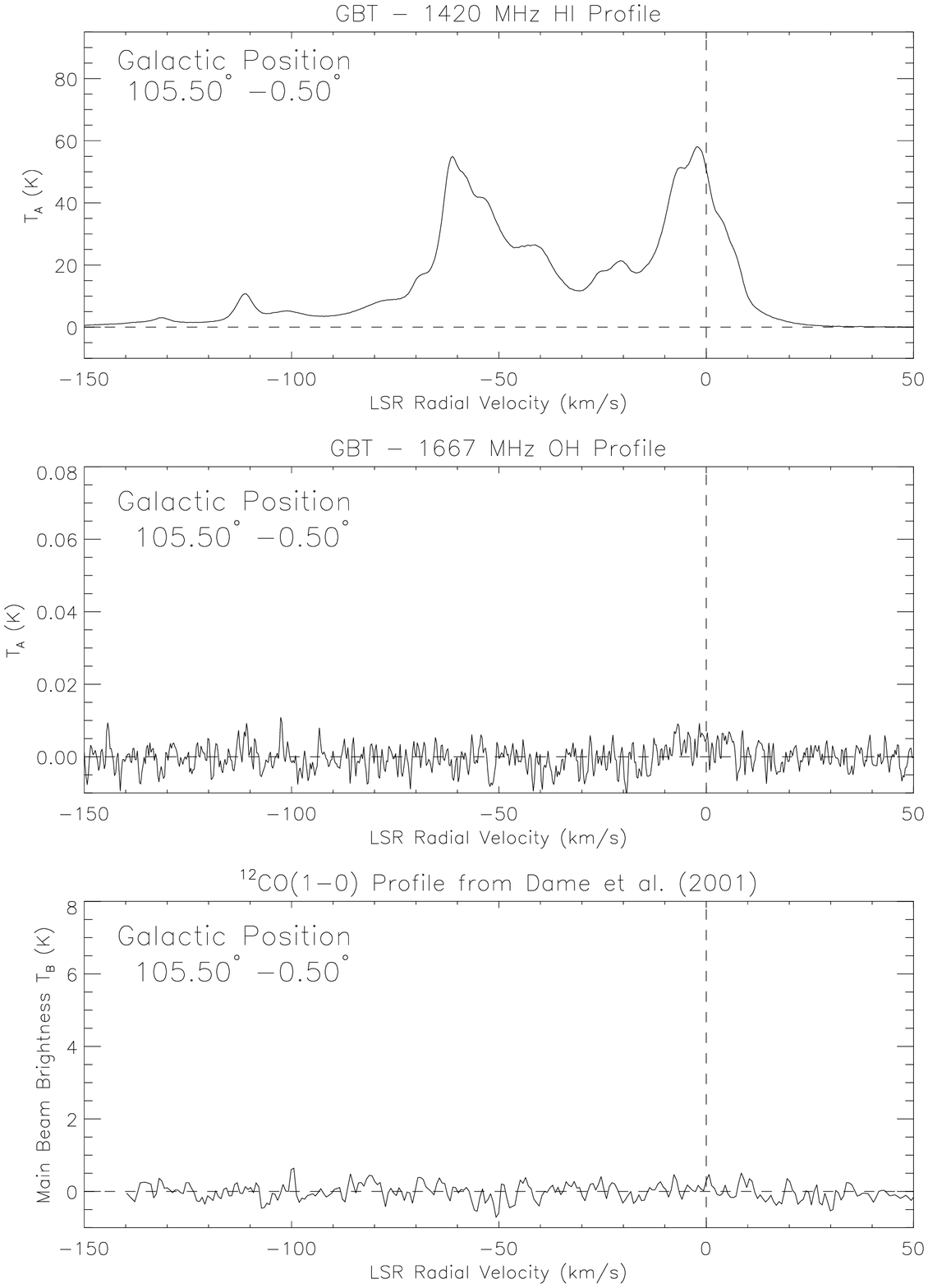}
\includegraphics[width=2.0in, angle=+0]{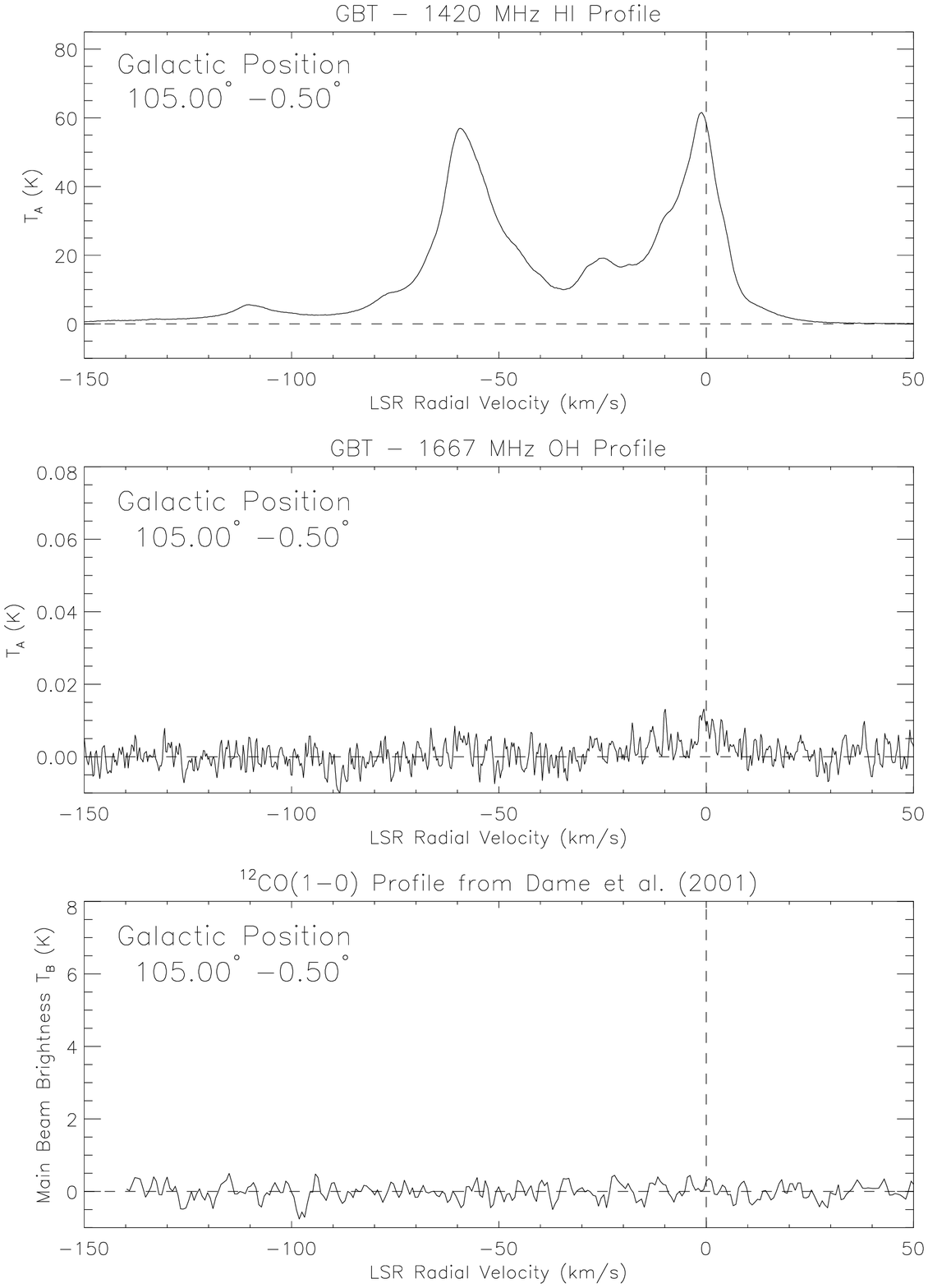}
\includegraphics[width=2.0in, angle=+0]{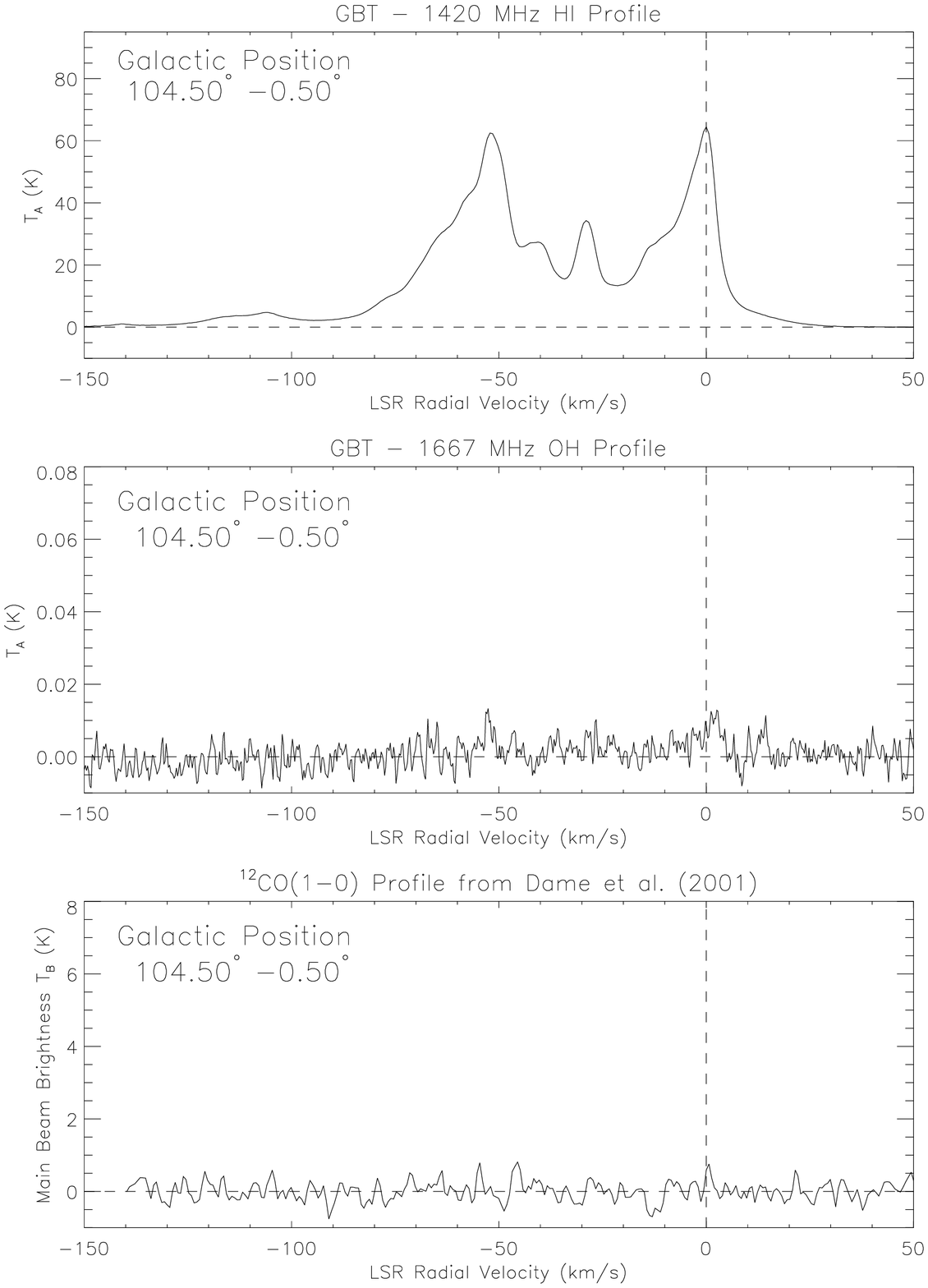}
\vspace{-0.1in}
\caption{\small h). As in Fig \ref{fig:stackedplots1}a, for $ = -0.50^{\circ}$.  \normalsize}
%\label{fig:stackedplotsB}
%\end{center}
\vspace{-0.1in}
\end{figure*}
%------------------------------------------------------
\setcounter{figure}{1}  % this will get incremented at the next statement
\begin{figure*}[h!]
%\epsscale{1.0}
%\vspace{-0.1in}
%\begin{center}
\includegraphics[width=2.0in, angle=+0]{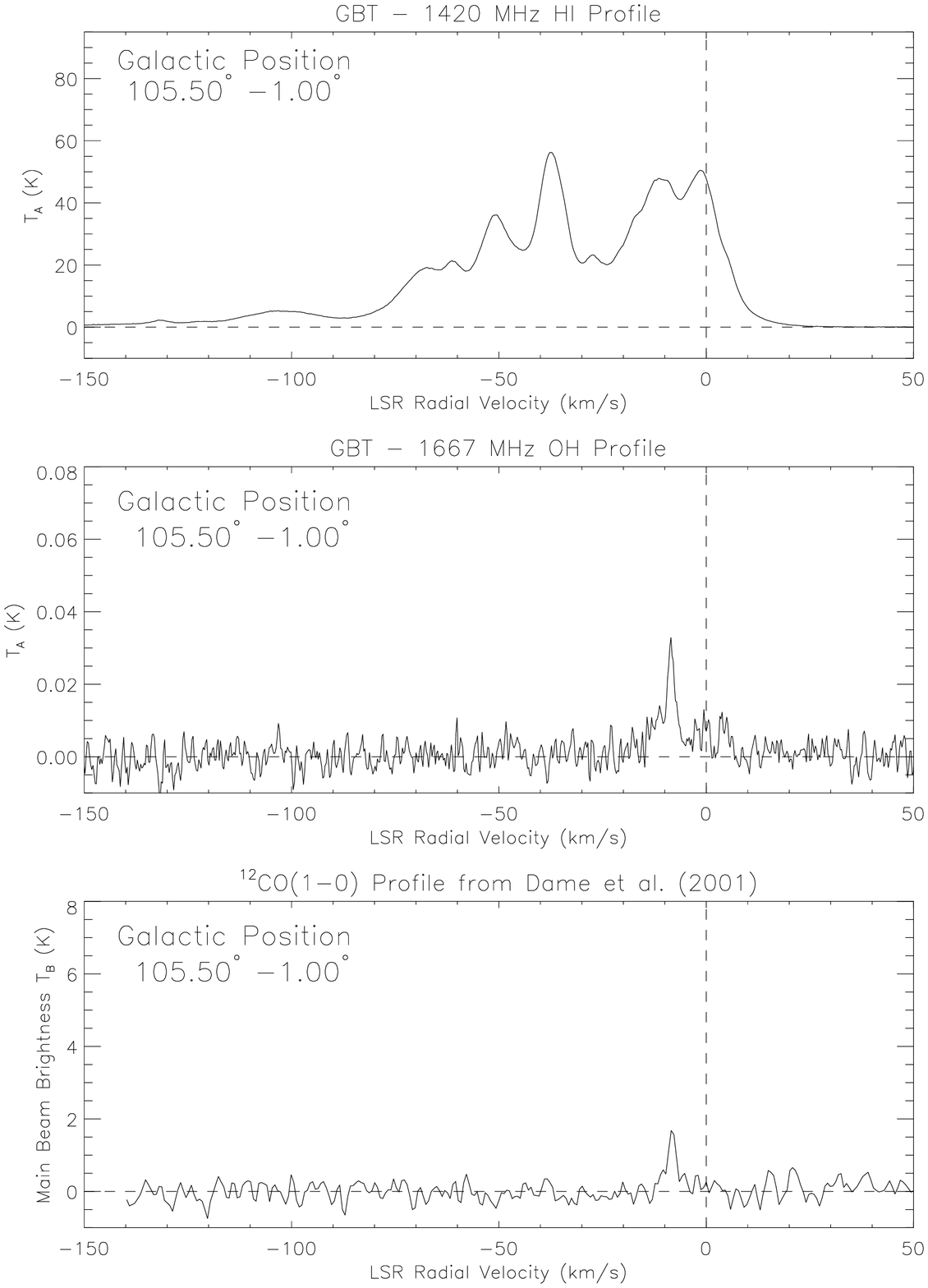}
\includegraphics[width=2.0in, angle=+0]{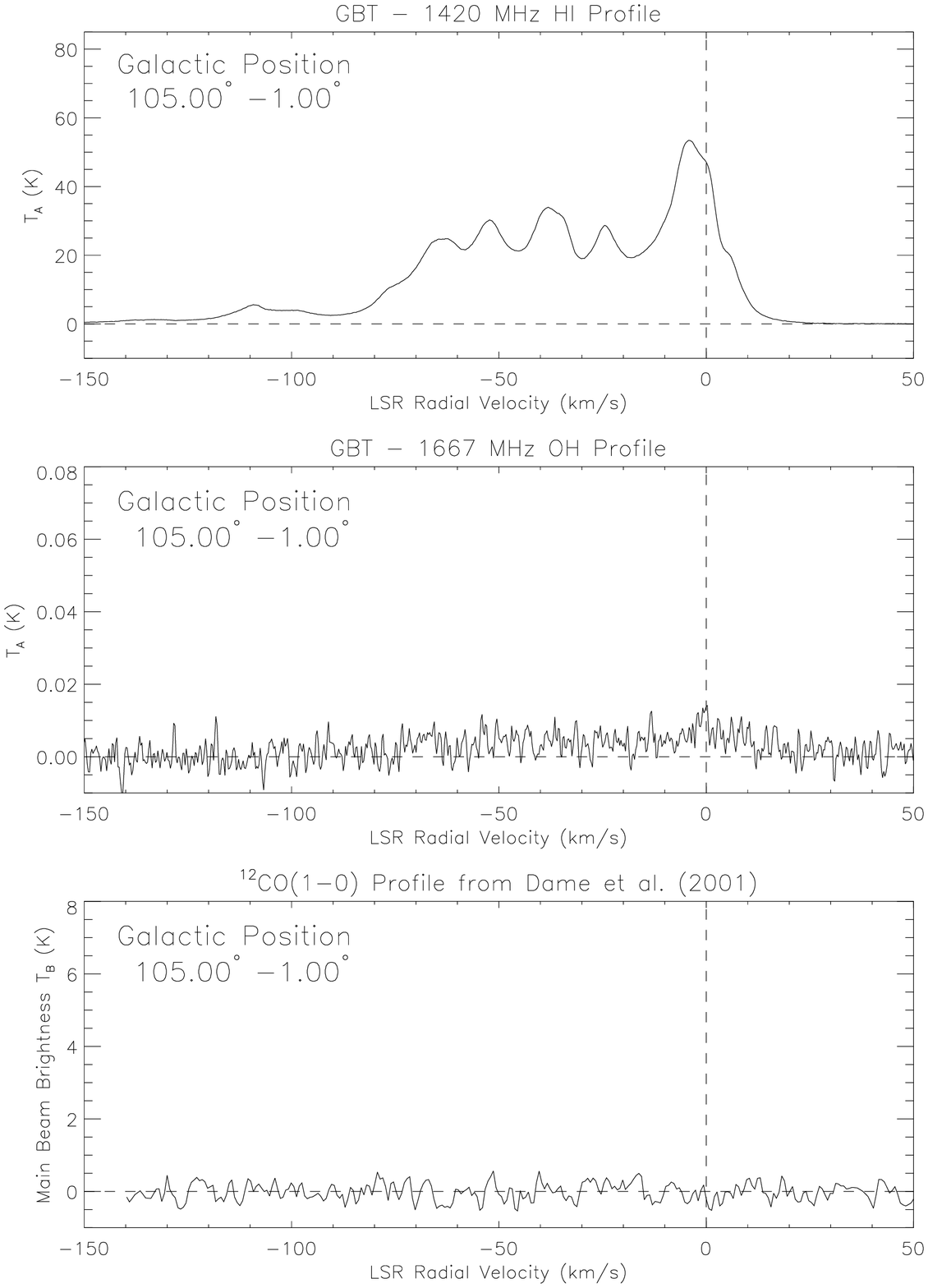}
\includegraphics[width=2.0in, angle=+0]{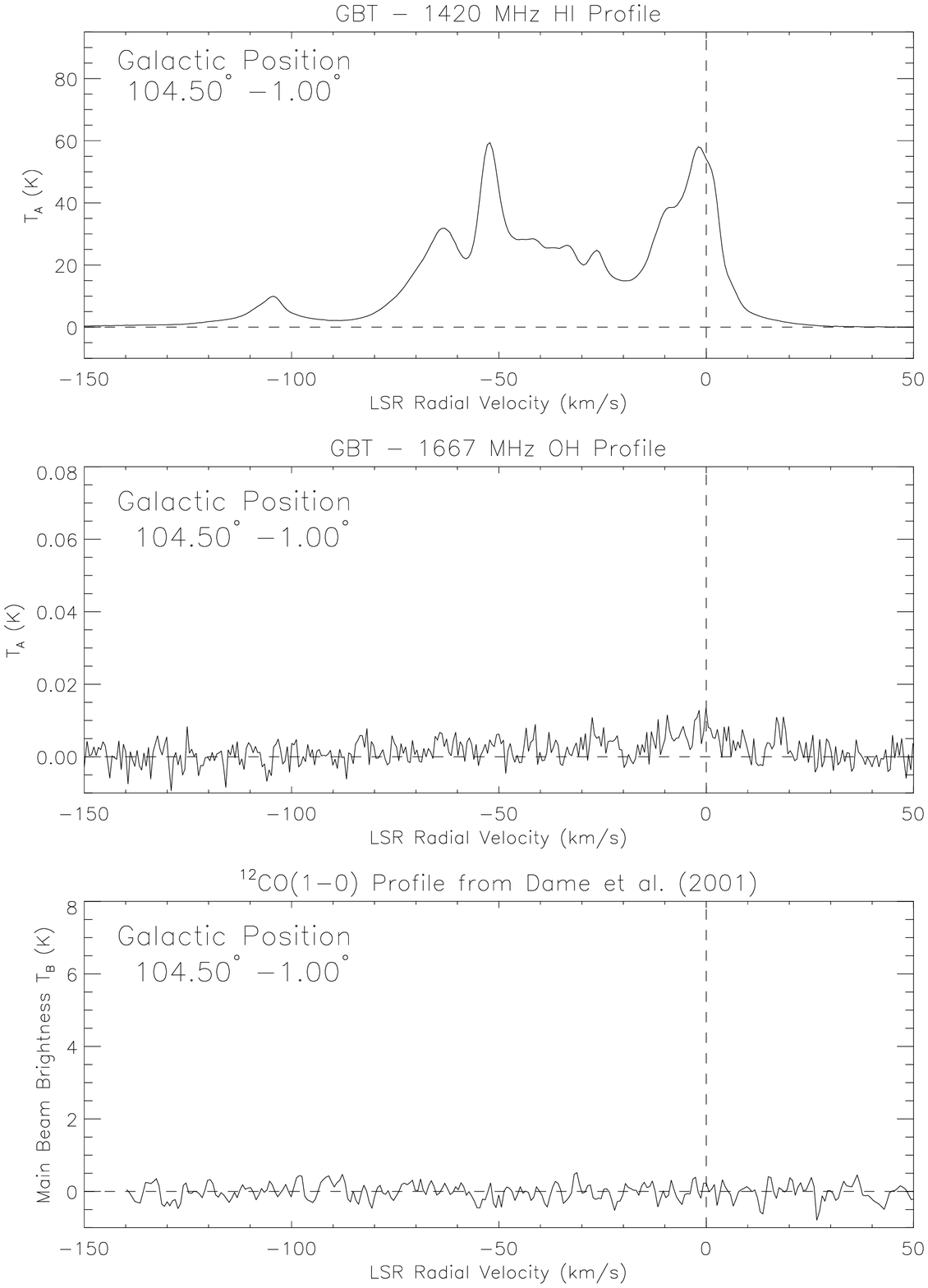}
\vspace{-0.1in}
\caption{\small i). As in Fig \ref{fig:stackedplots1}a, for $b = -1.00^{\circ}$. \normalsize}
%\label{fig:stackedplotsC}
%\end{center}
\vspace{-0.1in}
\end{figure*}
%--------------------------------
\end{document}

%% file: mycommands.tex
\newcommand{\HII}{\mbox{H\,{\sc ii}}}
\newcommand{\HI}{\mbox{H\,{\sc i}}}

\newcommand{\Htwo}{H$_{2}$}

\newcommand{\twCO}{$^{12}$CO}

\newcommand{\degrees}{$^{\circ}$}

\newcommand{\kmps}{\mbox{${\rm km\;s^{-1}}$}}

\newcommand{\lsim}{\mbox{$\mathrel{\vcenter{\hbox{\ooalign{\raise3pt\hbox{$<$}\crcr \lower3pt\hbox{$\sim$}}}}}$}}
\newcommand{\gsim}{\mbox{$\mathrel{\vcenter{\hbox{\ooalign{\raise3pt\hbox{$>$}\crcr \lower3pt\hbox{$\sim$}}}}}$}}